\documentclass[prd, reprint, superscriptaddress, 9pt, longbibliography]{revtex4-1}
\usepackage{gpkmac}

\begin{document}

    \title{Measurement-driven entanglement transition in hybrid quantum circuits}
    \author{Yaodong Li}
    \affiliation{Department of Physics, University of California, Santa Barbara, CA 93106, USA}

    \author{Xiao Chen}
    \affiliation{Kavli Institute for Theoretical Physics, University of California, Santa Barbara, CA 93106, USA}

    \author{Matthew P. A. Fisher}
    \affiliation{Department of Physics, University of California, Santa Barbara, CA 93106, USA}
    
    \date{January 23, 2019}

\begin{abstract}
In this paper we continue to explore ``hybrid" quantum circuit models in one-dimension with both unitary and measurement gates, focussing on the entanglement properties of wavefunction trajectories at long times, in the steady state.  
We simulate a large class of Clifford circuits, including models with or without randomness in the unitary gates, and with or without randomness in the locations of measurement gates, using stabilizer techniques to access the long time dynamics of systems up to 512 qubits.
In all models we find a volume law entangled phase for low measurement rates, which exhibits a sub-dominant logarithmic behavior in the entanglement entropy, $S_A = \alpha \ln |A| +  s |A|$, with sub-system size $|A|$.
With increasing measurement rate the volume law phase is unstable to a disentangled area law phase, passing through a single entanglement transition at a critical rate of measurement.
At criticality we find a purely logarithmic entanglement entropy, $S_A = \alpha(p_c) \ln|A|$,
a power law decay and conformal symmetry of the mutual information, with exponential decay off criticality.
Various spin-spin correlation functions also show slow decay at criticality. 
Critical exponents are consistent across all models, indicative of a single universality class.
These results suggest the existence of an effective underlying statistical mechanical model for the entanglement transition.
Beyond Clifford circuit models, numerical simulations of up to 20 qubits give consistent results.

    \end{abstract}

    \maketitle


   \tableofcontents        


\section{Introduction \label{sec1}}

Quantum many-body systems under unitary dynamics will generally thermalize~\cite{Deutsch1991, Srednicki1994, Calabrese_2005, Calabrese_2007, rigol2008, Kim2013, Mezei2017}.  But is thermalization inevitable?
Are there systems in which the thermalization of entanglement entropy is avoidable?
One example is many-body localization~\cite{Nandkishore2015, 2018arXiv180411065A}, in which entanglement growth is suppressed by strong quenched disorder.
Repeated local measurements provide an alternative approach for taming the growth of entanglement.
While unitary dynamics tends to increase entanglement, local measurements tend to disentangle.
When measurements are made continually, the steady-state wavefunction should exhibit non-maximal, and non-thermal, entanglement entropy~\cite{cao2018monitoring}.
If measurements are made as frequently as possible, the wavefunction will become localized in the Hilbert space near a trivial product state -- a quantum Zeno effect~\cite{Misra1977zeno}.
What happens in the intermediate regime when measurements are made at a small but finite rate?  Can the volume law scaling of entanglement entropy survive in the presence
of a non-zero rate of measurement?
These questions are pertinent to our basic understanding of quantum information dynamics.

Recently, in Refs.~\cite{nandkishore2018hybrid, nahum2018hybrid,li2018hybrid}, a prototypical (1+1)d circuit model with both unitary dynamics and projective measurements was introduced and explored.
Local unitary gates acted on all neighboring qubits, while
single (or two-) qubit measurement gates were sprinkled throughout the circuit,
with each space-time point occupied with probability $p$, representing the strength of the measurements.  In Ref.~\cite{nandkishore2018hybrid} it was argued that the volume law entangled phase is destroyed by arbitrary rare measurements, for all $p>0$, while the authors in Refs.~\cite{nahum2018hybrid,li2018hybrid}
presented arguments and numerical evidence for a stable volume law entangled phase,
separated from an area law entangled phase at a critical value of measurements, $p_c>0$.
Due to different approaches taken in these papers, a direct comparison was not immediate.


In this paper, we continue to investigate these hybrid circuit models with unitary-measurement dynamics.
Our goal is to further explore and characterize both the nature of the entanglement transition and the properties of the volume law entangled phase in the presence of weak measurements.
A central focus is on generic circuits with randomness in both the unitary gates and in the locations of the measurement gates.
The least constrained model we consider is a ``random Haar circuit'',
with 2-qubit unitaries taken from the Haar measure~\cite{mehta2004matrices, loggasrandommatrices} and single qubit measurements randomly scattered across the circuit~\cite{nandkishore2018hybrid, nahum2018hybrid}.
However, the 
high entanglement in the volume law phase poses formidable numerical challenges even in one dimension.  We thus will largely study  ``random Clifford circuits" with the  Haar unitaries replaced by random two-qubit Clifford unitaries, and the single qubit measurements restricted to the Pauli group~\cite{gottesman9604hamming, gottesman9807heisenberg, aaronson0406chp}.
Such Clifford circuits can be efficiently simulated on a classical computer, enabling us to  perform extensive large scale numerical studies.
We draw several conclusions from our data in the random Clifford circuit:
\envelope{itemize}{
\item
At long times, measurements reduce the entanglement entropy from maximal, and the steady-state entanglement fluctuates weakly over time and over circuit realizations, independent of the initial conditions.
These ``typical" steady states are non-thermal, qualitatively distinct from thermal states.

\item
The volume law phase persists when measurements are infrequent, consistent with results from Refs.~\cite{nahum2018hybrid, li2018hybrid}.
The algebraic structure of the Clifford dynamics provides a convenient framework for characterizing the entanglement structure of these wavefunctions, revealing an unusual scaling form of the entanglement entropy, namely $S_A = \alpha \ln |A| + s |A|$ for a contiguous subsystem $A$.
The sub-leading logarithm is exposed by analyzing the length distributions of the ``stabilizers" -- mutually commuting Pauli string (eigen)operators of the Clifford wavefunctions with unit eigenvalue.
The stabilizer distribution is ``bimodal", consisting of a power law distribution of ``short'' stabilizers that contribute to the logarithm, and ``long stabilizers'' with length $\ell \approx L/2$ giving the volume law piece ($L$ being the system size).
This logarithmic correction is conjectured to be a generic feature of volume law steady states in the presence of measurements.

\item
The  ``entanglement transition'', from volume law to area law states~\cite{nahum2018hybrid, li2018hybrid}, occurs when the weight 
under the ``long stabilizer" peak at $\ell \approx L/2$ vanishes continuously upon approaching $p_c$ from below.  Remarkably, the power law tail of ``short" stabilizers remains, implying a purely logarithmic form for the entanglement entropy right at the critical point, $p=p_c$.
The entanglement transition exhibits conformal symmetry of the mutual information at criticality, and we extract several critical exponents. In particular, we find that in all the models we study, the mutual information between two small regions separated by a large distance, $r$,  scales as $1/r^4$.
Off criticality the mutual information decays exponentially.
\item
We explore the fluctuations of certain spin-spin correlation functions across the transition, and find that they are enhanced at the critical point, mimicking the mutual information.  
}

We establish the generality of these results
by exploring models with imposed spatial symmetry constraints -- specifically 
Clifford circuits with the unitaries periodic in space and time (Floquet) and/or the measurement locations periodic in space and time.  All models are found to 
 exhibit a measurement-driven entanglement transition, with similar exponents and similar behavior of the stabilizer length distribution as in the random Clifford circuit.
 Apparently the randomness in the unitaries and measurement locations are inessential, with the remaining stochasticity in the measurement outcomes sufficient to account for the presence and universality of the entanglement transition.


Going beyond Clifford, we implement a full quantum simulation of more general circuit models for systems with size up to $L=20$ qubits.
Both random Haar circuits and (non-Clifford) Floquet circuits exhibit behavior consistent with their Clifford counterparts.  
We also explore models with (non-projective) ``generalized measurements'', 
with each and every qubit being measured at each time step,
and find evidence for an entanglement transition, with accessible exponents being consistent 
with the Clifford circuits.
Of particular interest is a space-time translationally symmetric Floquet model with generalized measurements, which exhibits an entanglement transition 
where the only stochasticity is in the results of the quantum measurements.


Motivated by the remarkable consistency between all of our different models, we conjecture that generic hybrid circuits have a volume law phase with logarithmic correction 
for weak enough measurements, and exhibit an entanglement transition in a single universality class.

Our paper is organized as follows.
In Sec.~\ref{sec2} we define the circuit models of interest.  Extensive numerical results for Clifford circuits are reported in Sec.~\ref{sec3} and \ref{sec4}.
In particular, Sec.~\ref{sec3} contains evidence for the phase transition in entanglement entropy, and allows characterization of the volume-law phase in terms of stabilizers. Sec.~\ref{sec4} is devoted to a detailed analysis of the critical behavior of the entanglement transition.
In Sec.~\ref{sec5}, we systematically explore Clifford circuit models with space and time symmetries imposed, either in the unitaries or the measurement locations -- or both.
In Sec.~\ref{sec6}, we consider more generic non-Clifford circuits, establishing complementary results via a full quantum simulation for smaller systems.
We close with discussions in Section~\ref{sec7}.

Finally, in Appendix~\ref{appA} we review Clifford circuits and define the stabilizer length distribution, and detail measurement and unitary Clifford {\it dynamics} -- beyond the steady state -- in Appendix~\ref{appB}.

\section{The circuit model \label{sec2}}

\begin{figure}[b]
    \includegraphics[width=.49\textwidth]{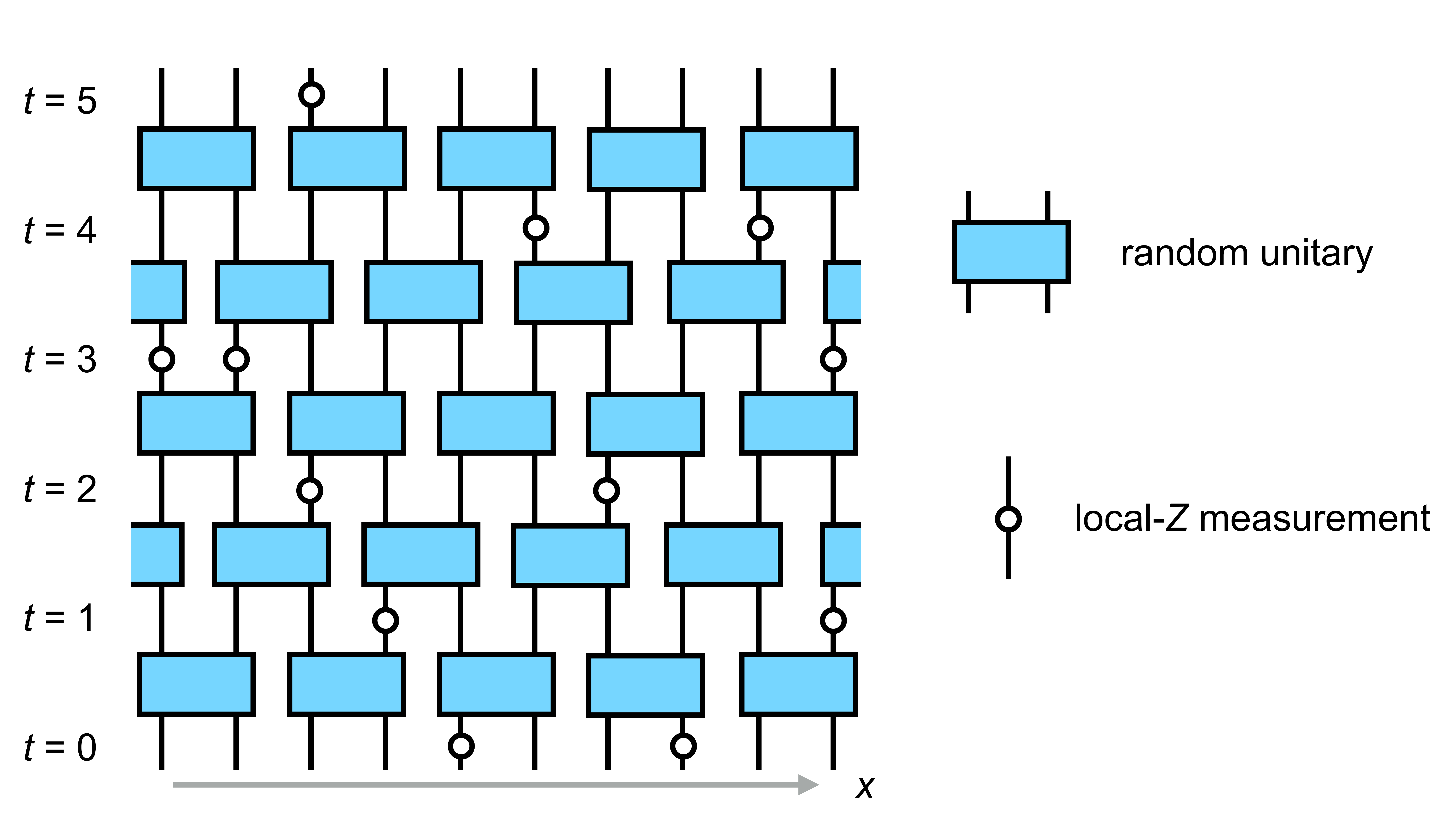}
    \caption{The random circuit model with random measurements. In this circuit, the unitaries are arranged in a brick-layer fashion, while the single qubit $Z$-measurements are positioned randomly in space and time.
    We depict the Poissonian arrangement in this figure, for which the measurements take place at each available space-time site independently with probability $p$.
    For a circuit with $L$ qubits and with depth $D$, there are $L D$ such available sites.}
    \label{fig:rand_cliff_circ}
\end{figure}

Consider first the prototypical quantum circuit model,
shown in Fig.~\ref{fig:rand_cliff_circ}, with 
$L$ qubits arranged on a one-dimensional chain.
The circuit dynamics is composed of two parts, as depicted
in Fig.~\ref{fig:rand_cliff_circ} and detailed below (in order), namely
(i) the background unitary evolution, and 
(ii) measurements made on selected qubits scattered throughout the system.
\envelope{enumerate}{
[(i)]
\item
The background unitary time evolution of the $L$-qubit wavefunction is determined by applications of local unitary gates which are arranged in a bricklayer pattern, {such that the geometry of the circuit} is periodic in both space and time.
The local unitaries act on neighboring pairs of qubits.
Each discrete time cycle of the circuit consists of two layers, and each layer has $L/2$ two-qubit unitary gates, acting on all the odd links in the first layer, and all the even links in the second.
We primarily consider circuits with periodic spatial boundary conditions, except in Appendix~\ref{appB} where circuits with open boundary condition are more convenient.

We define the depth of a circuit to be the number of unitary layers, and denote it by $D$. Therefore, a circuit with depth $D$ has $T = D/2$ time cycles.
The circuit as a whole can be regarded as a unitary transformation in the Hilbert space of many-body wavefunctions on $L$ qubits,
\envelope{eqnarray}{
    \label{eq:U_total}
    U_T = \prod_{t = 0}^{T-1} U(t),
}
where $U(t)$ is the time evolution operator for the $t$-th time cycle,
\envelope{eqnarray}{
    \label{eq:U(t)}
    U(t) =
    \(\prod_{x\ {\rm odd}} U_{(x, x+1), 2t+1} \)
    \(\prod_{x\ {\rm even}} U_{(x, x+1), 2t} \),
}
where $U_{(x,x+1), d}$ is the gate on link $(x,x+1)$ at depth $d$.
Under the action of a unitary gate, the wavefunction transforms as,
\envelope{eqnarray}{
    \label{eq:q_traj_U}
    \ket{\psi} \to  U_{(x,x+1), d} \ket{\psi},
}
so that the wavefunction at arbitrary time $T$ is $\ket{\psi(T)} = U_T \ket{\psi(0)}$.

\item
The full dynamics of the model is \emph{non-unitary}, wherein the space-time sheet of the unitary circuit is punctuated with measurements -- for simplicity chosen as single-qubit measurements.
In a circuit with depth $D = 2T$, there are $L \times D$ available space-time locations between unitary layers available for such measurements.
Measurements are made on a fraction $p$ of all these sites, chosen either randomly or deterministically.
{The parameter $p$ is thus the rate of measurement.}
In Sections \ref{sec2}--\ref{sec4} of this paper we will choose these sites randomly (Poisson distribution)
as depicted in Fig.~\ref{fig:rand_cliff_circ}, a model first proposed in Refs.~\cite{nandkishore2018hybrid, nahum2018hybrid}.
The unitary background is obtained by setting $p = 0$.

Under the action of a measurement the wavefunction transforms as,
\envelope{eqnarray}{
    \label{eq:q_traj_M}
    \ket{\psi} \to \frac{M_\alpha \ket{\psi}}{\lVert M_\alpha \ket{\psi} \rVert},
}
where $\{M_\alpha\}$ are a set of linear  ``generalized measurement'' operators satisfying $\sum_\alpha M^\dg_\alpha M_\alpha = 1$~\cite{nielsen2010qiqc}.
Under such a measurement, the process described by Eq.~(\ref{eq:q_traj_M}) is probabilistic, with outcome $\alpha$ happening with probability $p_\alpha = \bra{\psi} M^\dg_\alpha M_\alpha \ket{\psi}$.
Throughout much of the paper, and unless specified to the contrary, we will choose these
``generalized measurement" operators to be mutually orthogonal projectors,
that is $M_\alpha \rightarrow P_\alpha$, with $P_\pm = (1\pm Z)/2$ measuring the $Z$-component of the spin of individual qubits. 
Such projectors satisfy $P_\alpha P_\beta = \delta_{\alpha \beta} P_\alpha$ and $\sum_\alpha P_\alpha = 1$.
}

For a convenient initial wavefunction (unentangled, for example), once the realizations of each unitary and measurement gate are specified as well as the measurement outcomes, the many body wavefunction at any time step is determined, by following the transformations defined in Eqs.~(\ref{eq:q_traj_U}, \ref{eq:q_traj_M}).
This pure state time evolution is known as a \emph{quantum trajectory}~\cite{Wiseman1996}.
As emphasized in Refs.~\cite{nahum2018hybrid, li2018hybrid} the entanglement physics of interest to us will not be contained in the time evolution of the mixed state density matrix (appropriate when/if the measurement results are summed over, rather than tallied), which appears in more familiar treatments of open quantum systems~\cite{Breuer2002theory}.

While unitary gates generically increase entanglement, local measurements tend to reduce the entanglement entropy on average.
This competition is subtle since the effect of the unitary gates on the entanglement is strictly local and incremental~\cite{nahum2017KPZ}, while the measurement operators are expected to have some non-local effects on entanglement.
Moreover, this competition could lead to interesting entanglement dynamics at early times.
For example, in Ref.~\cite{nahum2018hybrid} the entanglement dynamics can be mapped to the first passage percolation~\cite{Hammersley1965, Chayes1986firstpassage, kesten1986aspects, kesten1987firstpassage} in certain limits, while in Ref.~\cite{li2018hybrid}, sublinear power-law growth of entanglement was observed at a critical measurement rate, in contrast to the linear growth in purely unitary circuits. 
Non-monotonic growth of entanglement can also occur in this type of circuit~\cite{nandkishore2018hybrid, LiMPAF2018Cat}.
However, in this paper we will primarily focus on the entanglement entropy of the late-time steady state, rather than its early-time dynamics.
We leave a detailed study of the latter to the future.

The primary quantity we use to characterize the steady state wavefunctions is the R{\'e}nyi entropy, defined as,
\envelope{eqnarray}{
    \label{eq:Renyi_EE}
    S_A^n = \frac{1}{1-n} \log_2 {\rm Tr} \(\rho_A\)^n, \quad \rho_A = {\rm Tr}_{\ovl{A}} \ket{\psi} \bra{\psi},
}
where $(A, \ovl{A})$ is a bipartition of the $L$-qubit system with $A$ being a contiguous subregion, and $\ket{\psi}$ is the pure state wavefunction we obtain by following the quantum trajectory.
A closely related quantity is the mutual information between two subregions,
\envelope{eqnarray}{
\label{eq:Renyi_MI}
    I^n_{A, B} = S^n_A + S^n_B - S^n_{A \cup B}.
}
The mutual information is guaranteed to be non-negative when $n \le 1$.

For a large part of the paper, we will consider Clifford circuits. 
In this case, all R{\'e}nyi entropies are equal to each other due to the flat entanglement spectrum~\cite{Klappenecker2002stabilizer, Linden2013stabilizer}, and we will drop the R{\'e}nyi index (the superscript $n$).

The generic circuit has three types of randomness:
(i) a random ensemble of unitary gates,
(ii) the random locations of the measurements, and
(iii) the intrinsic random outcome of each quantum measurement.  
We will mostly consider the mean values of the entanglement entropies, averaged over the various forms of randomness present in the circuit.
As we shall see in Sec.~\ref{sec3}, the distributions of the entanglement entropies in the steady state are narrow, so well represented by their averages.

        \section{The phase diagram \label{sec3}}

In this section we discuss the phase diagram of a generic circuit with random Clifford unitaries and random measurement placements.
Specifically, we consider circuits of the structure exactly as in Fig.~\ref{fig:rand_cliff_circ}, wherein the unitary gates are sampled from the uniform distribution on the two-qubit Clifford group ({see Appendix~\ref{appA}}), and the measurements are taken to be single-qubit Pauli-$Z$ measurements, namely $P_\pm = (1\pm Z)/2$, at random positions chosen independently with probability $p$ (the Poissonian fashion).
We shall refer to this specific model as the ``random Clifford circuit'', in short.

The primary motivation for studying the random Clifford circuits, rather than the more generic circuits with non-Clifford gates (e.g. random Haar unitaries), is numerical tractability.
On the single gate level, the random Clifford unitaries approximate the random Haar unitaries quite well, being known as a unitary 2-design~\cite{DiVincenzo2002hiding}.
Our expectation for the equivalence in terms of the entanglement physics is partially justified in Sec.~\ref{sec6}, where comparisons are made between the two circuits for small system sizes -- and consistency is found.

The simulability of Clifford circuits is a result known as the Gottesman-Knill theorem~\cite{gottesman9604hamming, gottesman9807heisenberg, nielsen2010qiqc, aaronson0406chp}.
As reviewed in Appendix~\ref{appA}, the methodology involves following the dynamics of ``stabilizers" --  mutually commuting and independent Pauli string operators -- that uniquely specify the wavefunction, and readily allow for calculation of the entanglement entropy~\cite{Fattal2004stabilizer, hamma2005entanglement1, hamma2005entanglement2, nahum2017KPZ}.
Clifford circuits have proven useful in the study of entanglement and operator dynamics in various contexts~\cite{nahum2017KPZ, nahum2018operator, chandran1501semiclassical}.

\subsection{The steady state}

\begin{figure}[t]
    \includegraphics[width=.49\textwidth]{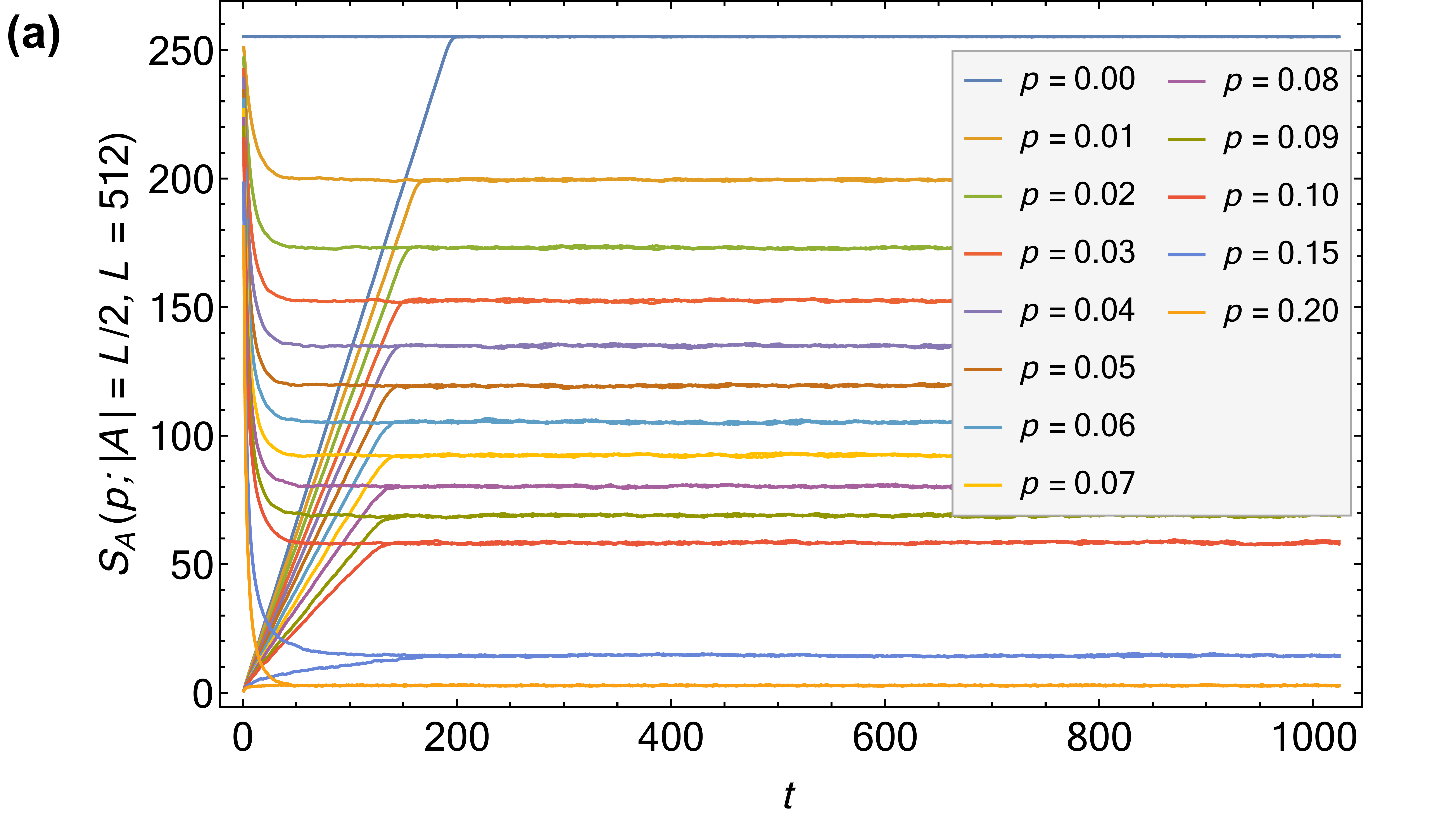}
    \includegraphics[width=.49\textwidth]{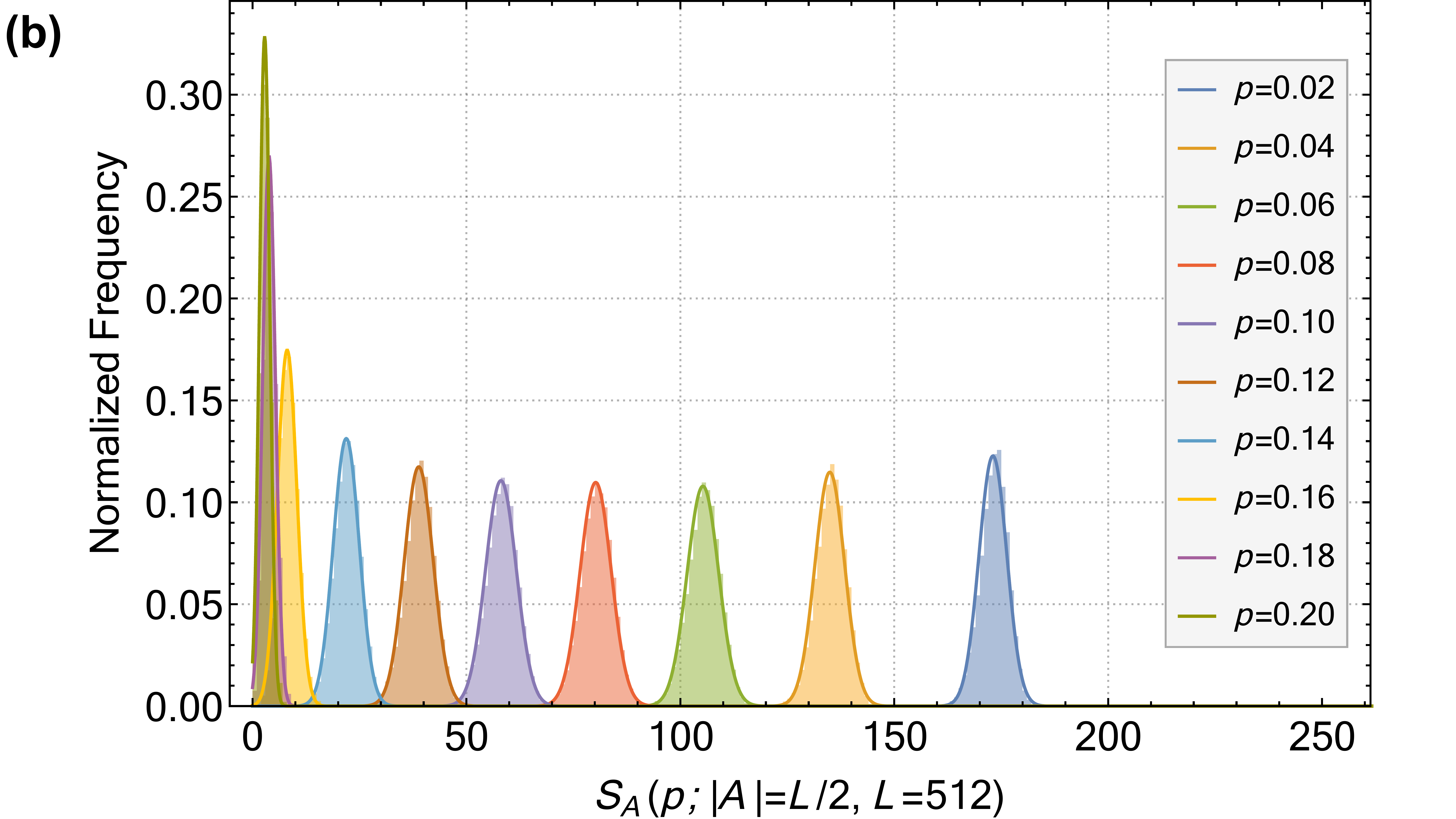}
    \caption{
    (a) Time dependence of the entanglement entropy $S_A$ with $|A| = L/2$ and $L=512$, in the random Clifford circuit averaged over circuit realizations, starting from either a maximally entangled state or a trivial product state.
    (b) Distribution function of $S_A$ for different circuit realizations and over time well after saturation. The solid lines are fits to a normal distribution.
    }
    \label{fig:bulk_EE}
\end{figure}

Given a circuit of a finite length $L$ of qubits, we are primarily interested in the late time behavior when $T \to \infty$.
In this infinite time (circuit depth) limit we expect the system to evolve into a steady state, characterized by a typical value of entanglement entropy that depends on the measurement rate $p$, but not the dynamics at finite times.
To check that this limit is well-defined, we compute the time dependence of the entanglement entropies starting from two types of initial states, namely,
\envelope{itemize}{
\item
    The trivial product state, $\prod_x \ket{0}_x$, which is a stabilizer state, i.e. the simultaneous eigenvector with eigenvalue 1 of its stabilizers
    $
        \mc{G} = \{Z_1, Z_2, \ldots, Z_L\}.
    $
\item
    The maximally entangled state, obtained by evolving the random Clifford circuit without measurements well after saturation. 
}

The results, averaged over circuit realizations, are plotted in Fig.~\ref{fig:bulk_EE}(a).
For all values of $p$ and for both choices of the initial state, the entanglement entropy saturates to a value that is determined solely by $p$.
We believe that this holds for an \emph{arbitrary} choice of the initial state.
Therefore, we can talk about the ``steady state'' for a given rate of measurement without referring to the initial state.  The steady state is thus a \emph{bulk property} of the circuit.

After saturation there are only minimal fluctuation in the entropies over time.
Moreover, the fluctuations are also small over different circuit realizations.
In Fig.~\ref{fig:bulk_EE}(b), we plot the distribution of the entanglement entropy taken from an ensemble of circuits, and over many time steps well after saturation.
Notice that the functions are sharply peaked for each $p$, and fit well to the Gaussian distribution.

We define $S_A(p; |A|, L)$ to be the late-time entanglement entropy of a subsystem with size $|A|$, when averaged over different circuit realizations, for a circuit with length $L$ and measurement rate $p$.  Given the (average) spatial translational symmetry this quantity only depends on the size (but not the location) of the subregion $A$.
In the following we will usually refer to this quantity as \emph{the entanglement entropy}, unless otherwise specified.

\begin{figure}[t]
    \centering
    \includegraphics[width=.5\textwidth]{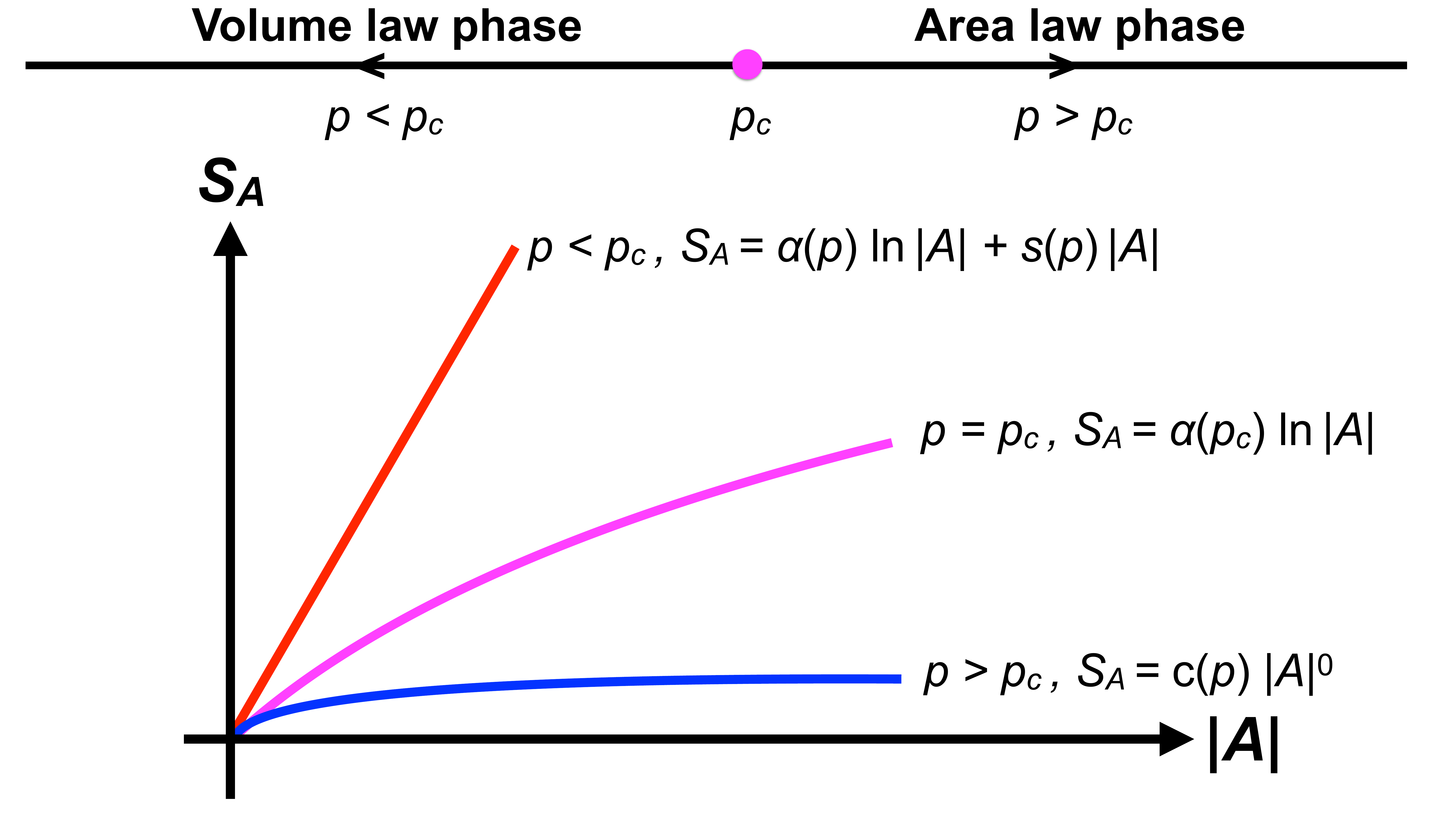}
    \caption{The phase diagram and scaling behavior of the entanglement entropy in both phases and at criticality.} 
    \label{fig:phase_diag}
\end{figure}

\begin{figure}[t]
    \includegraphics[width=.49\textwidth]{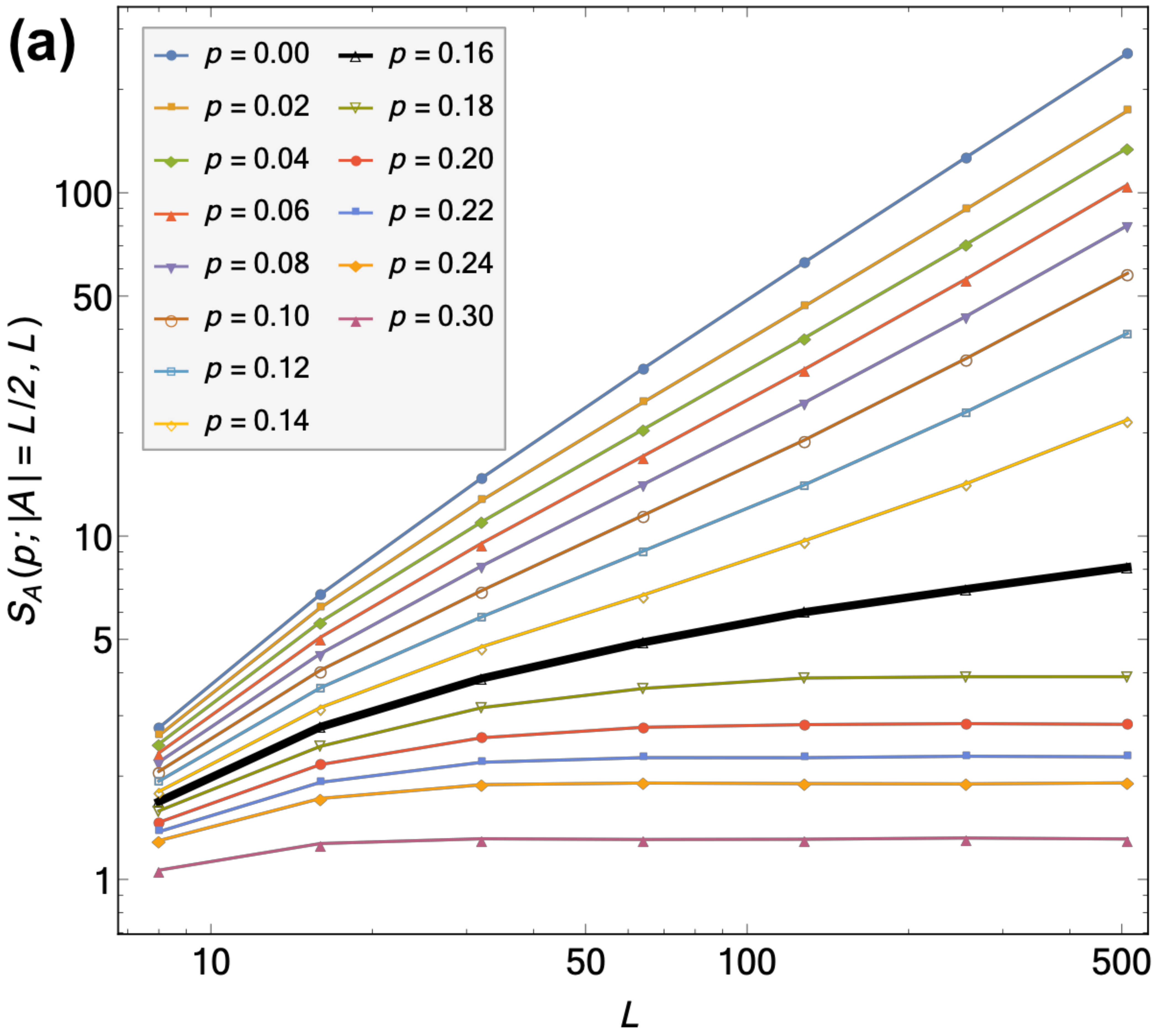}
    \includegraphics[width=.49\textwidth]{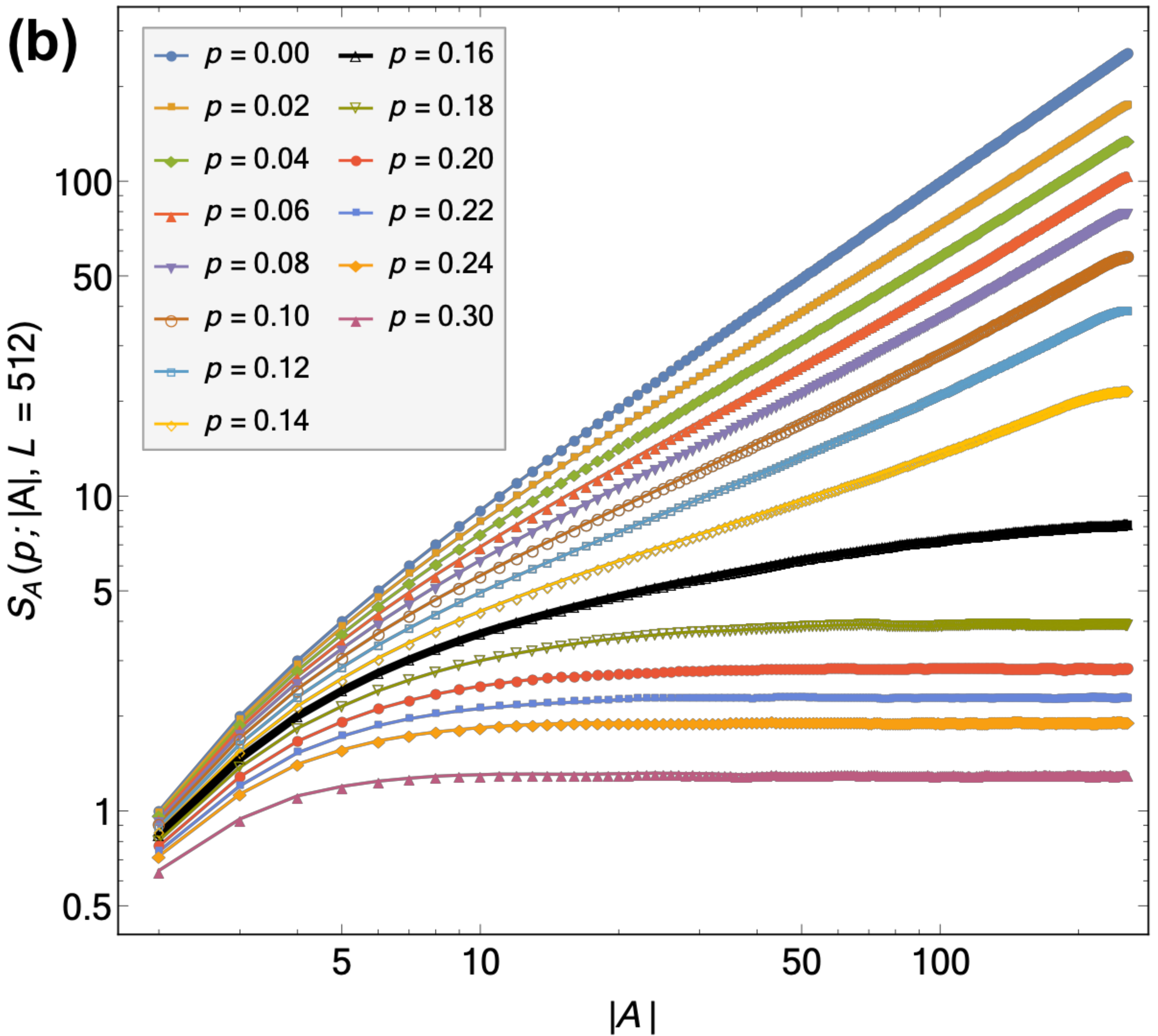}
    \caption{
    (a) Entanglement entropy $S_A(p; |A|, L)$ with fixed $|A|/L = 1/2$, as functions of $L$, for different values of $p$.
    (b) $S_A(p; |A|, L)$ with fixed $L = 512$, as functions of $|A|$, for different values of $p$.
    Both plots are on a log-log scale.
    {Notice that curves in (a) and (b) corresponding to the same value of $p < p_c$ has the same slope, $s(p)$ (see main text).}
    }
    \label{fig:EE_scaling}
\end{figure}

\subsection{The two phases}

Attempts have been made to map out the phase diagram~\cite{nandkishore2018hybrid, nahum2018hybrid, li2018hybrid}.
The limiting cases are easy to understand.
When $p \to 1$, the steady state is close to a trivial product state, and has area law entanglement entropy.
The other limit, $p \to 0$, corresponds to the random unitary circuit, where the steady state is characterized by maximal volume law entanglement entropy~\cite{nahum2017KPZ}.
The putative phase diagram is shown schematically in Fig.~\ref{fig:phase_diag}, which shows a volume law phase and an area law phase separated by some critical rate of measurement, $p_c$.
Whether $p_c$ is $0$ or finite was not agreed upon in earlier work.

Here our numerics for the random Clifford circuit supports a finite $p_c$, consistent with \cite{nahum2018hybrid, li2018hybrid}.
In Fig.~\ref{fig:EE_scaling}(a), we plot the entanglement entropy $S_A(p; |A| = a L, L)$ for different values of $p$ as functions of $L$, with a fixed $a = 1/2$.
We find qualitatively distinct behavior of $S_A$ below and above $p_c \approx 0.16$.
For $p < p_c$, the curves asymptote to straight lines of slope $1$ on a log-log scale, suggesting volume law scaling of the entanglement entropy, $S_A(p; |A| = a L, L) = s(p) L$.
For $p > p_c$, the curves are saturating to zero slope, suggesting an area law scaling, $S_A(p; |A| = a L, L) = c(p) L^0$.

In Fig.~\ref{fig:EE_scaling}(b), we plot $S_A(p; |A|, L)$ as a function of $|A|$ while fixing $L = 512$.
Similar scaling behavior is observed.

\subsection{Entanglement entropy from stabilizer distribution}

For Clifford circuits further information about the nature of the two phases 
can be revealed by examining the stabilizer distributions, as we now discuss.
We start by listing several results regarding the stabilizer formalism~\cite{gottesman9604hamming, gottesman9807heisenberg, nielsen2010qiqc, aaronson0406chp, nahum2017KPZ}.
These results are also reviewed in Appendix~\ref{appA}.
\envelope{enumerate}{
\item
A wavefunction $\ket{\psi}$ in the Clifford circuit of $L$ qubits is uniquely characterized by $L$ mutually commuting and independent Pauli string operators $\mc{G} = \{g_1, \ldots, g_L\}$ such that each one ``stabilizes'' the wavefunction, $g_i \ket{\psi} = \ket{\psi}$.

Elements of $\mc{G}$ are called \emph{stabilizers}.
Such a wavefunction is called a \emph{stabilizer state} or \emph{codeword}.
Only stabilizer states appear in the Clifford circuit.

Being Pauli string operators, the stabilizers have endpoints where they terminate.
Specifically, we define the \emph{left} and \emph{right endpoints} of a stabilizer to be
\envelope{eqnarray}{
    \label{eq:Pauli_lend}
    \mt{l}(g) &=& \min\{x : \text{$g$ acts non-trivially on site $x$} \}, \\
    \label{eq:Pauli_rend}
    \mt{r}(g) &=& \max\{x : \text{$g$ acts non-trivially on site $x$} \},
}
where $x$ is the coordinate of the site, which takes values in $\{1, 2, \ldots, L\}$.
For systems with periodic spatial boundary conditions, there is an arbitrariness in choosing the origin of the coordinate system, and there is no absolute distinction between left and right.
{However, we note that the functions $\mt{l}(g)$ and $\mt{r}(g)$ are well-defined once the origin is chosen and fixed, which we will always assume to be the case in the rest of the paper.}

\item
The choice of $\mc{G}$ is not unique.
For any stabilizer state, one can choose $\mc{G}$ such that there are exactly two stabilizer endpoints on each site,
\envelope{eqnarray}{
    \label{eq:clipped_gauge_condition}
    \rho_\mt{l}(x) + \rho_\mt{r}(x) = 2, \text{ for all sites $x$.}
}
We say $\mc{G}$ is in the \emph{clipped gauge}~\cite{nahum2017KPZ}.

Notice that $\mc{G}$ is \emph{not} uniquely fixed by this gauge condition.

\item
Within the clipped gauge, the entanglement entropy of a \emph{contiguous} subregion $A$ is given by half the number of stabilizers that cross either its left or right boundary,
\envelope{widetext}{
    \envelope{eqnarray}{
        \label{eq:EE_span}
        S_A = \frac{1}{2} \#\{ g \in \mc{G} : \(\mt{l}(g) \in A \text{ and } \mt{r}(g) \in \ovl{A} \) \text{ or } \(\mt{l}(g) \in \ovl{A} \text{ and } \mt{r}(g) \in A \)\}.
    }
}
With periodic spatial boundary conditions, the subregion $A$ can be either sites $\{x, x+1, \ldots, x+|A|-1\}$ when $x+|A| \le L+1$, or $x, x+1, \ldots, L, 1, 2, \ldots, x+|A|-(L+1)$ when $x+|A| > L+
1$.
In the clipped gauge, the entanglement entropy is given solely by the end positions of the stabilizers, and does not depend on their ``internal" contents.
}

\begin{figure}[b]
    \centering
    \includegraphics[width=.49\textwidth]{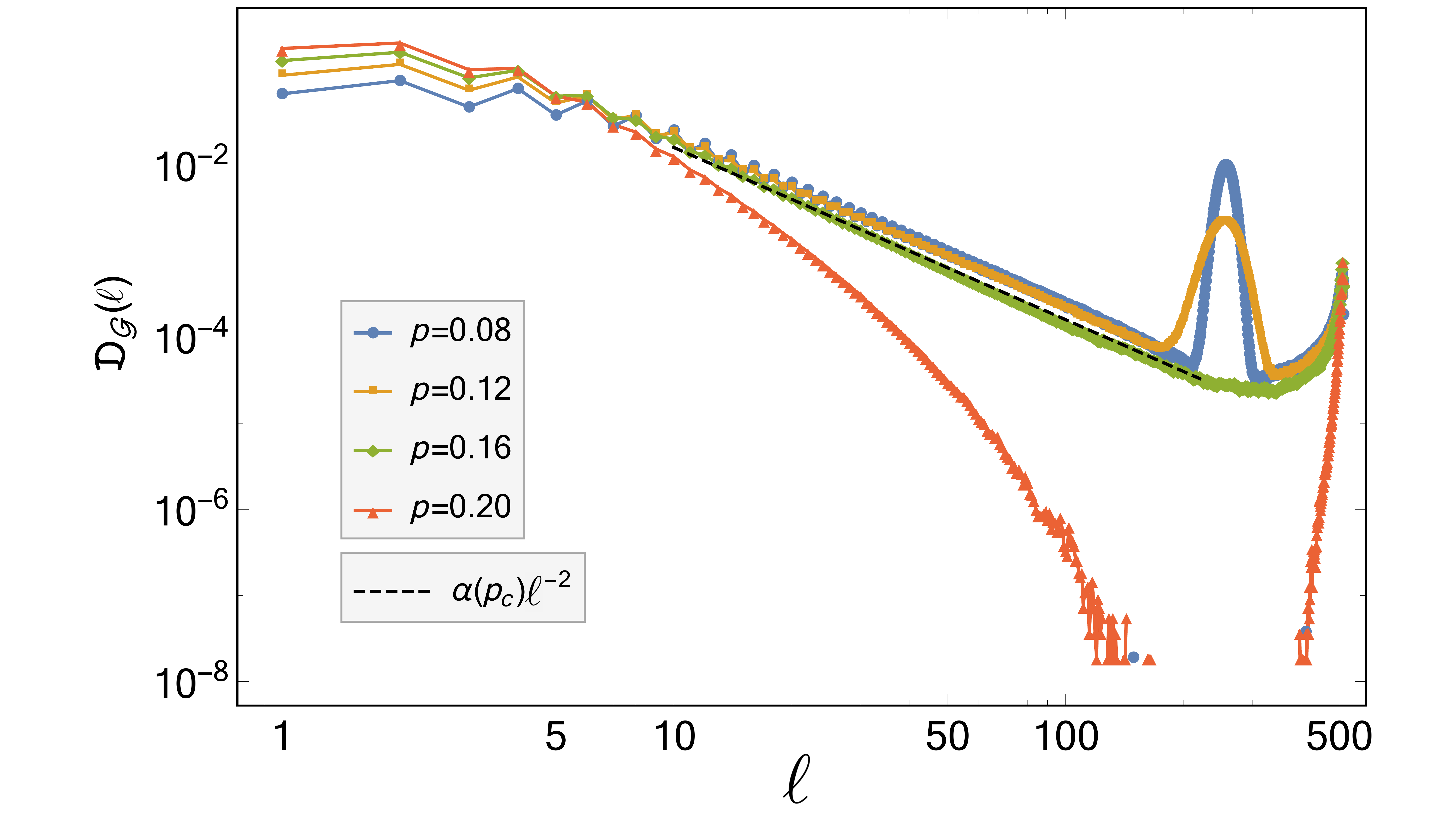}
    \caption{The normalized stabilizer length distribution $\mf{D}_{\mc{G}}(\ell)$ plotted on a log-log scale for a system with size $L = 512$.
    Here we take $\alpha(p_c) = 1.6$.}
    \label{fig:stab_len_dist}
\end{figure}

Consider the bigrams of stabilizer endpoints which encode the ``span'' of each stabilizer,
\envelope{eqnarray}{
    \mc{B}(\mc{G}) \equiv \{ \(\mt{l}(g_1), \mt{r}(g_1)\), \ldots, \(\mt{l}(g_L), \mt{r}(g_L)\) \}.
}
As shown in Appendix~\ref{appA}, for a given wavefunction this object is unique, provided  $\mc{G} = \{g_1, \ldots, g_L\}$ is in the clipped gauge.  Generally there may be many different choices of $\mc{G}$ that satisfy the (clipped) gauge condition, which all share the same bigram.
Nevertheless, the bigram fully characterizes the entanglement entropy of the wavefunction(s)  through the relation in Eq.~\eqref{eq:EE_span}, being insensitive to the gauge redundancy.

It is convenient to define the normalized stabilizer (spatial) distribution function,
\envelope{eqnarray}{
    D_\mc{G} (x, y) = \frac{1}{L} \overline{ \sum_{i=1}^L \delta_{\mt{l}(g_i), x} \delta_{\mt{r}(g_i), y}},
}
where the overline represents an ensemble average of the bigrams taken over different circuits and times.
We can also define the normalized stabilizer length distribution function,
\envelope{eqnarray}{
    \mf{D}_{\mc{G}}(\ell) = \frac{1}{L} \overline{ \sum_{i=1}^L \delta_{\mt{len}(g_i), \ell}},
}
where $\mt{len}(g_i) = \mt{r}(g_i) - \mt{l}(g_i)$.
The latter is the integral of the former,
\envelope{eqnarray}{
    \mf{D}_{\mc{G}}(\ell) = \sum_{x,y} \delta_{\ell, y-x} D_{\mc{G}}(x, y).
}

For circuits with periodic spatial boundary conditions, our numerics reveal (data not shown) that 
the spatial distribution of the stabilizers for a particular length $\ell$ is uniform, true at each
value of $\ell$ and $p$.  That is,
\envelope{eqnarray}{
    D_{\mc{G}}(x, y) = D_{\mc{G}}(x^\p, y^\p) \text{ if $y-x = y^\p - x^\p$.}
}
Thus, taking into account the geometric constraint that a stabilizer with length $\ell$ can only have its left endpoint in the range $(0, L-\ell)$, we have
\envelope{eqnarray}{
    \label{eq:rel_two_dist_fcns}
    D_{\mc{G}}(x, y) = \frac{\mathfrak{D}_{\mc{G}}(y-x)}{L - (y-x)} \approx \frac{\mathfrak{D}_{\mc{G}}(y-x)}{L},
}
where the last approximation applies when $y-x \lesssim L/2$.
These two distribution functions depend on each other through a simple relation, and one can be inferred from the other.

In Fig.~\ref{fig:stab_len_dist}, we plot the distribution function $\mathfrak{D}_{\mc{G}}(\ell) \approx D_\mc{G}(x, y) \times L$, where $\ell = y-x$, at different values of $p$, for fixed $L = 512$.
The distribution function is quite remarkable.
\envelope{itemize}{
\item
In the volume law phase $p < p_c$, the distribution is ``bimodal", namely a tail of ``short stabilizers'', which is checked to be independent of $L$ (data not shown), and a peak of ``long stabilizers'' at $\ell \approx L/2$
\footnote{
We also notice a small hump at $\ell \approx L$.
This part of the distribution is a boundary effect due to the periodic boundary condition, and the height of the hump decays as $1/L$ as we go to the thermodynamic limit.
Moreover, from Eq.~\eqref{eq:EE_span}, these long stabilizers of length $\sim L$ barely contribute to the entanglement entropy.
Thus we ignore this unimportant hump.
}.
On a log-log plot, the short stabilizer distribution for $p < p_c$ looks like a straight line with slope $-2$, corresponding to a power-law distribution $\mathfrak{D}_{\mc{G}}(\ell) \sim \ell^{-2}$.
The peak at $\ell \approx L/2$ has nonzero weight in the volume law phase, and the weight vanishes continuously as one approaches the critical point from $p < p_c$.
\item
In the area law phase, $p > p_c$, the power-law distribution of ``shorter" stabilizers becomes truly short-ranged.
}
The results in Fig.~\ref{fig:stab_len_dist} can be schematically summarized as,
\envelope{eqnarray}{
    \label{eq:len_dist_summ}
    \mathfrak{D}_{\mc{G}}(\ell) \sim
    \envelope{cases}{
        \alpha(p) \frac{1}{\ell^2} + s(p) \delta(\ell - L/2), \hspace{.08in} p < p_c \\
        \alpha(p) \frac{1}{\ell^2}, \hspace{1.12in} p = p_c \\
        \alpha(p) \frac{e^{-\ell/\xi}}{\ell^2}, \hspace{.95in} p > p_c \\
    }
}
where $\alpha(p)$ is the weight of the power law, which has weak dependence on $p$ or $L$, $s(p)$ is the weight of the peak, and $\xi$ is some finite length scale that cuts off the length of the stabilizers in the area law phase.

From the formula for entanglement entropy Eq.~(\ref{eq:EE_span}), we see that for a region $A$ with $1 \ll |A| \ll L$,
\envelope{eqnarray}{
    S_{A} &=& \frac{1}{2} \int_{x \in A} \int_{y \in \ovl{A}}
    \lz \theta(y-x) D_\mc{G}(x, y) \times L +
    \(x \leftrightarrow y\) \rz \nn
    &=& \frac{1}{2} \int_{x \in A} \int_{y \in \ovl{A}}
    \lz \theta(y-x) \mf{D}_\mc{G}(y - x) +
    \(x \leftrightarrow y\) \rz.
}
Combined with Eq.~\eqref{eq:rel_two_dist_fcns} and \eqref{eq:len_dist_summ}, we have
\envelope{eqnarray}{
    \label{eq:EE_from_stab_len_dist}
    S_A \sim \envelope{cases}{
        \alpha(p) \ln |A| + s(p) |A|, \hspace{.12in} p < p_c\\
        \alpha(p) \ln |A|, \hspace{.71in} p = p_c\\
        \alpha(p) \ln \xi . \hspace{.85in} p > p_c\\
    }
}

This scaling behavior is consistent with our findings in Fig.~\ref{fig:EE_scaling}.
When $p < p_c$, the two parts of the distribution contribute to the two terms separately: the volume law entanglement comes from the peak at $\ell \approx L/2$, while the logarithmic correction comes from the power law distribution of the ``shorter" stabilizers, which gets exposed at the critical point.

From the stabilizer length distribution, the existence of a phase transition is rather obvious.
The transition is accompanied by the vanishing of $s(p)$ as we approach $p_c$ from below, and by the divergence of $\xi$ as we approach $p_c$ from above.

        \section{Critical behavior \label{sec4}}

\subsection{Finite size scaling of entanglement entropy}

\begin{figure}
    \centering
    \includegraphics[width=.49\textwidth]{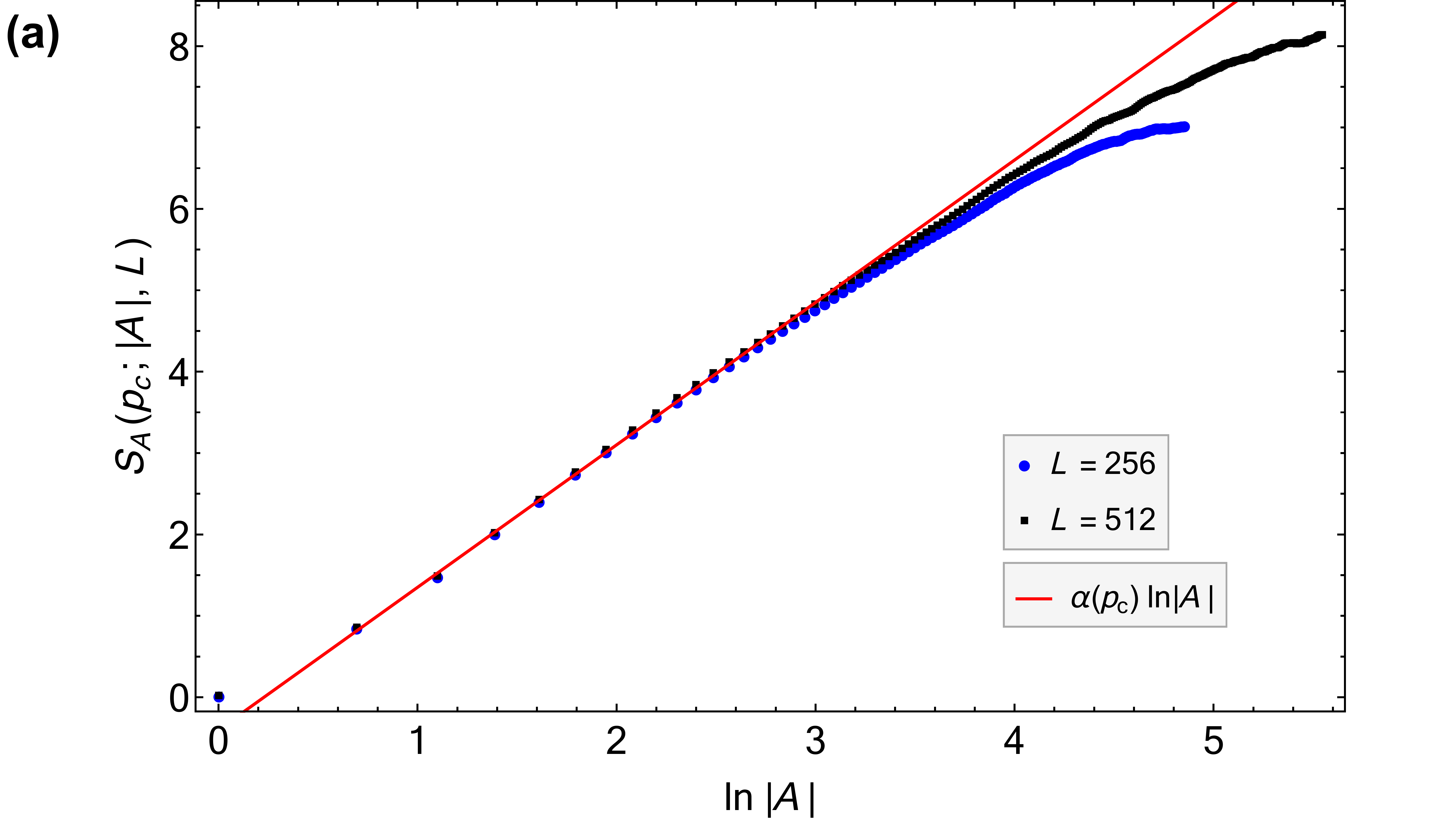}
    \includegraphics[width=.49\textwidth]{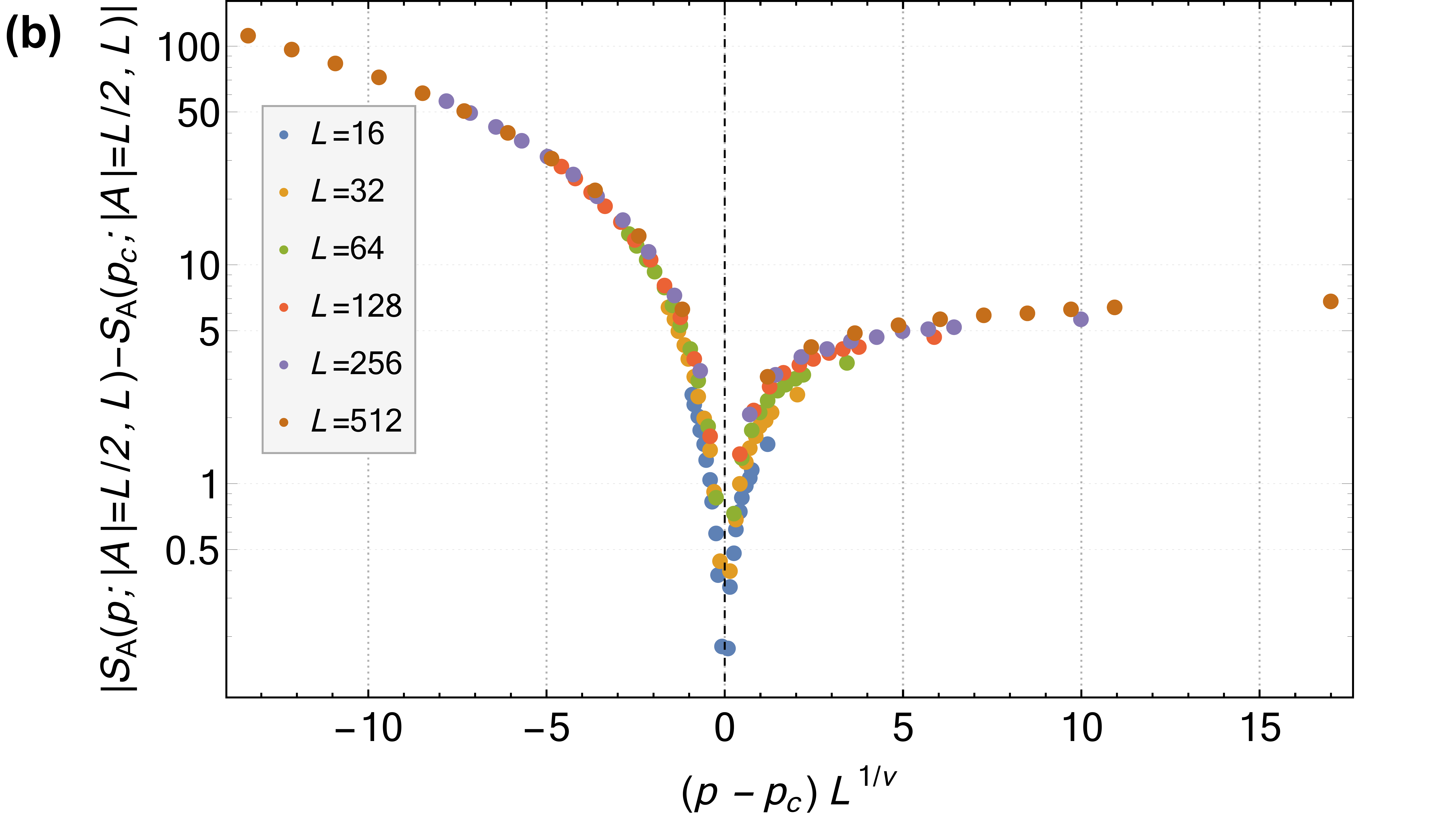}
    \caption{
    (a) Entanglement entropy at the critical point fits well to a purely logarithmic function, $S_A(p_c; |A|, L) \approx \alpha(p_c) \ln |A|$, where $\alpha(p_c) = 1.6$, plotted for $|A| < L/4$.
    (b) Collapsing the $S_A(p; |A| = L/2, L)$ data to the scaling form in Eq.~\eqref{eq:Delta_S_versus_F}, where we find $p_c = 0.16$ and $\nu = 1.3$.
    }
    \label{fig:EE_critical}
\end{figure}

As seen from Eq.~\eqref{eq:EE_from_stab_len_dist}, the inverse-square power law form of the stabilizer length distribution 
at $p=p_c$ implies that the entanglement entropy right at the critical point should vary logarithmically with sub-system size.
In Fig.~\ref{fig:EE_critical}(a) we plot $S_A(p; |A|, L)$ with fixed values of $L$ at $p_c$, and see that it indeed has the desired scaling form.
The coefficient of the logarithmic function matches well to that of the inverse square power law, $\alpha(p_c)$, as expected.

To further probe the entanglement transition, we consider a finite size scaling form for $S_A(p; |A| = a L, L)$,
\envelope{equation}{
\label{eq:EE_scaling_form}
S_A(p; |A| = aL, L) = \alpha(p_c) \ln L + F \( (p-p_c) L^{1/\nu} \).
}
In order to match on to Eq.~\eqref{eq:EE_from_stab_len_dist} in the thermodynamic limit, the function $F$ must be proportional to $L$ when $p < p_c$, and cancel the $\ln L$ term when $p > p_c$.
Therefore $F(x)$ has the following asymptotics,
\envelope{eqnarray}{
    F(x) \approx \envelope{cases}{
        |x|^\nu,        \hspace{.85in} x \to -\infty \\
        \text{const},   \hspace{.75in} x = 0 \\
        - \alpha(p_c) \nu \ln |x|. \hspace{.26in} x \to +\infty
    }
}
Therefore, from Eq.~\eqref{eq:EE_from_stab_len_dist} we identify $s(p)$ with $(p_c-p)^\nu$
for $p<p_c$, and $\xi$ with $|p-p_c|^{-\nu}$ having the meaning of the correlation length.

This scaling form appeared in Refs.~\cite{nahum2018hybrid, vasseur2018rtn}.
In Ref.~\cite{vasseur2018rtn} this formula follows if/when the entanglement entropy can be mapped to the change of the free energy caused by the insertion of two boundary condition changing operators in a 2d classical spin model. 
These two operators are inserted at the boundaries of the subsystem $A$ and the free energy cost for them can be represented as the logarithm of the two point correlation function.
Deep within the two phases, the volume law and area law scalings of the entropy are consistent with the free energy of a domain wall connecting the two boundaries of $A$ in the ordered and disordered phases of the classical spin model, with finite and zero surface tensions, respectively.
The logarithmic correction in the volume phase would be accounted for by the
contributions to the free energy due to capillary wave fluctuations of the interface in the
ordered phase of the spin model~\cite{buff1965capillary, weeks1977capillary}.
Right at the critical point the two point correlation function of the boundary condition changing operator decays as a power law.  
Thus, upon taking logarithms,  the coefficient $\alpha(p_c)$ in the entanglement entropy has the meaning of twice the scaling dimension of the boundary condition changing operator.

In order to put Eq.~\eqref{eq:EE_scaling_form} into a conventional finite size scaling form, we will subtract out the critical entropy to cancel out the $\ln L$ term, and fit our entanglement entropy data to the scaling form,
\begin{equation}
\label{eq:Delta_S_versus_F}
    \lvert S_A(p; |A| = aL, L) - S_A(p_c; |A| = aL, L) \rvert = \tilde{F}\((p-p_c) L^{1/\nu} \).
\end{equation}
In Fig.~\ref{fig:EE_critical}(b) we plot the left hand side of Eq.~\eqref{eq:Delta_S_versus_F} (with $a=1/2$) versus $(p-p_c)L^{1/\nu}$ for values of $p$ both below and above $p_c$, choosing the exponent $\nu =1.3$ to give the best scaling collapse.  
The quality of the data collapse supports the existence of a diverging correlation length $\xi \sim |p-p_c|^{-\nu}$ and the validity of the scaling hypothesis near criticality.

Notice that in Ref.~\cite{li2018hybrid} a different scaling form was used for data collapse, and a different $\nu$ was found.

\subsection{Mutual information and correlations near criticality}

\begin{figure}
    \centering
    \includegraphics[width=.49\textwidth]{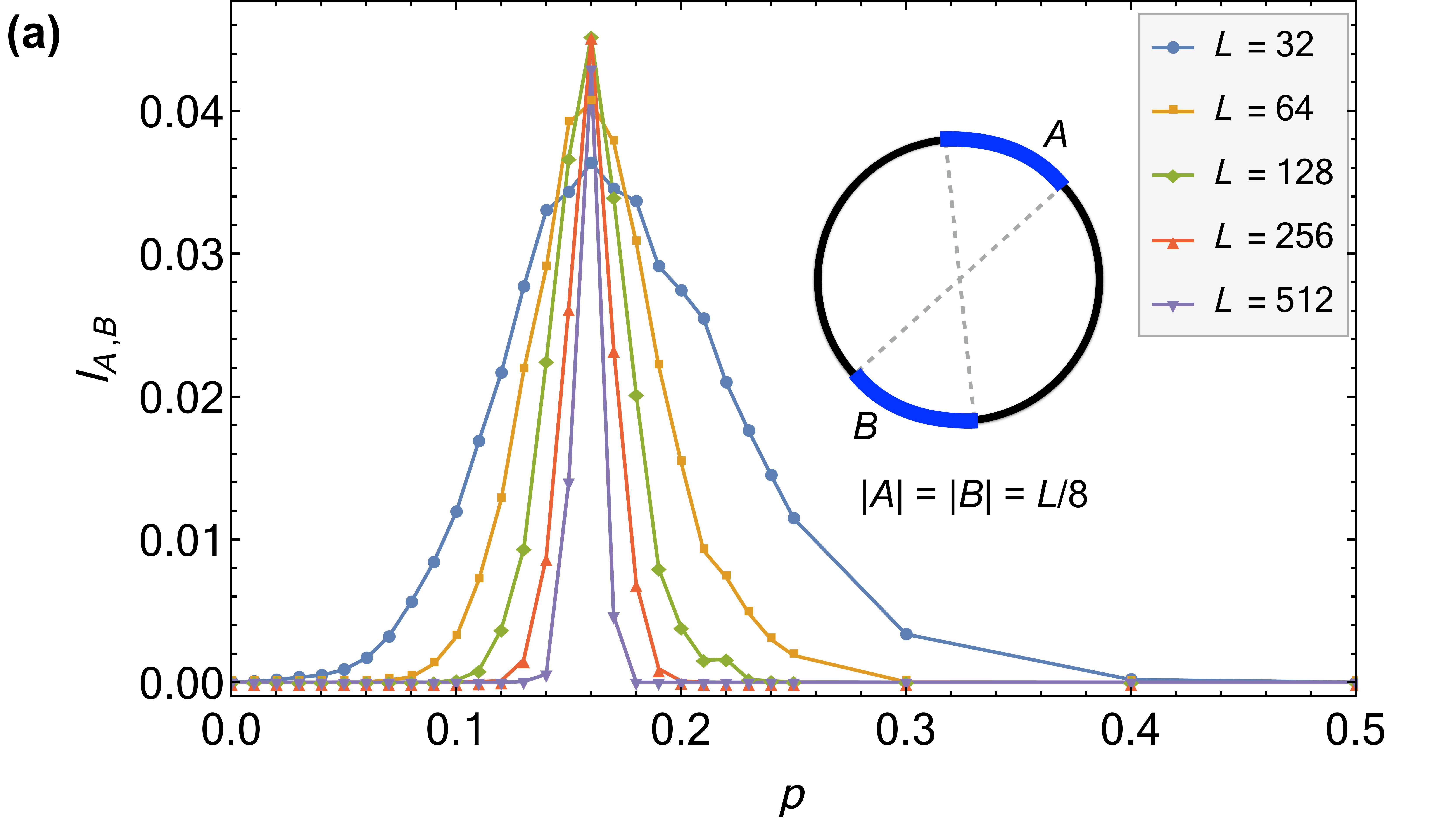}
    \includegraphics[width=.49\textwidth]{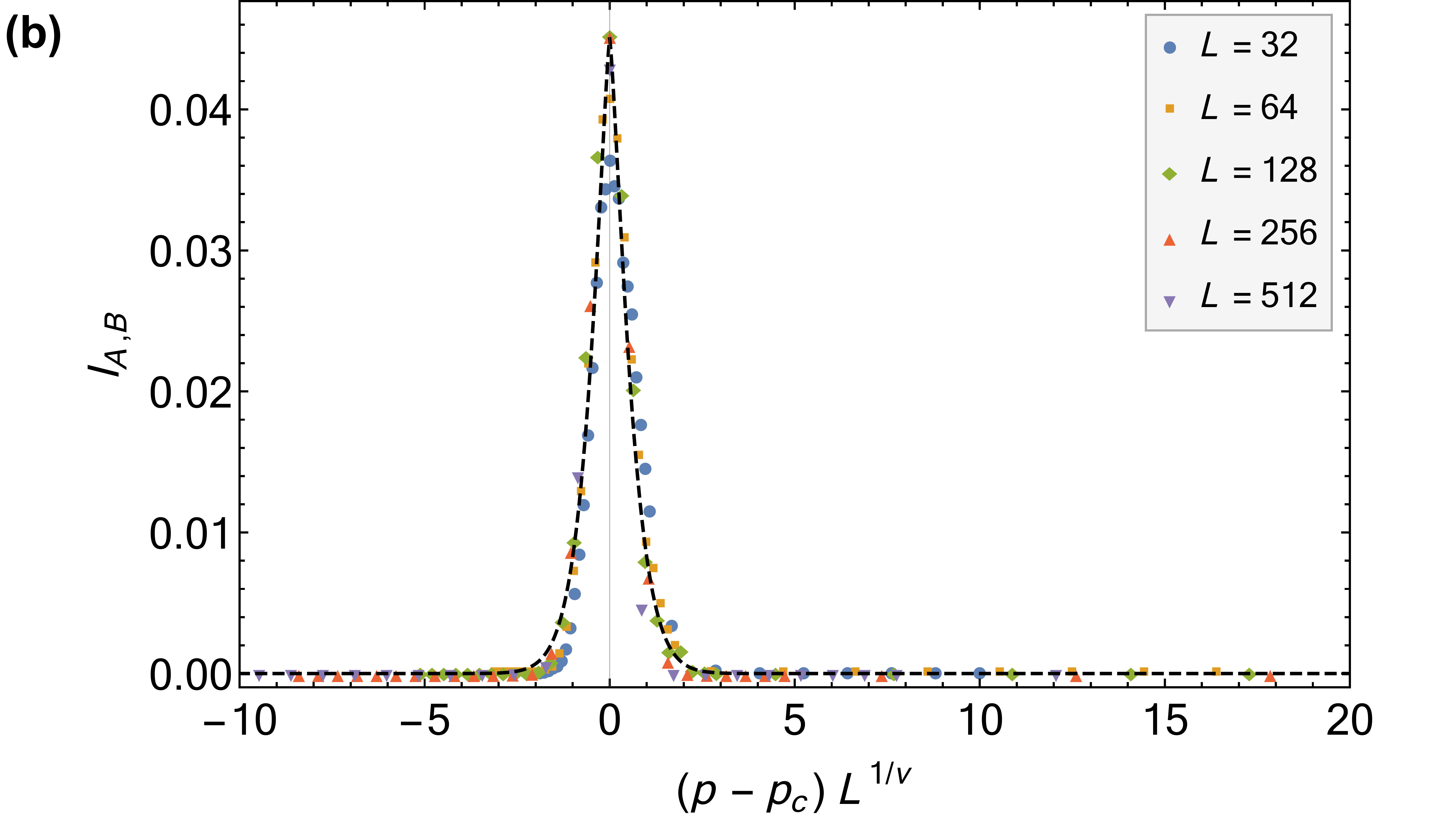}
    \caption{(a) The mutual information, $I_{A,B}$, with region sizes $|A| = |B| = L/8$ and separation $r_{A, B} = L/2$, as shown in the inset.
    (b) Data collapse of the curves in (a), where we have taken $\nu = 1.3$. Dashed lines show the function $f(x) = e^{- c |x|^\nu}$ where $c \approx 1.7$.}
    \label{fig:mutual_info}
\end{figure}

The bipartite mutual information $I_{A, B}$ is one convenient measure of correlations between two disjoint regions $A$ and $B$.
Loosely speaking, it is the entanglement shared only between $A$ and $B$, but not with any third party.
We will first focus on the mutual information when the two regions $A$ and $B$, of size $|A|=|B|=L/8$, are antipodal in the system with periodic boundary conditions, their centers separated by $r_{A, B}= L/2$.
In both phases, away from criticality, we expect the mutual information to fall off exponentially with the system size, varying as $I_{A, B} \sim \exp(-L/\xi)$, much like the behavior of correlation functions in conventional finite temperature transitions away from the critical point.
Right at criticality we expect $I_{A, B}$ to be enhanced due to the longer range correlation~\cite{nahum2018hybrid}.

In Fig.~\ref{fig:mutual_info}(a), we plot the mutual information $I_{A, B} (p; |A| = |B| = L/8, r_{A, B}=L/2,L) $
as a function of $p$ for different system sizes.
The mutual information has a peak at $p=p_c$, which gets sharper with increasing system sizes, as we expect.
Moreover, the height of the peak saturates to a constant that is independent of $L$,
which is consistent with the conformal symmetry discussed in the next subsection.

In Fig.~\ref{fig:mutual_info}(b), we attempt a data collapse with the following finite size scaling form,
\envelope{equation}{
    I_{A, B}(p; |A| = |B| = L/8, r_{A, B} = L/2, L) 
    = f\((p-p_c) L^{1/\nu} \),
}
where $f(x) \propto e^{-c |x|^\nu}$, and $c$ is a non-universal constant.
The collapse is with high quality, and the data fits well to the predicted functional form of $f(x)$.

\begin{figure}[t]
    \centering
    \includegraphics[width=.49\textwidth]{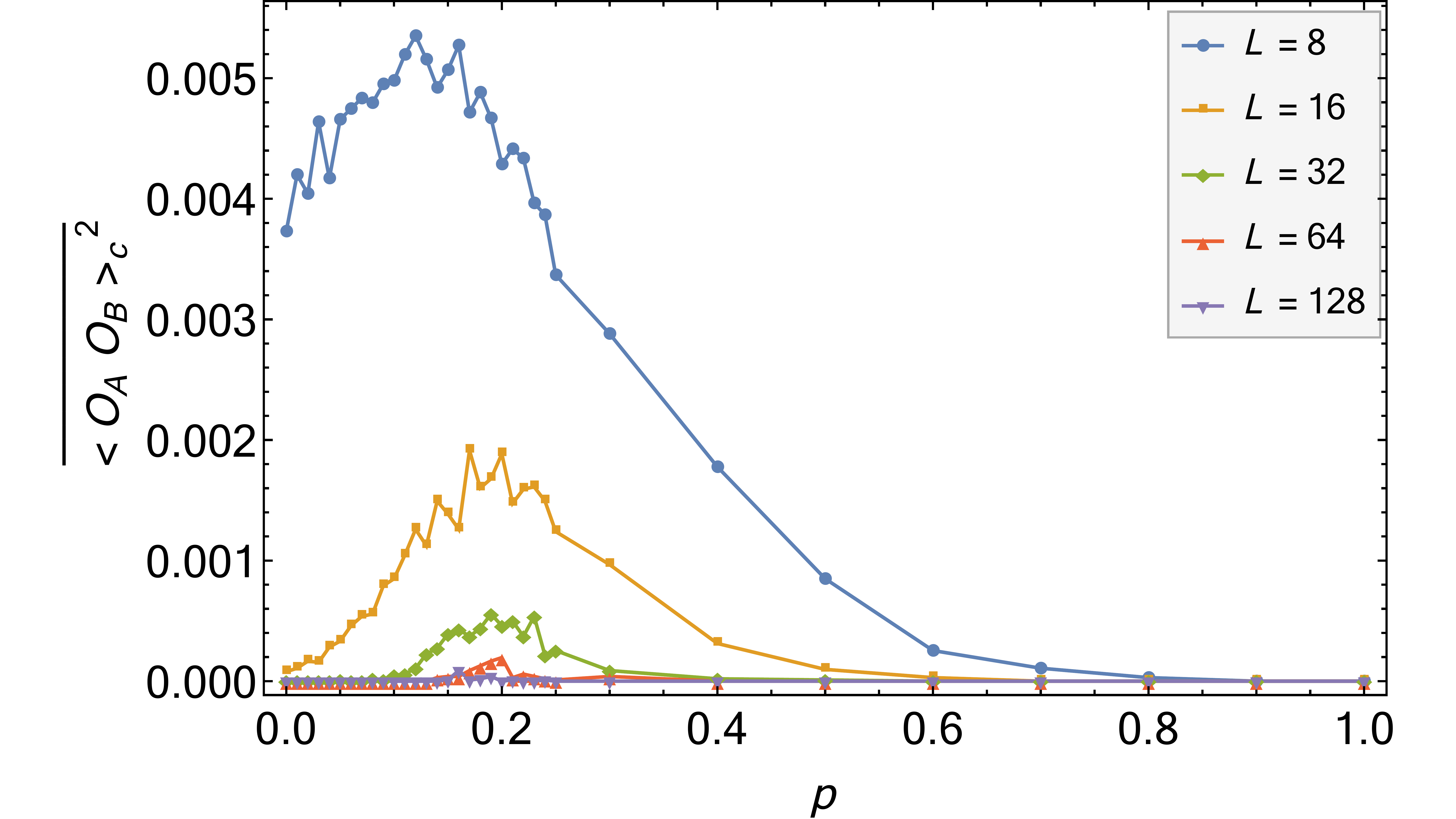}
    \caption{The squared correlation function for two regions $A$ and $B$, as shown in Fig.~\ref{fig:mutual_info}.}
    \label{fig:fluc_corr_fcn}
\end{figure}

The von Neumann mutual information serves as an upper bound on the fluctuation of connected correlation functions between two disjoint regions $A,B$~\cite{WVHC},
\envelope{eqnarray}{
\label{eq:MI_bound}
    I_{A, B} \ge \frac{1}{2} \frac{ \lvert \avg{\mc{O}_A \mc{O}_B}_c \rvert^2}{\lVert \mc{O}_A \rVert^2 \lVert \mc{O}_B \rVert^2},
}
where  $\avg{...}_c$ denotes the connected correlation function, $\mc{O}_A$ and $\mc{O}_B$ are operators on $A$ and $B$, respectively, and $\lVert ... \rVert$ is the operator norm.
For the purpose of illustration, we take $A$ and $B$ to be the same antipodal subregions as above with $|A| = |B| = L/8$, and the operators to be
\envelope{eqnarray}{
    \mc{O}_A = \sum_{x \in A} Z_x, \quad \mc{O}_B = \sum_{x \in B} Z_x.
}
In Fig.~\ref{fig:fluc_corr_fcn} we plot the averaged value of $|\avg{\mc{O}_A \mc{O}_B}_c|^2$ as a function of $p$.
Notably, the curves all show a peak at $p_c$, which gets sharper as $L$ is increased.

We emphasize that the average squared correlation function is only obtained by examining the quantum trajectories one by one, and cannot be written as the expectation value of any operator,
\envelope{eqnarray}{
    \ovl{\avg{\mc{O}_A \mc{O}_B}_c^2} \neq \text{Tr} \(\rho \, \mc{O}_{A\cup B}\).
}
Indeed, since $\rho$ is the infinite temperature density matrix for arbitrary $p > 0$, it does not contain any information about the entanglement phase transition~\cite{nahum2018hybrid,li2018hybrid}.

\subsection{Emergent conformal symmetry at criticality}
\label{sec:CFT}

\begin{figure}[t]
    \centering
    \includegraphics[width=.49\textwidth]{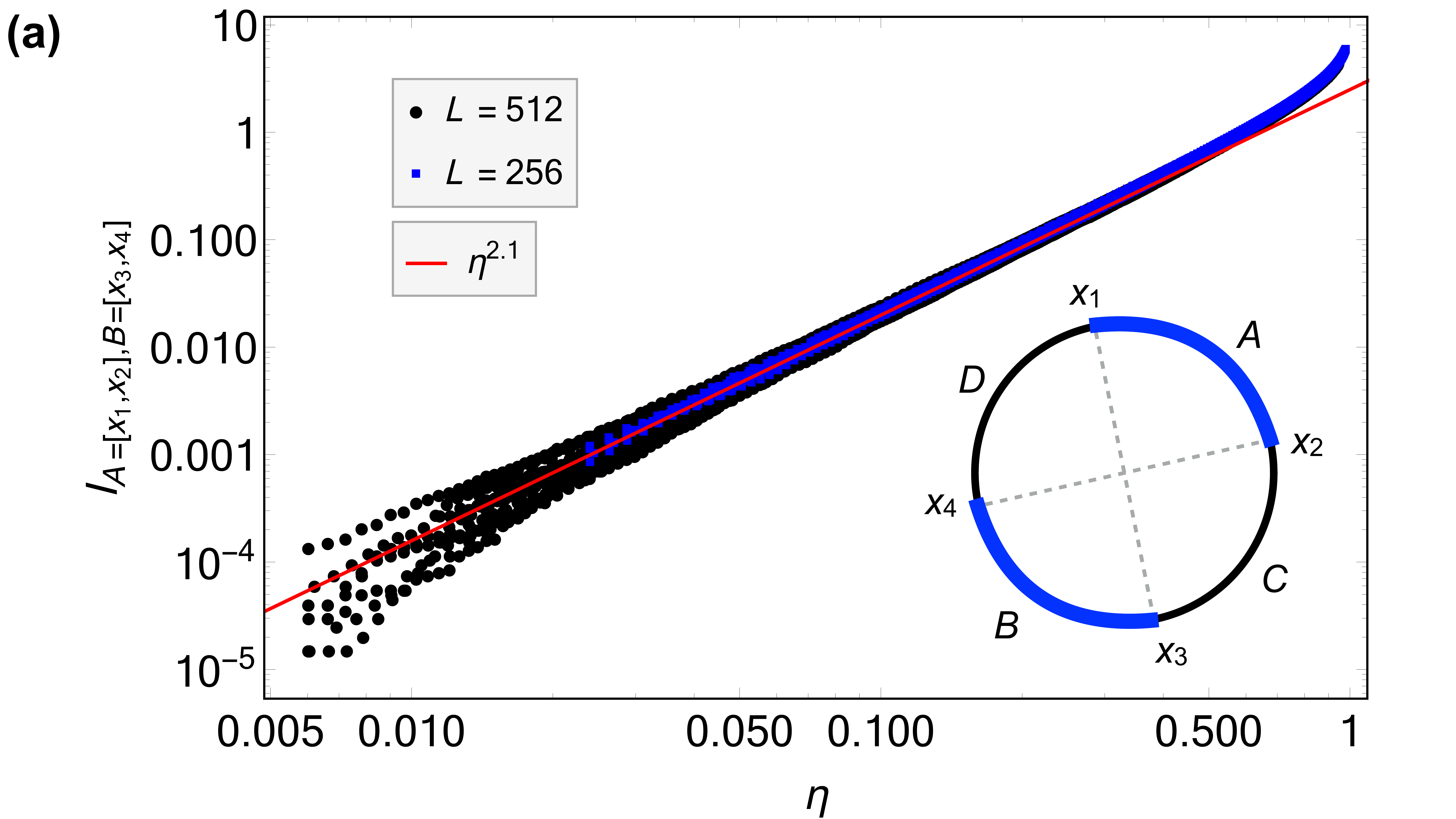}
    \includegraphics[width=.49\textwidth]{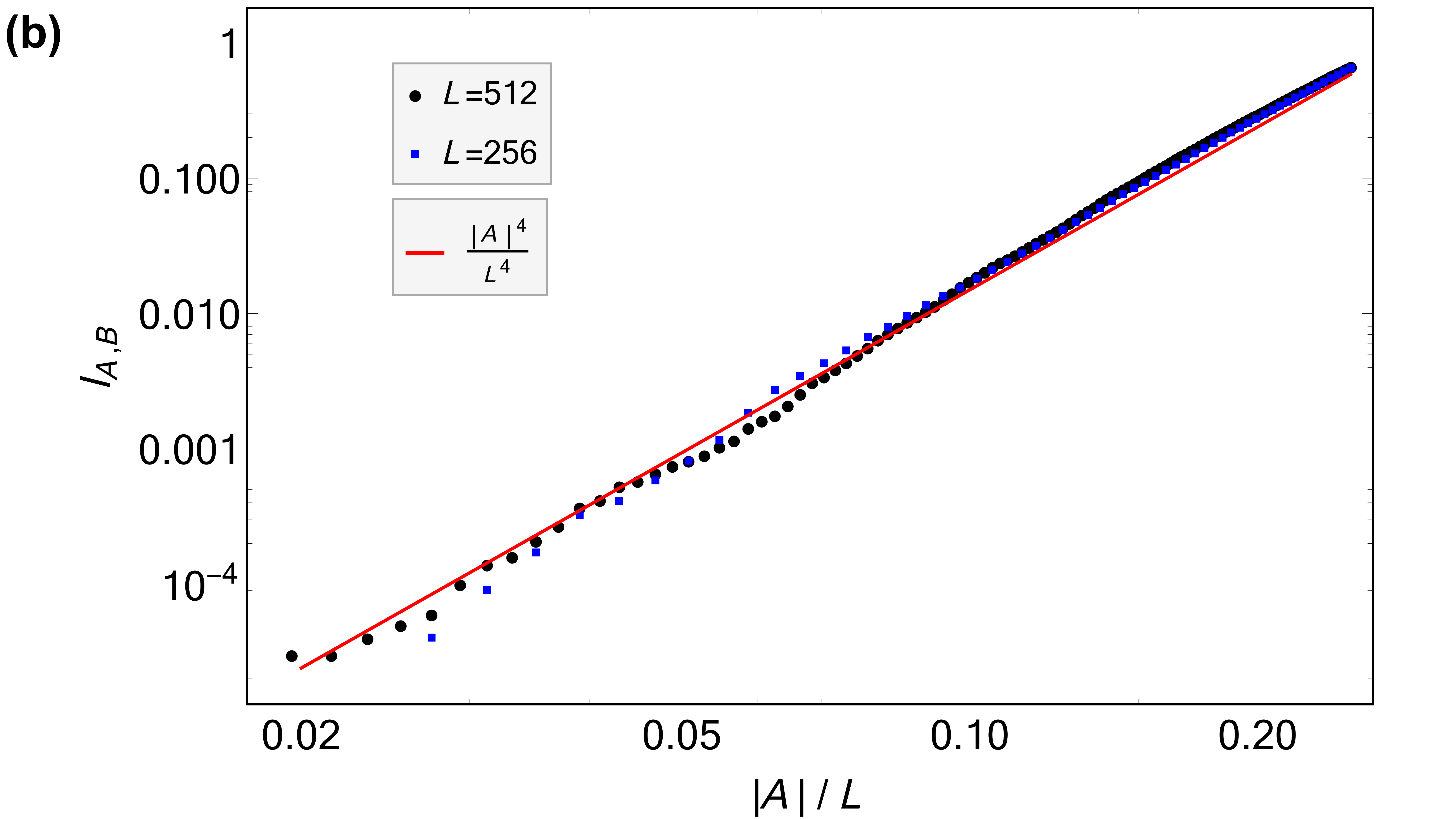}
    \caption{
    (a) Data collapse for the mutual information, $I_{A, B}$, at $p_c$ as a function of the cross ratio $\eta$, on a log-log scale. The red line corresponds to $\eta^{2.1}$.
    (b) Fitting $I_{A, B}$ at $p_c$ to Eq.~(\ref{eq:I_vs_AoverL}), where we vary $|A| = |B|$ but keep $r_{A,B} = L/2$ fixed.
    The red line shows the function $(|A|/L)^4$.
    }
    \label{fig:conf_symm}
\end{figure}

In 1d equilibrium quantum critical systems, the entanglement entropy and mutual information {of the ground state} show universal scaling behaviors, as predicted by conformal field theories (CFT)~\cite{Calabrese:2009qy}.
The logarithmic scaling of the entanglement entropy and the diverging correlation length suggest that our non-unitary entanglement transition might likewise be described by some appropriate conformal field theory~\cite{vasseur2018rtn, nahum2018hybrid}.

To check for such possible underlying conformal symmetry, we compute the mutual information between two disjoint intervals, whose size and locations can be varied.
Let $A = [x_1, x_2]$, $B = [x_3, x_4]$, $C = [x_2, x_3]$, $D = [x_4, x_1]$ be a partition of the system.
In a conventional conformal field theory the mutual information between $A$ and $B$ is related to a 4-point correlation function of boundary condition changing operators, $I_{A,B} = F \( \avg{\phi(x_1) \phi(x_2) \phi(x_3) \phi(x_4)} \)$.
As a direct consequence of the conformal symmetry, it is a function only of the cross ratio~\cite{DiFrancesco:1997nk}, i.e., 
\envelope{eqnarray}{
    \label{eq:I_eta}
    I_{A,B} = f(\eta), \text{ where } \eta \equiv \frac{x_{12} x_{34}}{x_{13}x_{24}},
}
where $x_{ij}$ is taken as the chord distance, $x_{ij} = \frac{L}{\pi} \sin\( \frac{\pi}{L}|x_i - x_j|  \)$ because of the periodic boundary condition.

We numerically compute the mutual information for a sequence of choices for the partition such that the cross ratio takes value across several orders of magnitude.
In Fig.~\ref{fig:conf_symm}(a), we plot the mutual information versus the cross ratio at the critical point.
We find that the data points lie on a single curve, confirming the prediction of CFT.
In the limit $\eta \ll 1$, we find $I_{A, B} \propto \eta^{\Delta}$, where $\Delta \approx 2$.

One interesting regime is when $A$ and $B$ are distant sites, $|A| = |B|=1 \ll r_{A, B} \ll L$.
Here $\eta \propto r_{A, B}^{-2}$, so that, 
\envelope{eqnarray}{
\label{eq:pc_corr_power_law}
I_{A, B} \propto r_{A, B}^{-2\Delta}.
}
Since the left and right boundaries of $A$ (or $B$) are close, one can apply the operator product expansion (OPE) to simplify the 4-point correlation function, and the mutual information can now be viewed as the sum of 2-point correlation functions between operators that appear in the OPE.
The dominant term comes from the operator with lowest scaling dimension, which can now be identified with $\Delta$ in the putative underlying CFT.

We can also consider another regime where $\eta \ll 1$.
Let $|A| = |B| = a L$, with $a \ll 1$ and $r_{A,B} = L/2$, so that $\eta \propto a^2$.
We thus have, 
\envelope{eqnarray}{
    \label{eq:I_vs_AoverL}
    I_{A, B} \propto \eta^\Delta \propto a^{2\Delta} = \(\frac{|A|}{L} \)^{2\Delta},
}
as verified in Fig.~\ref{fig:conf_symm}(b) with $\Delta =2$, and confirming the result in Fig.~\ref{fig:mutual_info} where the height of the peak saturates to a constant with increasing $L$.
This setup will prove useful in extracting $\Delta$ in other models.

To summarize, the numerical results strongly support an emergent conformal symmetry at the critical point, and open up the possibility of an underlying CFT description.


        \section{Circuits with Symmetry\label{sec5}}

In previous sections we have been focusing on stochastic circuit models which have three types of randomness present:
(i) spatial and temporal randomness in the unitary gates,
(ii) spatial and temporal randomness in the positions of the measurements, and
(iii) stochasticity in the measurement outcomes.
Due to (i) and (ii) these models are quite generic, with no imposed symmetries or constraints (excepting the Clifford constraints).
In this section we consider simple Clifford circuit models which have additional constraints imposed, 
involving space or time translational symmetry.
In all examples considered we find the existence of a phase transition 
sharing similar critical exponents with the random Clifford circuit.
Remarkably, this is true even for our most constrained model which
has both space and time translational symmetry in the unitary gates and the measurement locations  (spatially uniform Floquet) -- the only remaining stochasticity being the measurement outcomes.
This indicates the ubiquitous and universal character of the entanglement transition in hybrid unitary-measurement systems.

\subsection{Floquet circuits with randomly located measurements}
\begin{figure}[t]
    \centering
    \includegraphics[width=.49\textwidth]{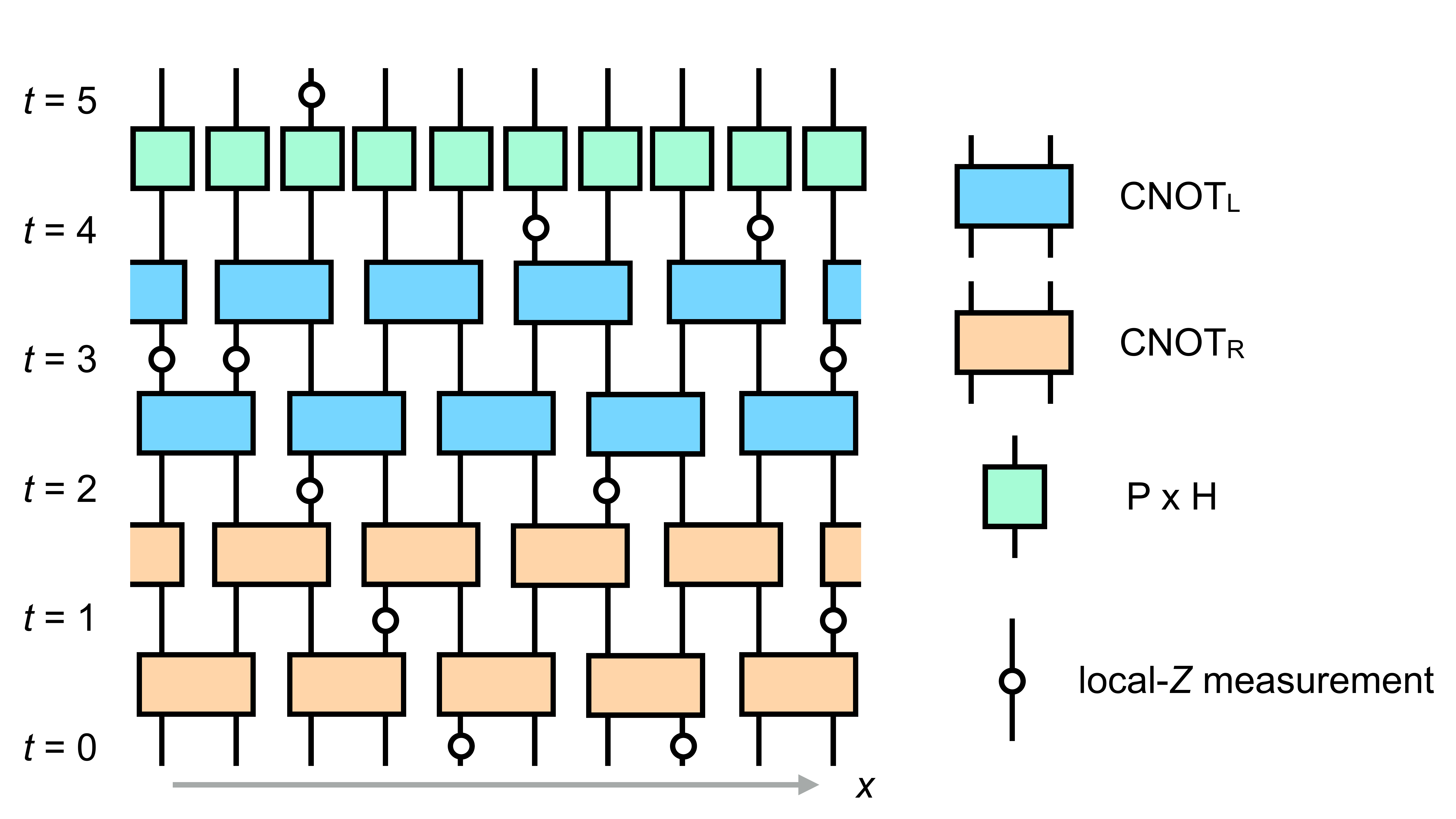}
    \caption{
    The Floquet Clifford circuit model within one time period.
    Measurements are made at random locations between each adjacent unitary layer.
    The $\text{CNOT}_{\text{L/R}}$ gate is the controlled-NOT gate with the left/right qubit as the control, and $\text{P}$ and $\text{H}$ are the phase gate and the Hadamard gate, respectively (see Appendix~\ref{appA}).
    }
    \label{fig:floquet_random_circ}
\end{figure}
\begin{figure}[t]
    \centering
    \includegraphics[width=.49\textwidth]{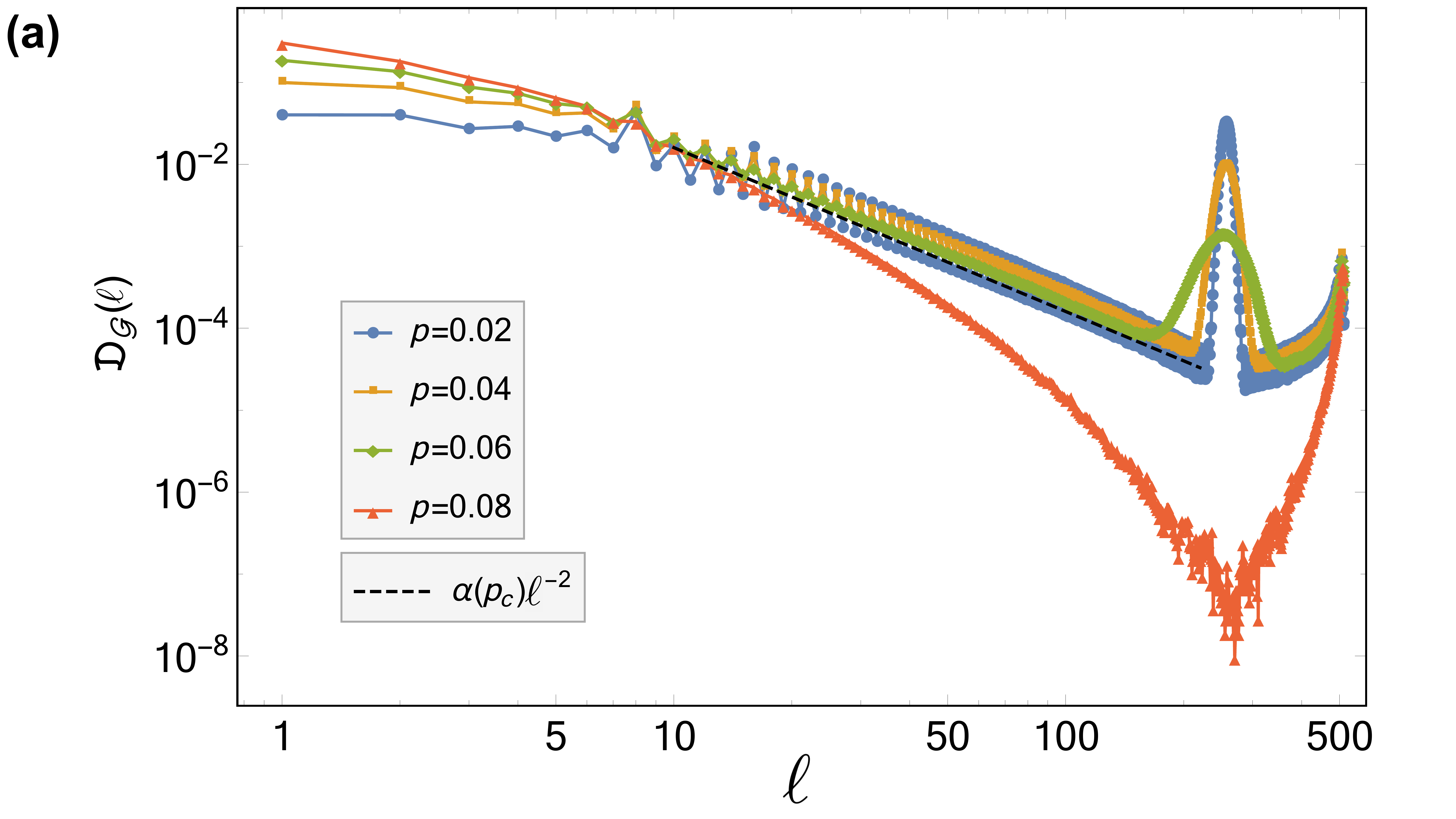}
    \includegraphics[width=.49\textwidth]{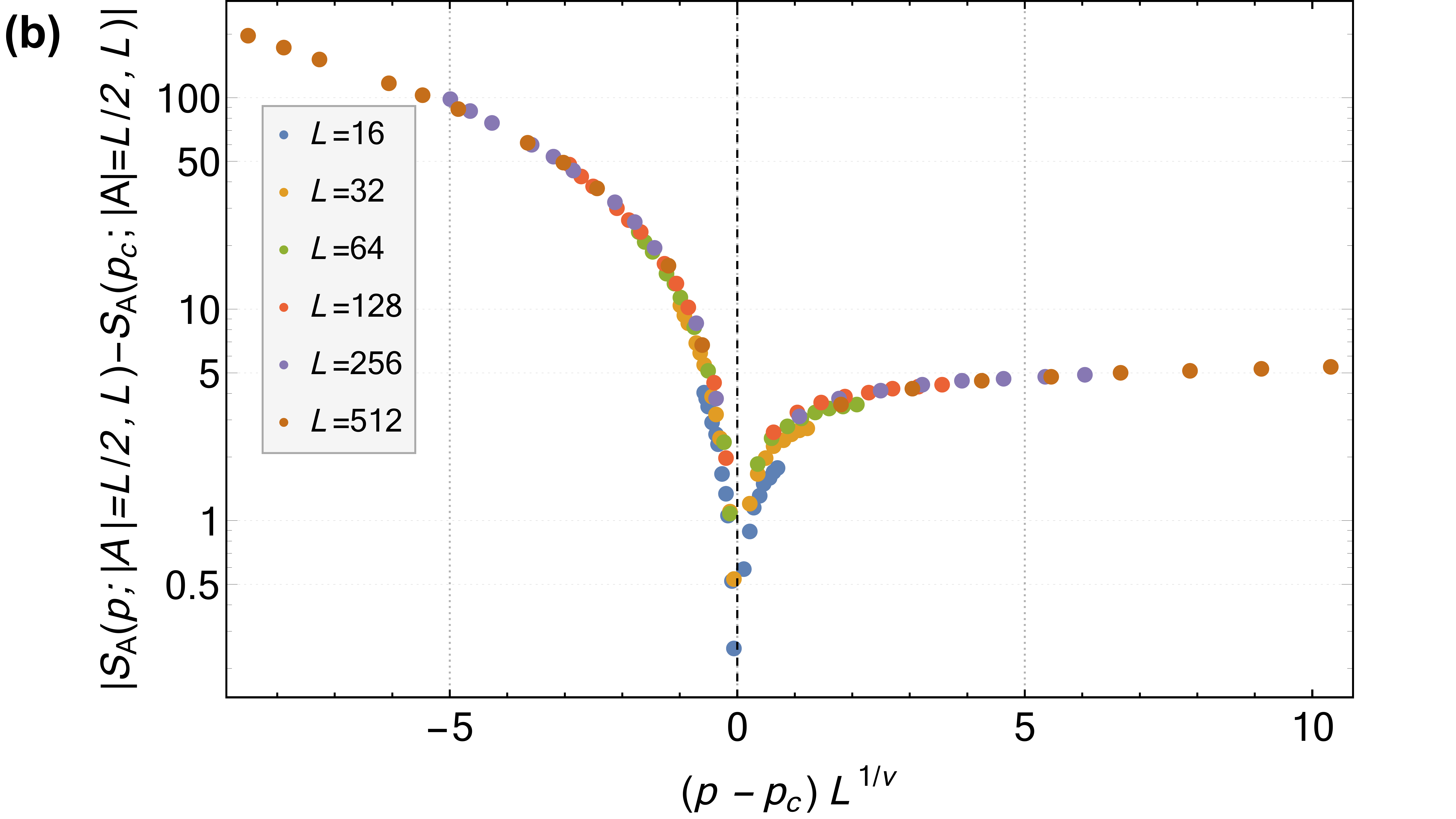}
    \includegraphics[width=.49\textwidth]{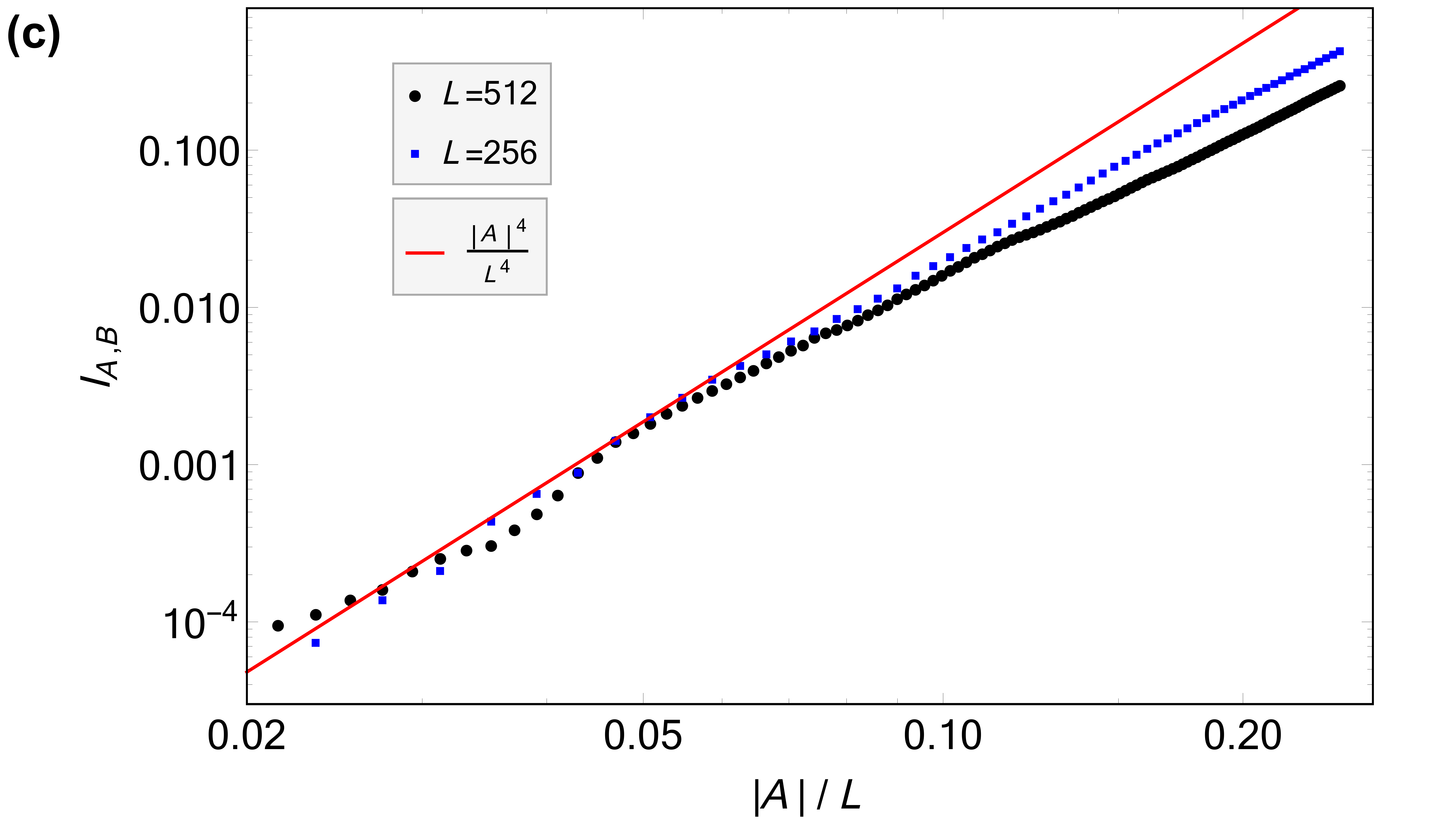}
    \caption{
    Numerical data for the circuit in Fig.~\ref{fig:floquet_random_circ}.
    (a) The normalized stabilizer length distribution for $L=512$, where $\alpha(p_c) = 1.6$.
    (b) Collapsing the $S_A(p; |A| = L/2, L)$ data to the scaling form in Eq.~\eqref{eq:Delta_S_versus_F}, where we set $p_c = 0.075$ and $\nu = 1.3$.
    (c) Mutual information at $p_c$ for the geometry as in Fig.~\ref{fig:conf_symm}(b). We can similarly extract the exponent $\Delta \approx 2$ from the data with $|A| / L \ll 1$.
    }
    \label{fig:floq_rand_result}
\end{figure}

Unitary circuit models without measurements are naturally adapted for mimicking systems with periodic drive~\cite{Kim2014,chalker2017analytic, chalker2018analytic, prosen2017analytic, prosen2018analytic}.
In such circuits, the unitary gates are periodic in time, but could be either random or regular in space. 
As for unitary Hamiltonian dynamics, there is a notion of chaos in such Floquet circuits, as diagnosed by the entanglement growth~\cite{ZhangKimHuse_2015,Bertini2018}, the operator growth (and butterfly effect in out-of-time-order correlator)~\cite{Chen2016}, the level spacing statistics and the spectral form factor~\cite{hosur:2015ylk,chalker2017analytic,prosen2018analytic,chalker2018analytic}, etc; familiar examples include the kicked Ising model, which will be discussed in the next section. 
The temporal randomness is not essential for the development of chaos.

Here we first examine the measurement-driven entanglement transition in Floquet Clifford circuits where the unitary background has both spatial and temporal translation symmetries, but the measurements are still made at random positions, as shown in Fig.~\ref{fig:floquet_random_circ}.
We choose the Floquet Clifford unitaries  to be ``chaotic'', 
having a recurrence time that is exponential in the system size
and maximal entanglement at shorter times.   For the Clifford gates shown in Fig.~\ref{fig:floquet_random_circ}
we check that this holds by examining small system sizes (data not shown).

For the circuit in Fig.~\ref{fig:floquet_random_circ} the results 
for our numerical simulation are shown in Fig.~\ref{fig:floq_rand_result}.
The stabilizer length distribution shown in Fig.~\ref{fig:floq_rand_result}(a) has a behavior very similar to that of the random Clifford circuit,
clearly indicating the existence of both a phase transition and of $S_A = \alpha \ln|A| + s |A|$ scaling of the entanglement entropy in the volume law phase. 
The coefficient of the critical logarithmic entropy, $\alpha(p_c)\approx 1.6$, as extracted from the stabilizer length distribution, is close in value to that of the random Clifford circuit. 
Moreover, we can fit the entanglement entropy data near the transition
with the finite-size scaling form in Eq.~\eqref{eq:Delta_S_versus_F} using the same critical exponent $\nu \approx 1.3$, and find a reasonable collapse (see Fig.~\ref{fig:floq_rand_result}(b)).  Finally, from the mutual information at criticality for the geometry as in Fig.~\ref{fig:conf_symm}(b), we can extract the exponent $\Delta \approx 2$ (see Fig.~\ref{fig:floq_rand_result}(c)), consistent with the random Clifford circuit results.       

\begin{figure}[t]
    \centering
    \includegraphics[width=.49\textwidth]{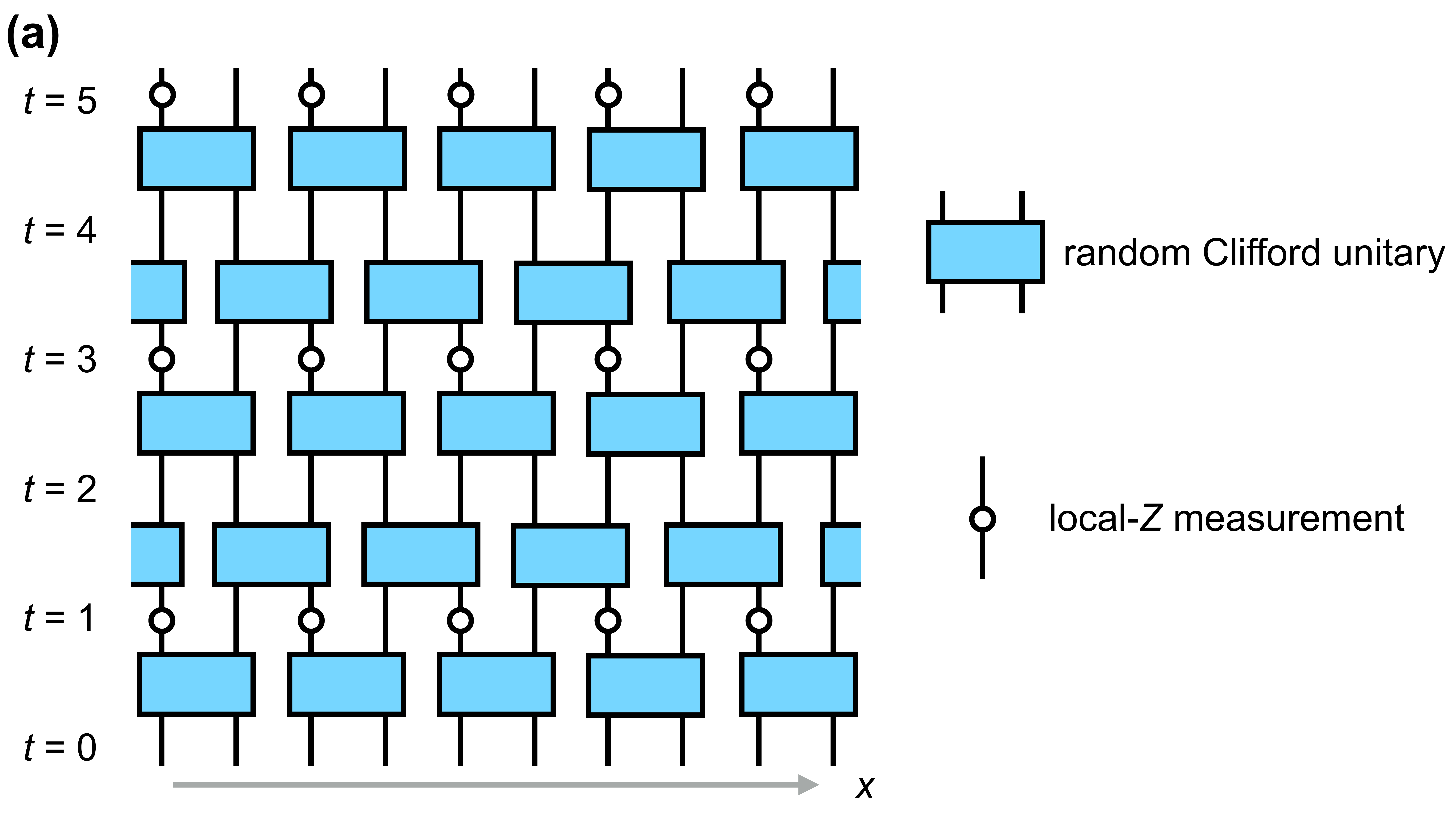}
    \includegraphics[width=.49\textwidth]{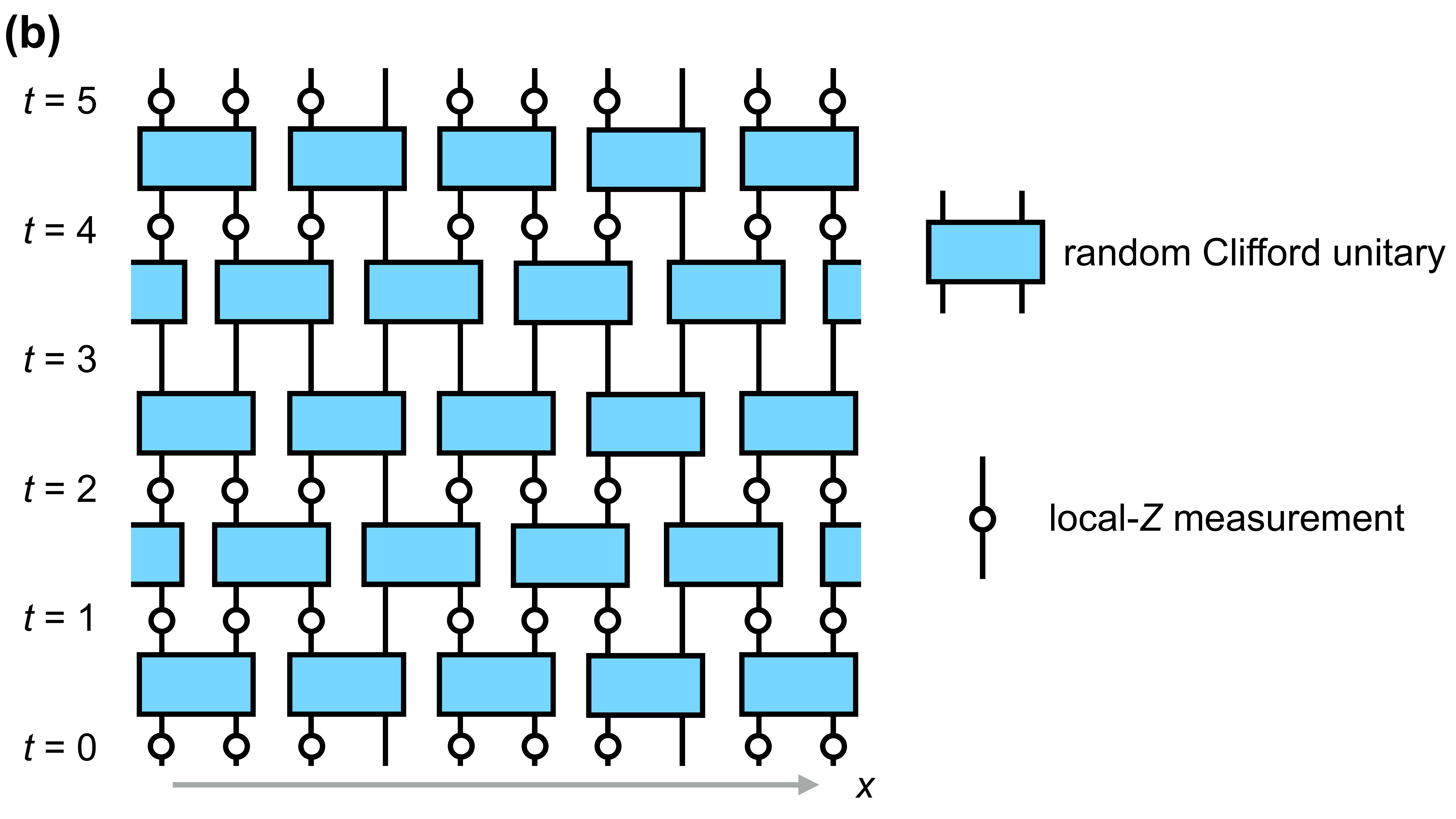}
    \caption{
    Two examples of circuits with random Clifford unitaries but quasi-periodic measurements, for (a) $p < 0.5$, and (b) $p > 0.5$.
    }
    \label{fig:rand_qp_circ}
\end{figure}

\subsection{Random unitary circuit with periodic measurements} 

\begin{figure}
    \centering
    \includegraphics[width=.49\textwidth]{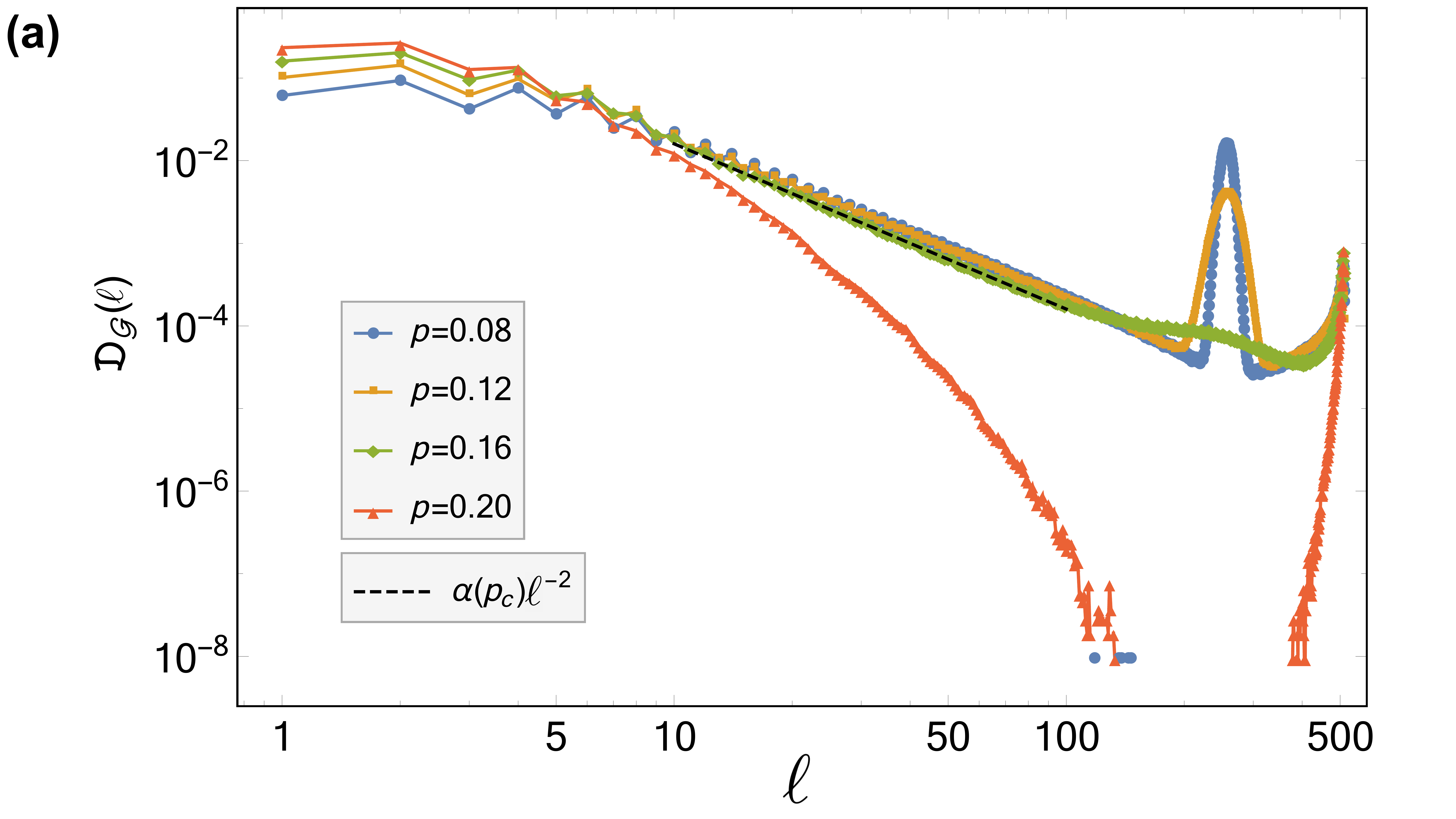}
    \includegraphics[width=.49\textwidth]{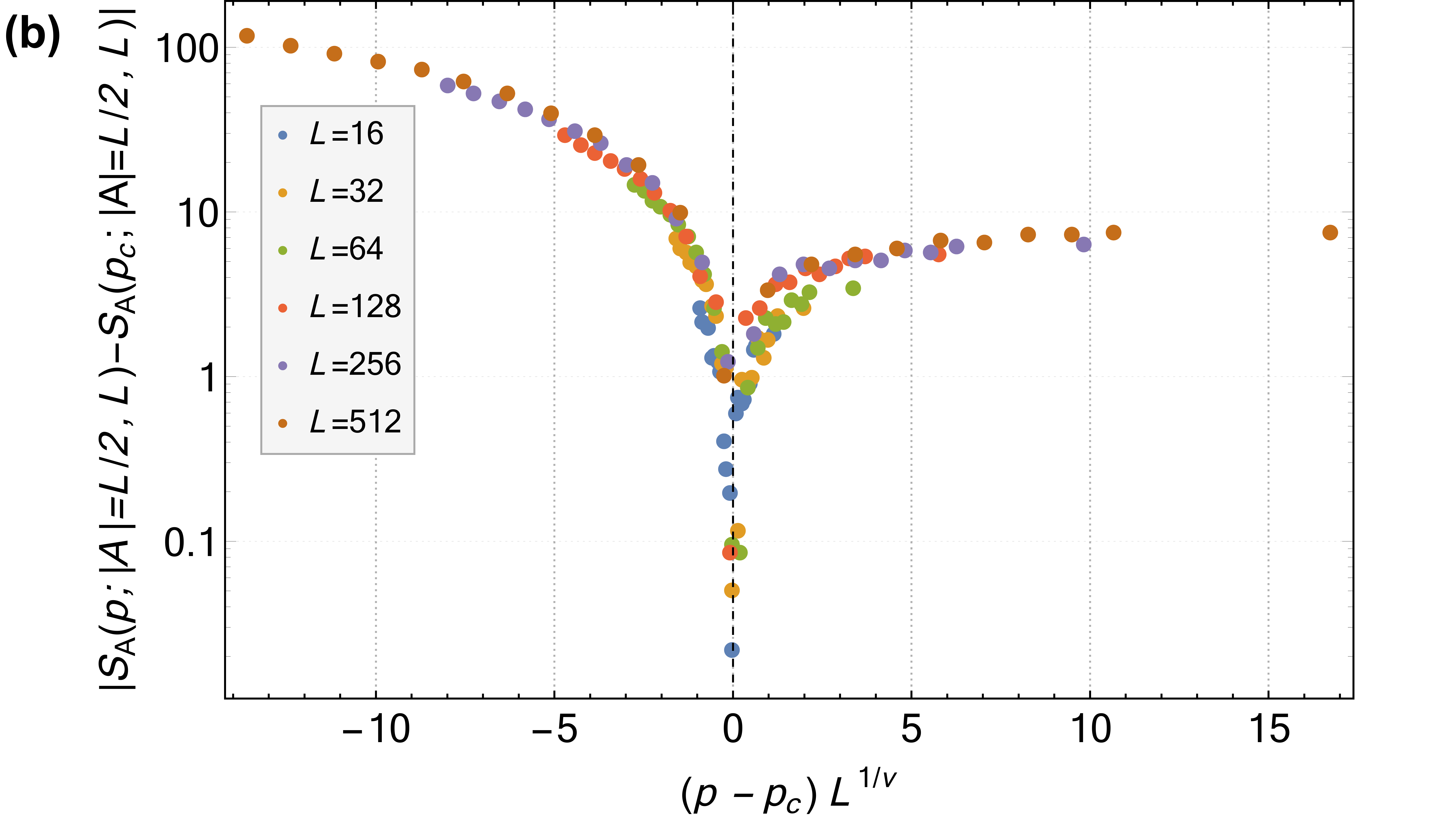}
    \includegraphics[width=.49\textwidth]{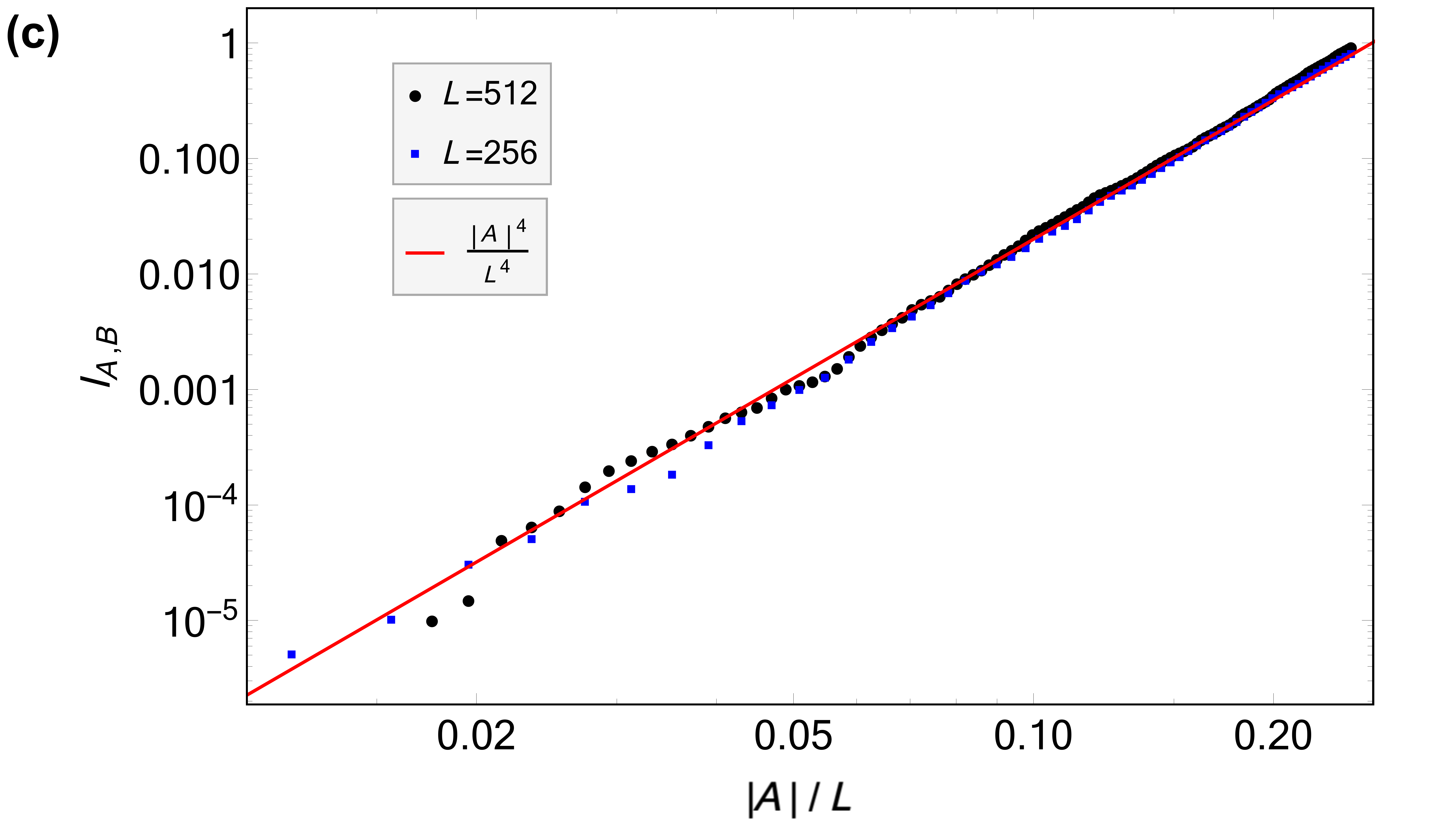}
    \caption{Numerical data for the circuit in Fig.~\ref{fig:rand_qp_circ} with periodically located  measurement gates.
    (a) The normalized stabilizer length distribution for $L=512$, where $\alpha(p_c) = 1.6$.
    (b) Collapsing the $S_A(p; |A| = L/2, L)$ data to the scaling form in Eq.~\eqref{eq:Delta_S_versus_F}, where $p_c = 0.162$ and $\nu = 1.3$.
    (c) Mutual information at $p_c$ for the same geometry as in Fig.~\ref{fig:conf_symm}(b), where we identify $\Delta \approx 2$.
    }
    \label{fig:rand_qp_result}
\end{figure}

We next consider a circuit in which the measurements are arranged (quasi-)periodically, while the background unitary circuit is still composed of random Clifford unitaries, as illustrated in 
Fig.~\ref{fig:rand_qp_circ}.  Specifically, at a fixed measurement rate $p$, for each spacetime site $(x, d)$ a measurement is made if and only if
\envelope{equation}{
        \left\lfloor x \sqrt{p} \right\rfloor < \left\lfloor (x+1) \sqrt{p} \right\rfloor, \text{ and }
        \left\lfloor d \sqrt{p} \right\rfloor < \left\lfloor (d+1) \sqrt{p} \right\rfloor,
}
where $\left\lfloor r \right\rfloor$ is the largest integer that is not greater than $r$.

In Fig.~\ref{fig:rand_qp_result}, we plot the numerical results for this circuit, and observe behavior that is essentially the same as in the earlier models -- both the random and Floquet Clifford circuit models with
randomly located measurements. 
Evidently, eliminating the randomness in the locations of the measurements 
does not change the existence -- or universality class -- of the entanglement transition.

\subsection{Circuits with space-time translational symmetry \label{sec:5C}}

\begin{figure}
    \centering
    \includegraphics[width=.49\textwidth]{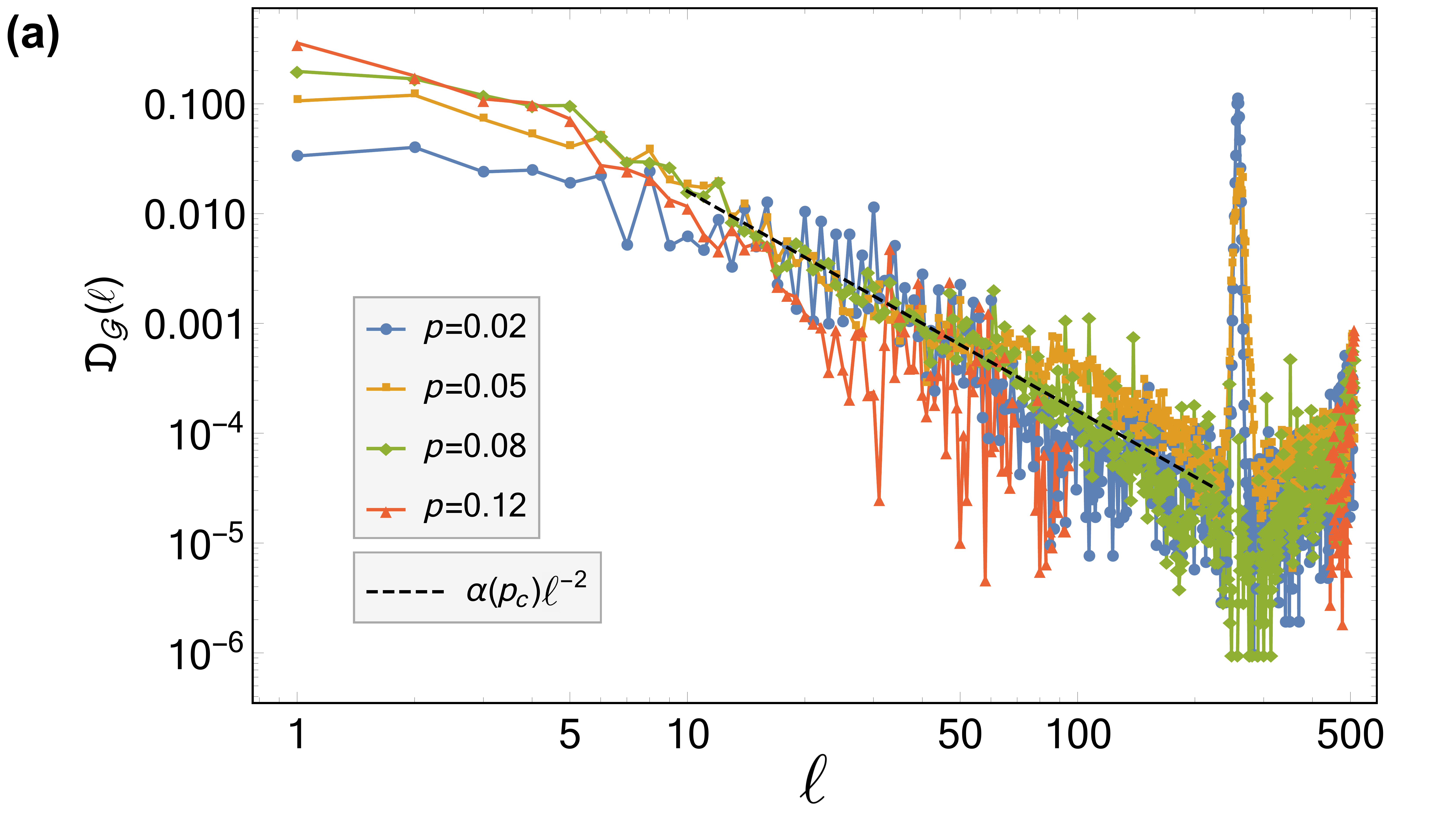}
    \includegraphics[width=.49\textwidth]{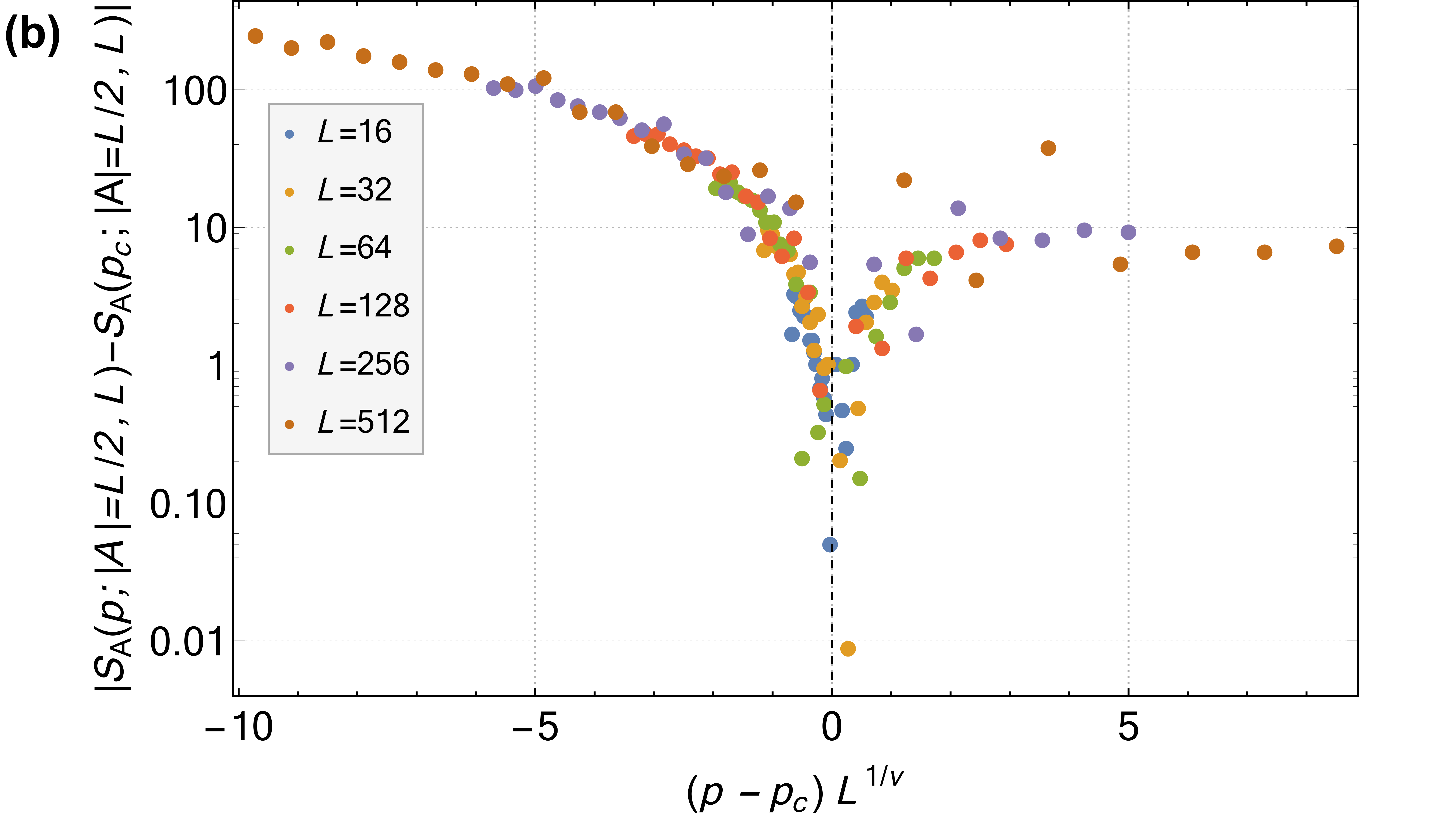}
    \caption{Data for a Clifford circuit with space-time translational symmetry, as defined in subsection~\ref{sec:5C}.
    (a) The normalized stabilizer length distribution for $L=512$, where $\alpha(p_c) = 1.6$.
    (b) Collapsing the $S_A(p; |A| = L/2, L)$ data to the scaling form in Eq.~(\ref{eq:EE_scaling_form}), where $p_c = 0.08$ and $\nu = 1.3$.
    }
    \label{fig:floq_qp_result}
\end{figure}

Lastly, we consider a circuit with translational symmetry in space and time for both the unitaries
and measurement positions.  The only remaining stochasticity is in the 
randomness in the outcome of a measurement, which is intrinsic to quantum mechanics.

In our circuit we superpose the Floquet unitary background in Fig.~\ref{fig:floquet_random_circ} with the quasi-periodic measurement pattern in Fig.~\ref{fig:rand_qp_circ}.  Numerical results are shown in Fig.~\ref{fig:floq_qp_result}.
As compared to our earlier models, we once again find essentially the same stabilizer length distribution indicative of two phases and an entanglement transition.  Moreover,
the critical exponents $\nu = 1.3$ and $\alpha(p_c) = 1.6$ at the entanglement transition are the same as in the other models.

The significant fluctuations in Fig.~\ref{fig:floq_qp_result} are due to the lack of averaging --  since we have only a single circuit in this case there is no ensemble averaging.
Moreover, for Clifford circuits with Pauli measurements, the measurement outcomes are represented by the signs of the stabilizers, and do not affect the entanglement structure or the mutual information.
Thus, the randomness in the measurement outcomes has no effect on the quantum information quantities here, and we have an almost deterministic Clifford circuit.
The only type of averaging available is as a function of time.

        \section{Beyond Clifford \label{sec6}}

In this section we explore the transition in qubit systems beyond the stabilizer formalism.

\subsection{Random Haar circuit}

\begin{figure}[t]
\centering
\includegraphics[width=.45\textwidth]{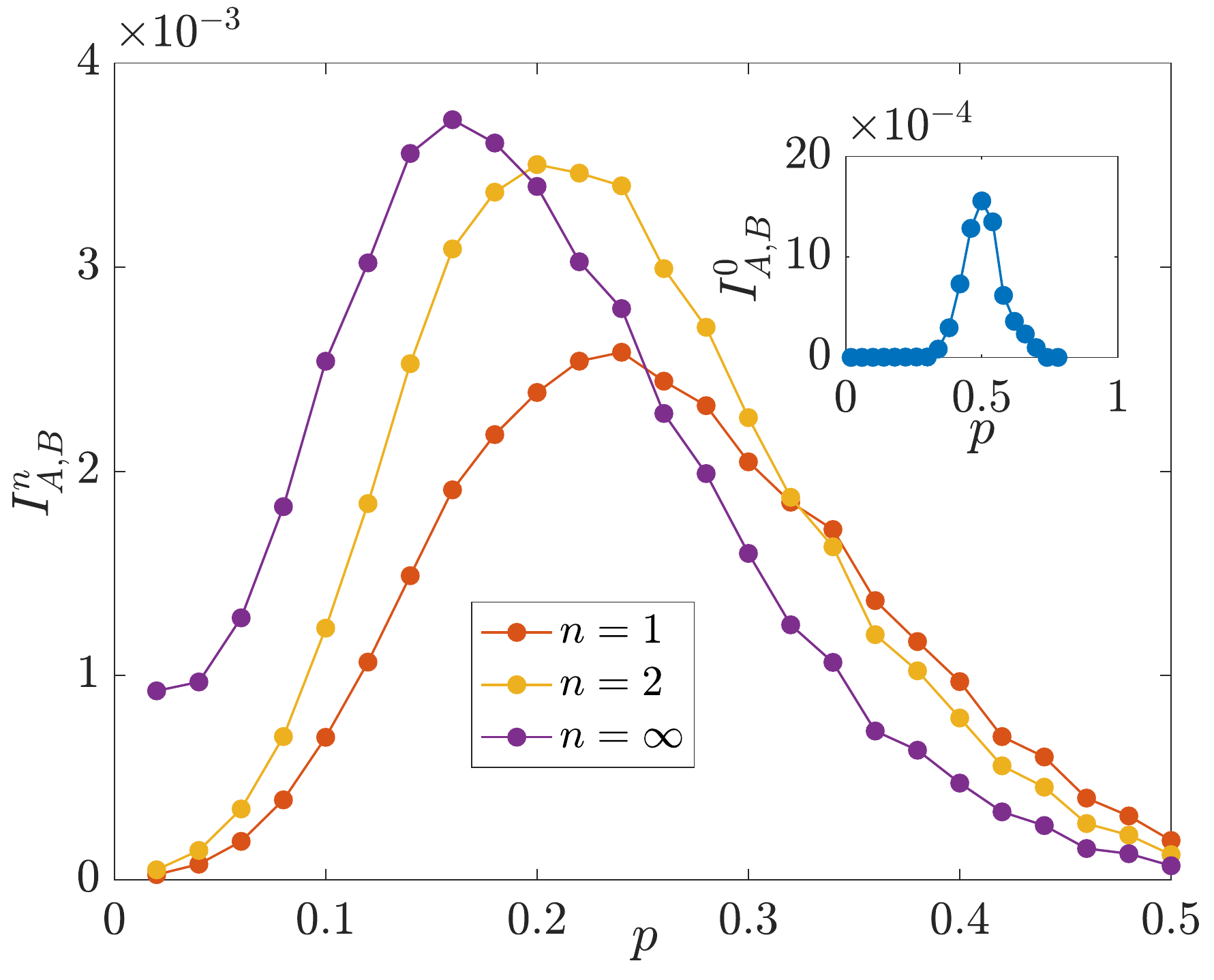}
\caption{ The mutual information for the random Haar circuit with projective measurements. In the numerical simulation the two regions $A$ and $B$ have size $|A|=|B|=1$ and are antipodal in a system with periodic boundary conditions of size $L=20$.
Here the regions $A$ and $B$ are single sites.} 
\label{fig:Haar_MI}
\end{figure}

\begin{figure}[t]
\centering
 \subfigure[]{\label{fig:Haar_Z_d} \includegraphics[width=.4\textwidth]{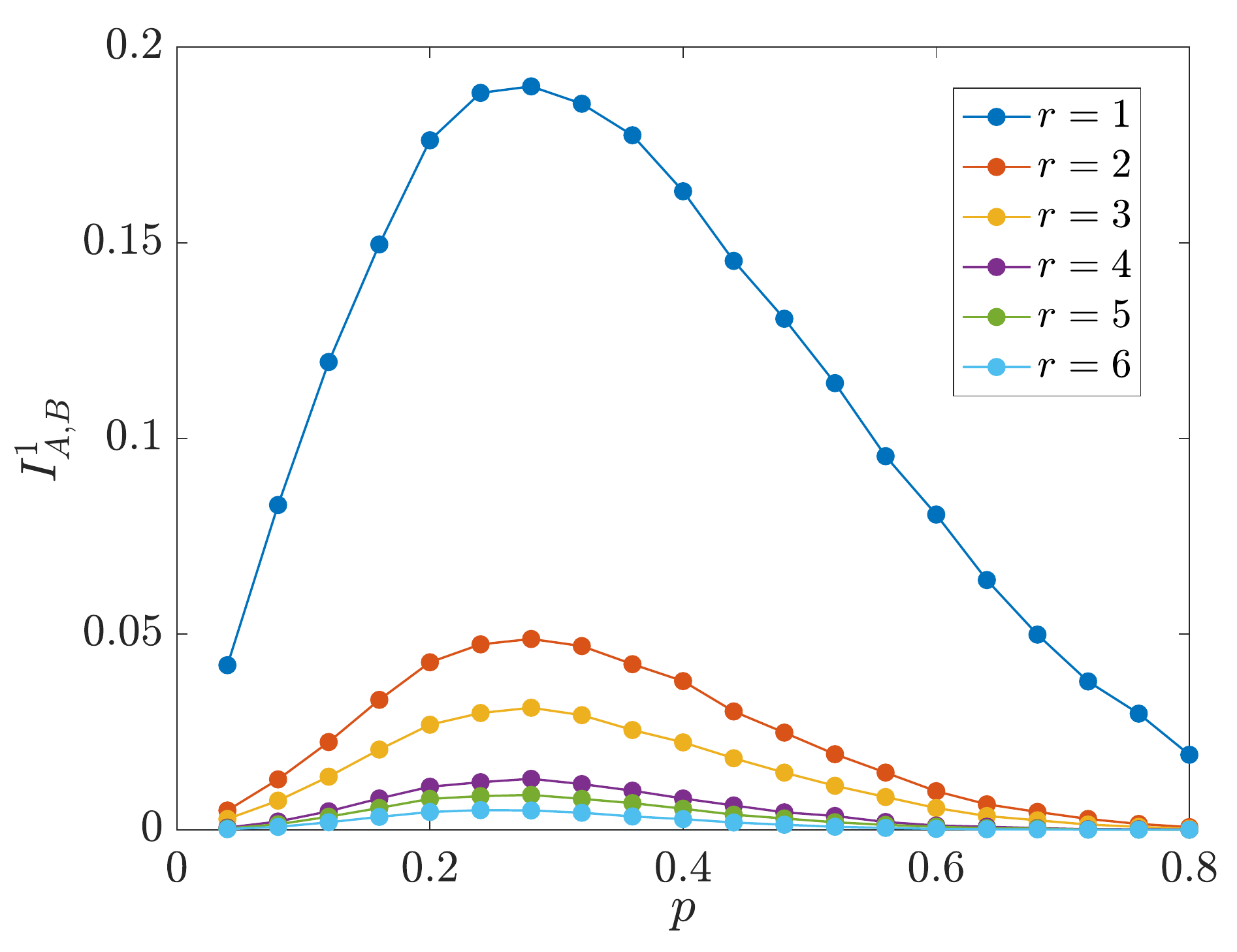}} 
 \subfigure[]{\label{fig:Haar_corr_d} \includegraphics[width=.4\textwidth]{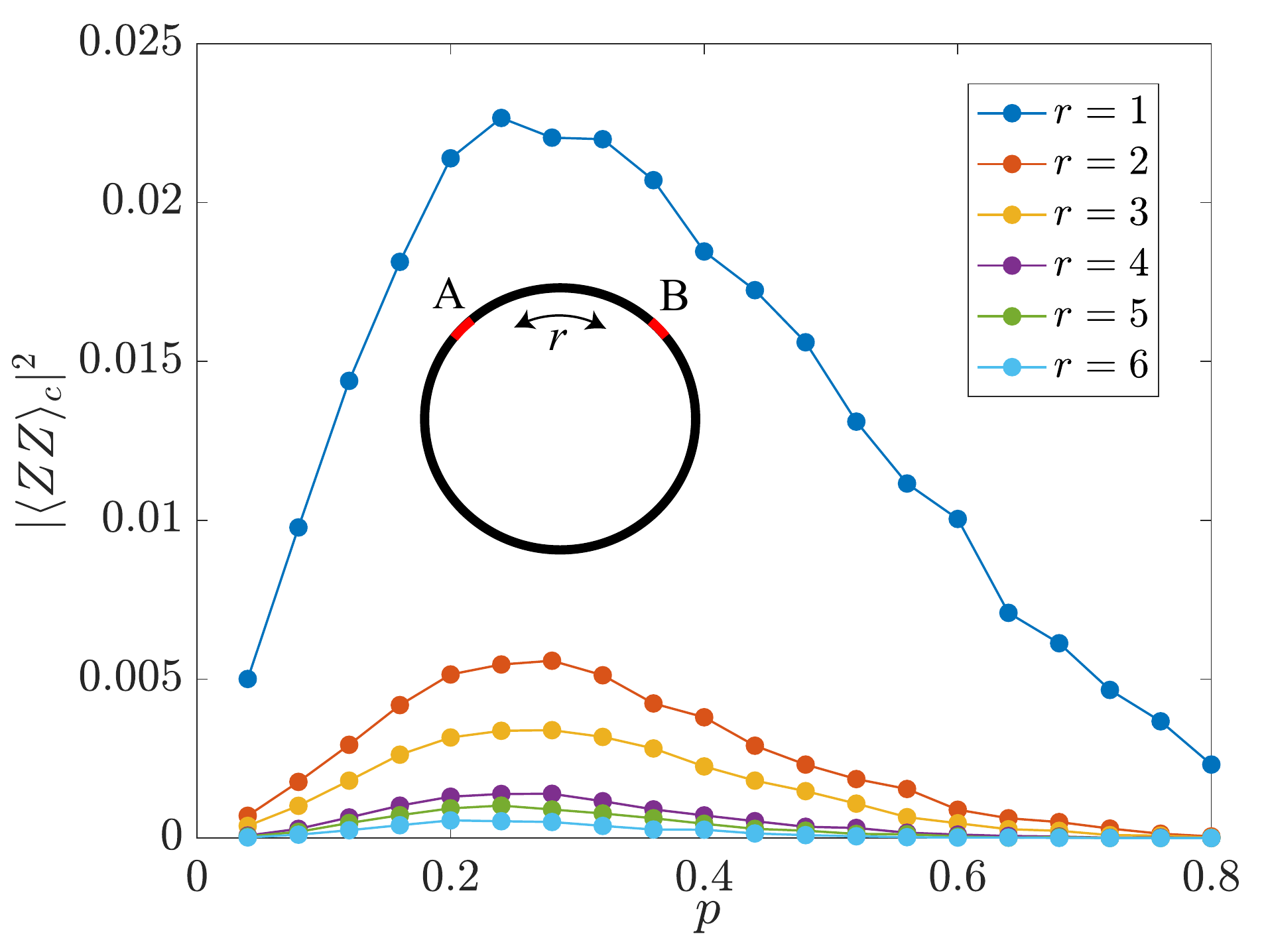}}
\caption{ Comparison between (a) the mutual information and (b) the squared correlation function, in the random Haar circuit with projective measurements.
In the numerical calculation $A$ and $B$ are separated by distance $r$ with $|A|=|B|=1$ (see the inset).} 
\label{fig:Haar_corr}
\end{figure}

Consider the random Haar circuit with the structure 
shown in Fig.~\ref{fig:rand_cliff_circ}, where each rectangle now represents a two qubit gate which is a $4\times 4$ matrix chosen randomly and independently from the Haar measure of the unitary group~\cite{mehta2004matrices, loggasrandommatrices, mezzadri2006haar}.
Without measurements, this is a minimal model to study operator dynamics and chaos propagation in systems with small onsite Hilbert space and local interaction~\cite{nahum2018operator,keyserlingk2018operator}.
With measurements, it is the most generic model in which the unitary-measurement dynamics can be addressed.

\subsubsection{Random Haar circuit with projective measurements}

We first consider the random Haar circuit with projective measurements.
As in Fig.~\ref{fig:rand_cliff_circ}, the single site projective measurements, taken to be $P_\pm = (1\pm Z)/2$, are introduced on each site independently with probability $p$. 
This model is closest in spirit to the random Clifford circuit studied in Sec.~\ref{sec3} and \ref{sec4}, with which comparisons should be made.

As for the Clifford circuits, we use mutual information between two antipodal regions 
(in a system with periodic boundary conditions) to diagnose the putative phase transition.
This approach is particularly useful for small systems with $L=20$, where it is hard to distinguish between volume law and area law scaling behavior by directly looking at the entanglement entropy.
The numerical results, where the two regions are taken to be single sites, are shown in Fig.~\ref{fig:Haar_MI}.

We notice that the mutual information for all R{\'e}nyi indices show a peak, signifying the existence of a transition.
Within the Haar circuit, R{\'e}nyi entropies and the mutual information can depend on the R{\'e}nyi index $n$, and we discuss them separately.
For 
$I^0_{A, B}$, the peak is located at $p_c=0.5$, as predicted by the percolation mapping~\cite{nahum2018hybrid} (see the inset of Fig.~\ref{fig:Haar_MI}).
This situation is different for $I^n_{A,B}$ with $n \ge 1$, whose peaks are located at $p$ much smaller than $0.5$, and there is no obvious mapping to percolation \footnote{
    Notice that for R{\'e}nyi indices greater than $1$, there is no subadditivity of entanglement, and the mutual information is not necessarily non-negative, although in our data the mean values are never negative.
}.
While these peaks are rather broad due to finite size effects, they sit close to one another, suggesting that $p_c$ is independent of $n$ for $n \geq 1$ -- i.e. there is a single transition (instead of a different transition for each $n$).

As discussed in Sec.~\ref{sec4}, the fluctuation in the connected correlation function is upper bounded by the mutual information. 
We consider the following quantity in this model,
\envelope{eqnarray}{
    \ovl{\lvert \avg{\mc{O}_A \mc{O}_B}_c\rvert^2}, \text{ where } \mc{O}_A = Z_{1} \text{ and } \mc{O}_B = Z_{r + 1},
}
and the distance $r$ is varied.
In our numerical calculations shown in Fig.~\ref{fig:Haar_corr}, we find that it takes a similar form as $I^1_{A,B}$ and has a peak at the corresponding $p_c$.

\subsubsection{Random Haar circuit with generalized measurements}

\begin{figure}[t]
\centering
 \subfigure[]{\label{fig:Haar_weak_MI} \includegraphics[width=.4\textwidth]{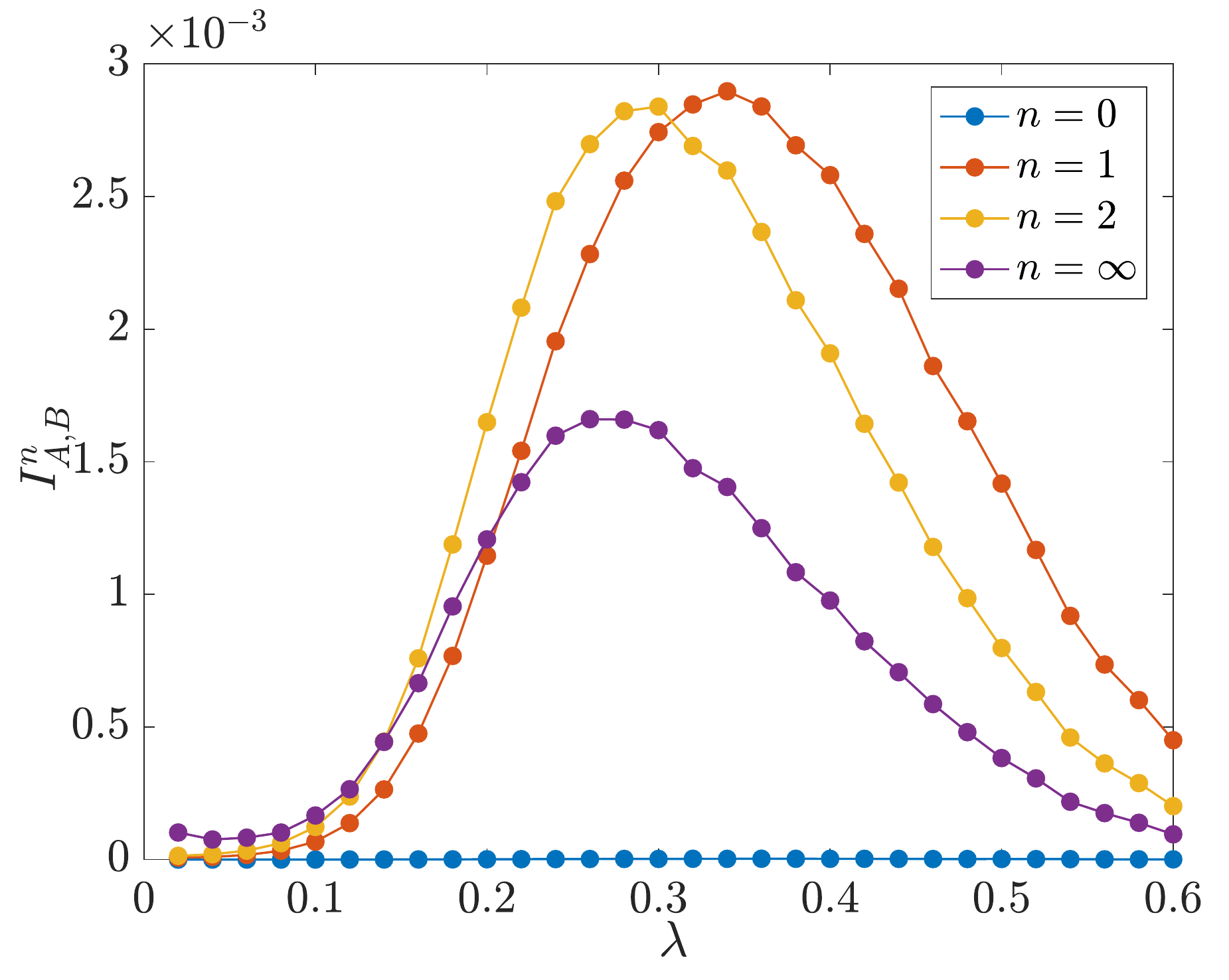}} 
  \subfigure[]{\label{fig:Haar_weak_corr_d} \includegraphics[width=.4\textwidth]{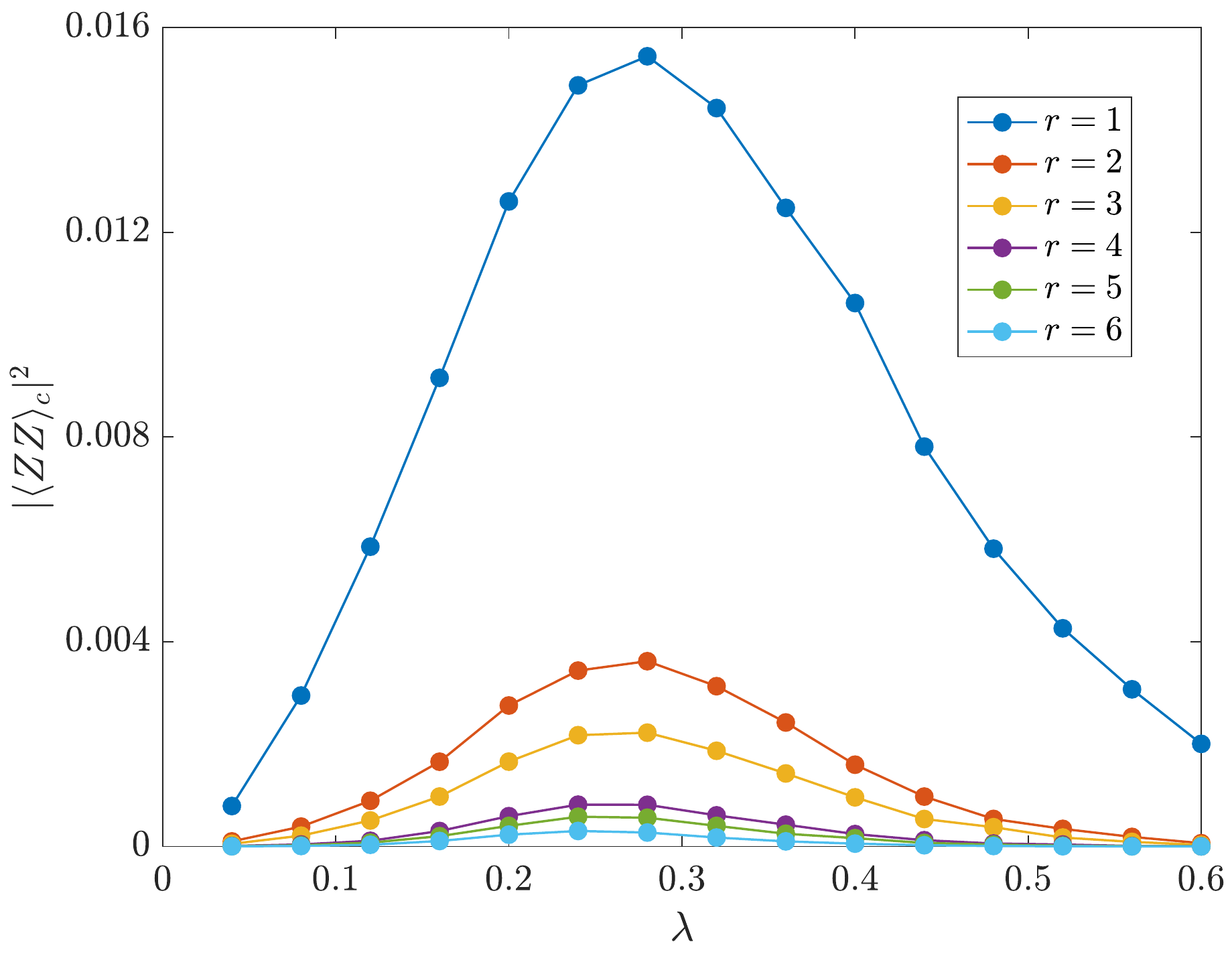}} 
\caption{Data for the Haar unitary circuit with generalized measurements. (a) The mutual information, $I^1_{A,B}$, where $A$ and $B$ are antipodal in the system with periodic boundary conditions. (b) The squared correlation function as a function of $\lambda$. Here the two intervals $A$ and $B$ are separated by distance $r$ (see the inset of Fig.~\ref{fig:Haar_corr_d}). In both (a) and (b), we take $L=20$ and $|A|=|B|=1$. } 
\label{fig:Haar_weak}
\end{figure}

Projective measurements can be generalized to measurements that model imperfect measuring devices, known as ``generalized measurements'' or ``weak measurements''~\cite{nielsen2010qiqc}.  Here, the coupling between the system and the measuring device is weak, and less information ($\le$ one bit) is extracted from the system by one such measurement. 
We consider a model in which the single site measurement gates in Fig.~\ref{fig:rand_cliff_circ} are taken to be generalized measurements with operators,
\begin{align}
 M_{\pm}=\frac{1\pm\lambda Z}{\sqrt{2(1+\lambda^2)}}.
\end{align}
These measurement operators satisfy the required completeness relation, ${M}_+^\dag {M}_+ +{M}_-^\dag {M}_- = 1$.
The parameter $\lambda$ represents the measurement strength: in the limit $\lambda\to 0$, the system and the measuring device are totally decoupled and $M_\pm$ acts trivially on the wavefunction, while in the limit $\lambda\to 1$, it becomes a projective measurement.
For simplicity, we take the measurement rate $p=1$ so that the generalized measurements are uniformly applied to each and every qubit in the circuit.
Notice that these generalized measurements do not have a Clifford counterpart.

In Fig~\ref{fig:Haar_weak_MI}, we present results for $I^n_{A,B}$, where we find a peak for $n\geq 1$.
The closeness of the peaks again suggests a single phase transition, as in the Haar circuit with projective measurements.
Compared to the projective measurement case we note that here there is no phase transition in $S_A^0$ -- as long as $\lambda<1$, $S_A^0$ obeys a volume law.
Moreover, we compute the squared correlation function and find a peak close to $\lambda_c$ (see Fig.~\ref{fig:Haar_weak_corr_d}).

Despite the uniformly imposed generalized measurements, the wavefunctions are not completely disentangled as long as $\lambda < 1$.  Moreover, the volume law phase is stable for $\lambda < \lambda_c$.

\subsection{Floquet Ising circuits}

\begin{figure}[t]
\centering
 \includegraphics[width=.5\textwidth]{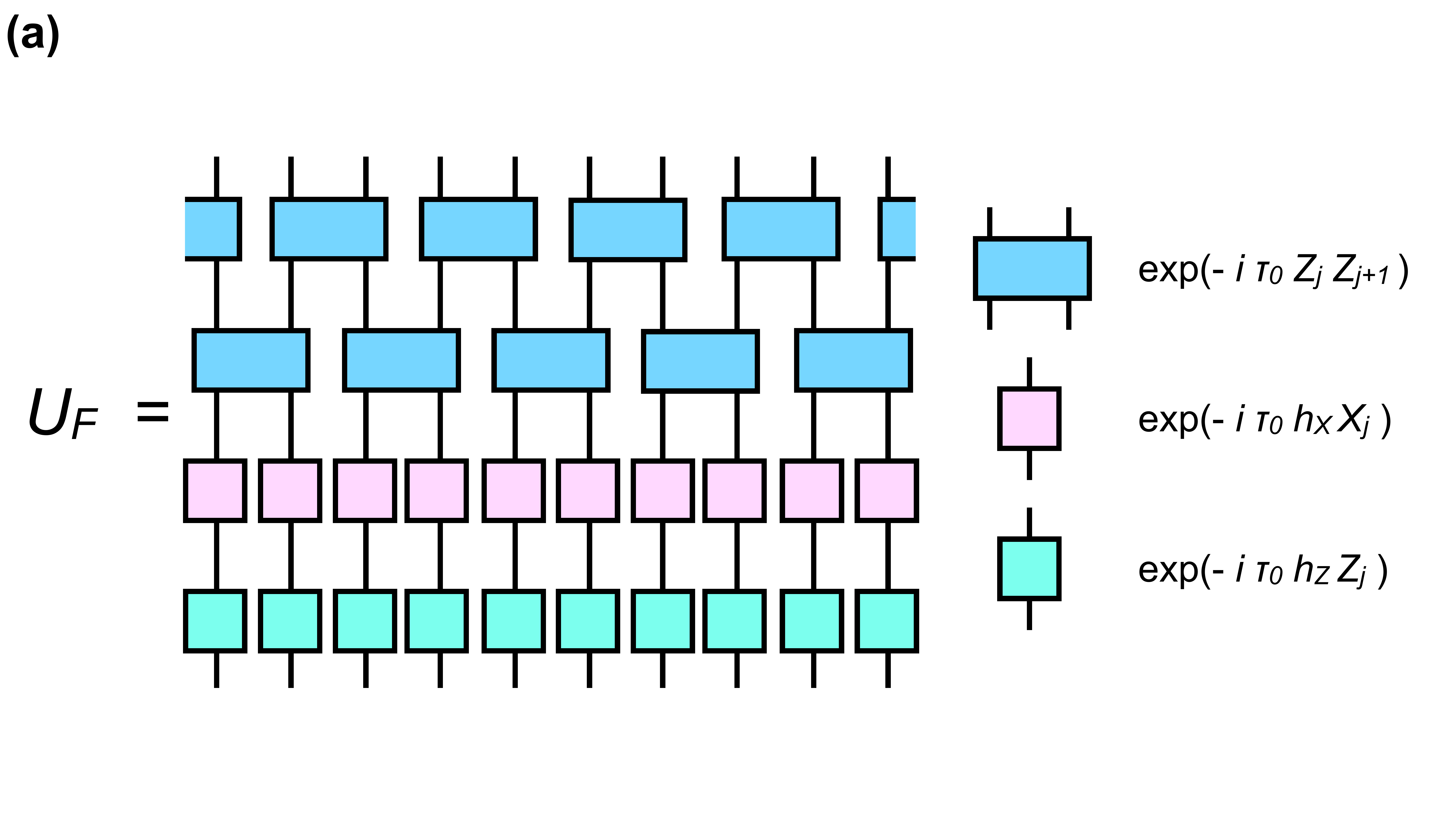}
 \includegraphics[width=.5\textwidth]{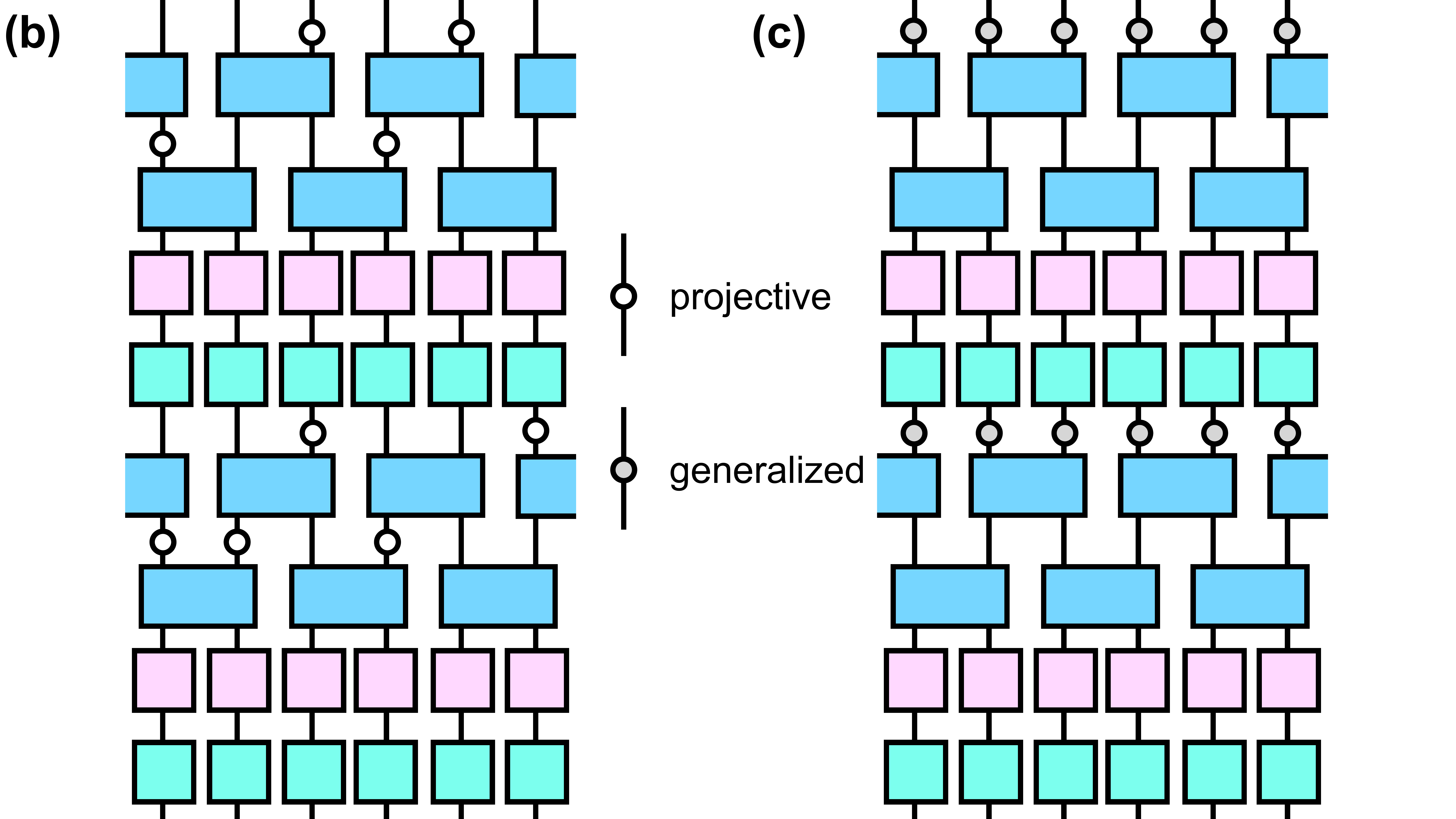}
\caption{ (a) The Floquet operator is specified by a quantum circuit. (b) The projective measurements are introduced in the circuit after each two-qubit gate layer with probability $p$. (c) The generalized (weak) measurements are applied uniformly with $p=1$ in the circuit after each Floquet operator, ${U}_F$.}
\label{fig:Floq_circuit}
\end{figure}

As a 
generalization of the Floquet Clifford circuits from Sec.~\ref{sec5}, we consider a Floquet Ising spin chain model with the following Floquet operator,
\begin{align}
{U}_F=\exp[-i \tau_0 {H}_Z]\exp\left[-i\tau_0 H_X\right],
\label{flo_op}
\end{align}
where
\begin{align}
&{H}_X=  h_X  \sum_{j=1}^{L}  X_j, \nonumber\\
&{H}_Z=\sum_{j=1}^{L-1} Z_j Z_{j+1}+ h_Z \sum_{j=1}^L  Z_j.
\label{floq}
\end{align}
The Floquet operator defines a one-dimensional periodically driven system with period $T=2\tau_0$.
This Floquet model is integrable when $h_Z = 0$. 
We will focus on the generic non-integrable case with $h_Z \neq 0$.
The circuit in Fig.~\ref{fig:Floq_circuit}(a) represents a particular discretization of the Floquet operator that we adopt.
For the special parameter set, $(\tau_0, h_X, h_Z)=(\pi/4, 1, 1)$, the discretized Floquet operator falls within the Clifford group.
Without measurements, the Floquet circuit has both temporal and spatial translational symmetries, and no randomness is present.

 \subsubsection{Floquet Ising circuit with projective measurements}
 
\begin{figure}[t]
\centering
\includegraphics[width=.45\textwidth]{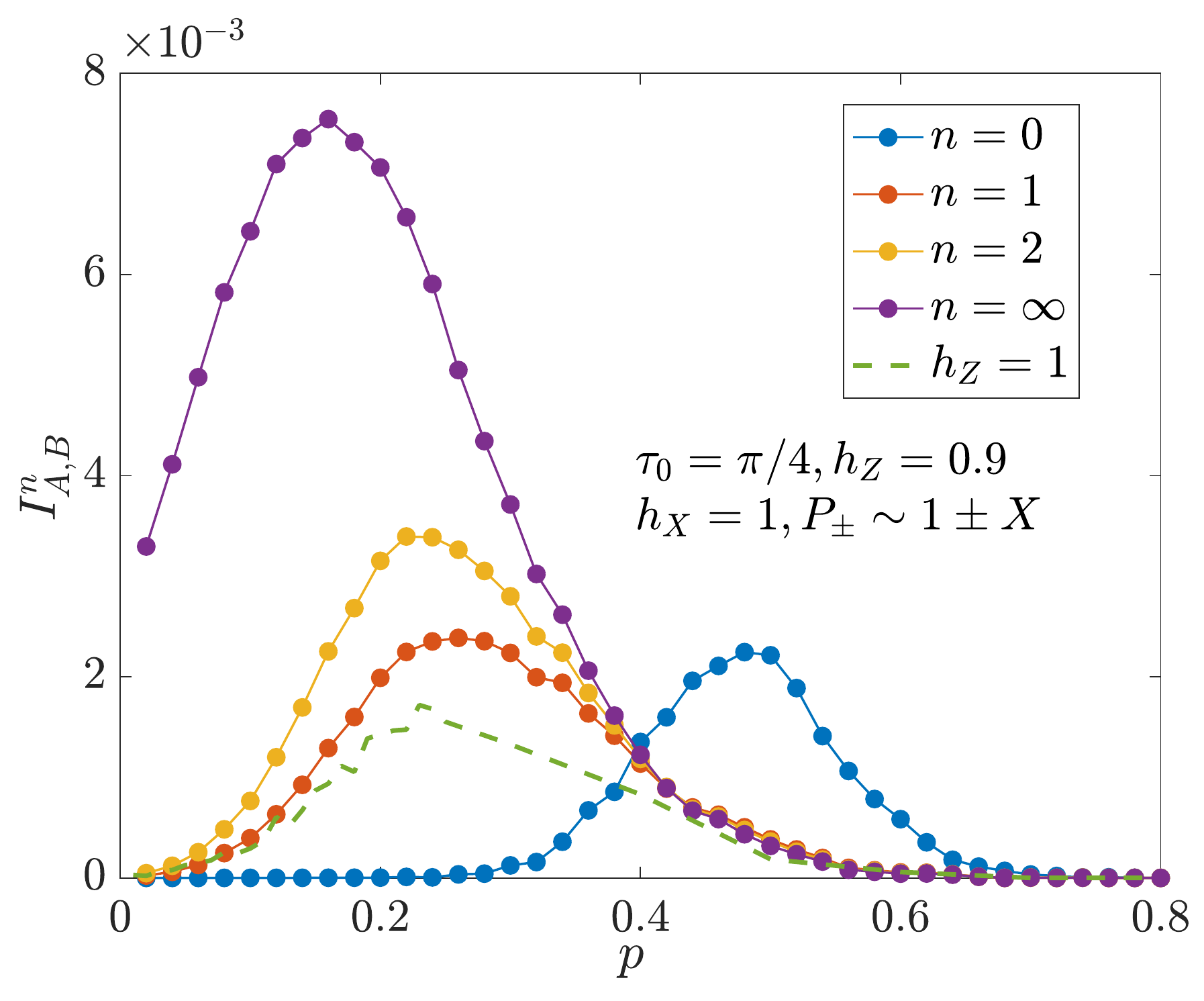}
\caption{ Mutual information $I^n_{A,B}$ for the Floquet spin chain model with projective measurements. $A$ and $B$ are antipodal in the periodic boundary condition. We take $L=20$ and $|A| = |B| = 1$.}
\label{fig:Floq_projective}
\end{figure}


We introduce projective measurements in the Floquet circuit (see Fig.~\ref{fig:Floq_circuit}(b)), taking the measurement gates to be $ P_\pm = \frac{1}{2} \(1 \pm X\)$.
The single site projective measurements are applied randomly in the same fashion as in Fig.~\ref{fig:floquet_random_circ}.

In Fig.~\ref{fig:Floq_projective} we show data for the mutual information as a function of $p$.
Here we have taken the parameter $h_Z=0.9$, with the rest of the parameters the same as the Clifford parameters.
There is a peak in $I^n_{A,B}$, with the location of the peak depending weakly on the R{\'e}nyi index, which we identify as $p_c$.
Again, the data supports the existence of the entanglement transition.

The dashed line in Fig.~\ref{fig:Floq_projective} shows the data for $h_Z = 1.0$, i.e. the Clifford limit in which there is no $n$ dependence.  The Clifford curve is close to the $n=1$ curve
for the non-Clifford circuit and gives a consistent estimation of $p_c$.
This comparison further justifies using the Clifford circuits as a convenient stand-in
for more generic (non-Clifford) quantum circuits.



 \subsubsection{Floquet Ising circuit with generalized measurements}

\begin{figure}[t]
\centering
\includegraphics[width=.49\textwidth]{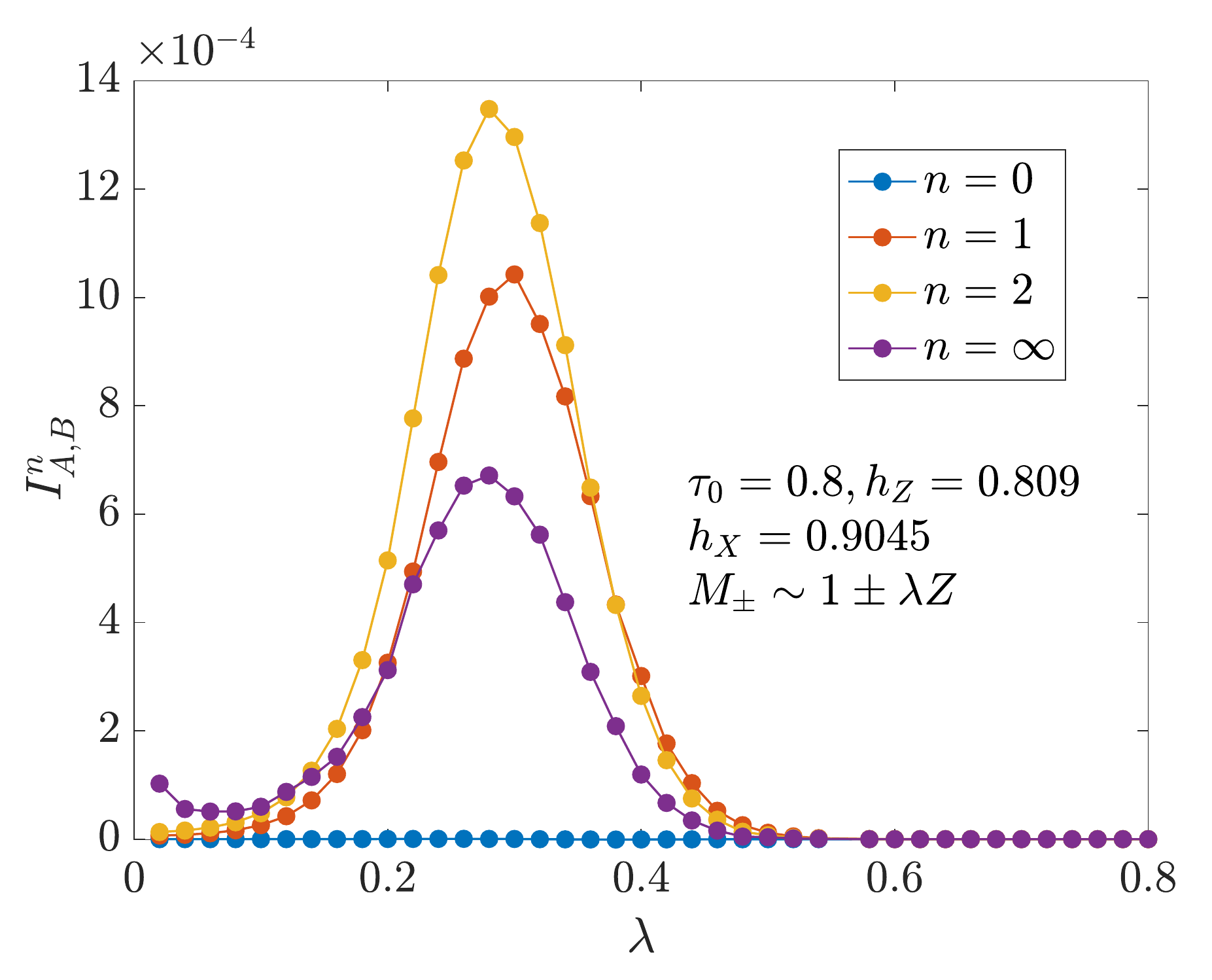}
\caption{ Mutual information $I^n_{A,B}$ for the Floquet spin chain model with generalized measurements. $A$ and $B$ are antipodal in a system of size $L=20$ with periodic boundary conditions, while $|A|=|B|=1$. The Floquet parameters are chosen as $(\tau_0,h_X,h_Z)=(0.8,0.9045,0.809)$~\cite{Kim2014}.
}
\label{fig:Floq_weak}
\end{figure}

We next introduce generalized measurements in the Floquet spin chain model,
again taking the measurement rate $p=1$, so that the generalized measurements are uniformly applied at each and every site after ${U}_F$ (see Fig.~\ref{fig:Floq_circuit}(c)).
The result for the mutual information is presented in Fig.~\ref{fig:Floq_weak}.
Once again, the presence of the peak is indicative of an entanglement transition.
As in the random Haar circuit with generalized measurements, there is no phase transition in $S_A^0$.



\subsection{Various properties at criticality}
\begin{figure}[b]
\centering
 \includegraphics[width=.49\textwidth]{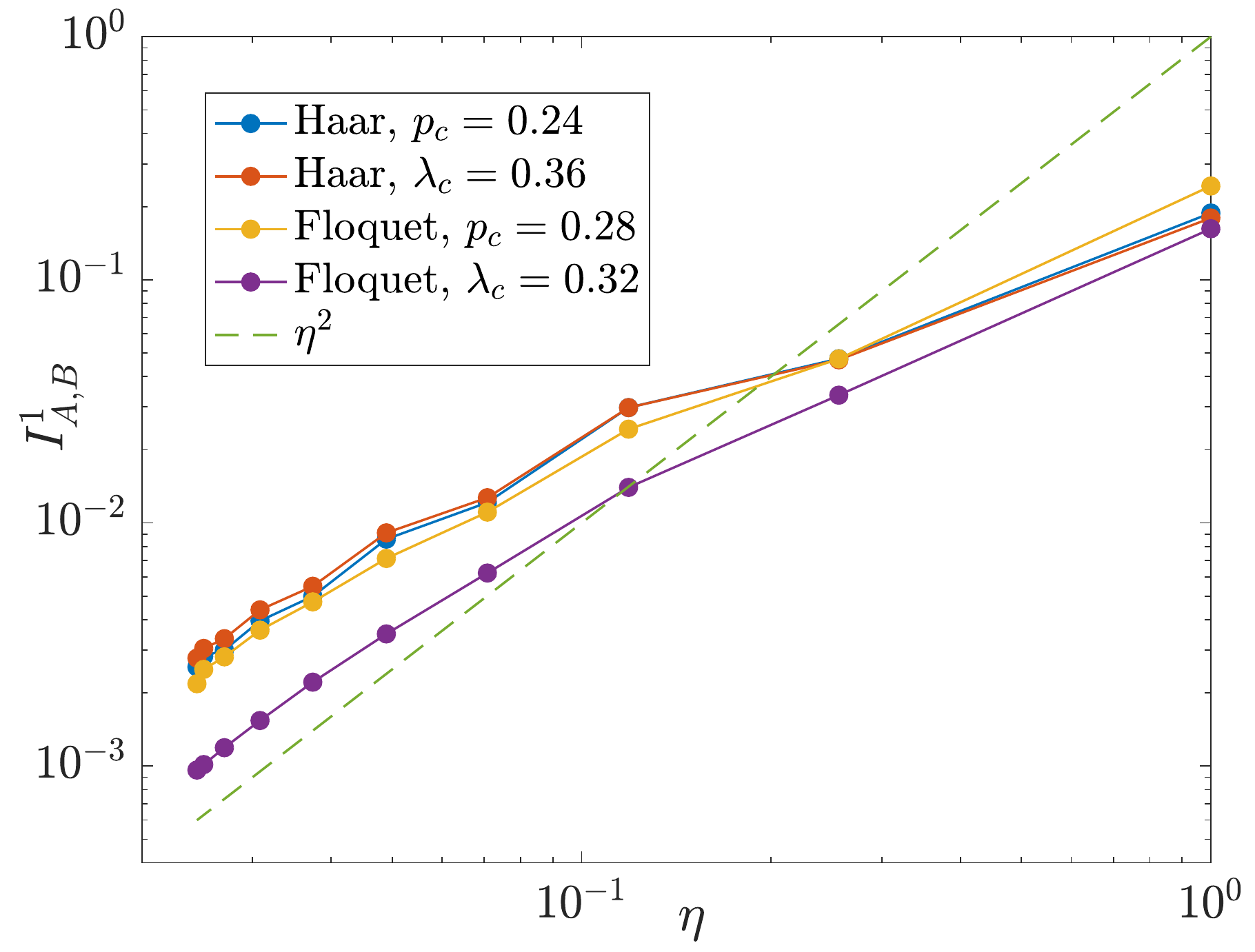}
\caption{ The mutual information $I_{A,B}^1$ for the four non-Clifford models studied in Section VI, each at their respective critical points, plotted versus the cross ratio, $\eta$, on a log-log scale.
Here, the critical values, $p_c$ and $\lambda_c$ were determined by the peak location
of $I_{A,B}^1$ when $r_{A,B} = L/2 = 10$.}
\label{fig:MI_eta}  
\end{figure}
\subsubsection{The location of $p_c$}
The previous numerical results for random Haar circuit and Floquet Ising model suggest that $p_c$ is independent of the  R\'enyi index $n$ when $n>1$.
This result can be further supported by the following inequality for R\'enyi entropies,
\begin{align}
S^{\infty}_A \leq S^n_A \leq \frac{n}{n-1}S^{\infty}_A,
\end{align}
where the second inequality holds when $n>1$.
Since $S^n_A$ is bounded on both sides by $S^{\infty}_A$, in the thermodynamic limit, the scaling behavior of $S^n_A$ ($n>1$) must be the same at any $p$.
This indicates that the transition for $S^n_A$ with $n>1$ occurs at the same $p_c$ and the critical exponent $\nu$ should also be the same.
However, the coefficient $\alpha$ in $S^n_A(p_c; |A|, L) =\alpha(p_c)\ln |A|$ at the critical point could depend on $n$. 
\subsubsection{Scaling of mutual information}
As shown in Sec.~\ref{sec:CFT}, for the Clifford circuits we were able
to extract the operator scaling dimension of a (putative) underlying CFT
from the scaling of mutual information at criticality.
Here, we attempt the same for the four non-Clifford models considered in this Section.
To this end, we compute $I_{A,B}$ with fixed $|A|=|B|=1$, varying the distance $r_{A,B}$ between the two sites.
In this case the cross ratio varies as $\eta \propto r_{A,B}^{-2} \ll 1$.

In Fig.~\ref{fig:MI_eta} we plot the mutual information as a function of the cross ratio ${\eta}$, which is defined in Eq.~\eqref{eq:I_eta} for a system with periodic boundary conditions.
At small values of ${\eta}$,  the mutual information for all four models varies as a power law, $I^1_{A,B} \propto {\eta}^\Delta$ with $\Delta \approx 2$, consistent with the Clifford circuit results (see Figs.~\ref{fig:conf_symm}(b), \ref{fig:floq_rand_result}(c), and \ref{fig:rand_qp_result}(c)).

        \section{Discussion \label{sec7}}

\subsection{Summary}

In this paper we have investigated a broad class of hybrid quantum circuit models
constructed by interleaving unitary and measurement gates, the latter breaking the circuits unitarity.   Under the circuit dynamics we have followed quantum trajectories of the qubits, focussing on the entanglement properties of the evolving pure state wavefunction at late times (in the steady state).   Entanglement generated by the unitary gates competes with the
disentanglement from the measurements.
As established numerically, upon varying the frequency of measurements, $p$, the phase diagram has two stable phases -- a volume law entangled phase when measurements are rare/weak ($p< p_c$), and an area law entangled phase when measurements are frequent/strong ($p > p_c$).
These two phases are separated by a critical point at $p=p_c$, with associated universal scaling properties.

The entanglement entropy in the volume law phase has a remarkable sub-leading correction that is logarithmic 
in the sub-system size, $S_A = \alpha(p) \ln |A| +  s(p) |A|$,
as we established by analyzing the length distribution of stabilizers used to simulate our Clifford circuits.   The coefficient of the logarithm is non-universal throughout the volume law phase, but vanishes in the absence of measurements, $\alpha(0)=0$.
 The coefficient of the linear piece in the entanglement entropy, $s(p)$, smoothly vanishes as one approaches the phase transition from the volume law phase, scaling as $s(p) \sim \xi^{-1} \sim (p_c - p)^\nu$
 with a universal correlation length exponent $\nu \approx 1.3$.   
At the critical point, the logarithmic scaling of the entanglement entropy survives, with a universal coefficient given by $\alpha(p_c)  \approx 1.6$.
Moreover, 
the mutual information between two sites was found to decay as a power law of the distance at the critical point, $r^{-2\Delta}$ with exponent $\Delta \approx 2$,
while the bipartite mutual information for more general geometries depends only on the cross ratio, as expected for a conformal field theory (CFT).   Together with the
logarithmic entanglement at $p_c$, this suggests the possible existence of an
underlying CFT description.

It should be emphasized that these results were established by considering a large class
of quantum circuits, both with and without Clifford gates.
In addition to generic random models with no symmetries, 
we also explored circuits with space-time translational symmetries of the unitary dynamics and/or the measurement gate locations.   In all cases we found stable volume law phases with a logarithmic correction, and similar critical exponents as in models without those symmetries.


\subsection{Conjectures beyond numerics}

Our findings suggest a remarkable degree of universality, both at the phase transition 
and in the properties of the volume law entangled phase.
We thus propose the following conjectures for \emph{local circuit models in 1d}:
\envelope{enumerate}{
\item
In circuits with generic background unitary dynamics and homogeneous arrangement of measurements, there exists a stable volume law entangled phase when measurements are rare.
\item
In the volume law phase, the entanglement entropy always has a logarithmic correction.
\item
There is a continuous phase transition separating the volume law and area law phases {of the von Neumann and higher R\'{e}nyi entanglement entropies},
with critical properties  in the same universality class as the models explored in this paper (including both Clifford and Haar circuits)
\footnote{{We have explicitly excluded the zeroth R\'{e}nyi entropy in our conjecture, which is quite singular and appear to be in a different universality class; see Sec.~\ref{sec:discuss_nature_of_transition} for more discussions.}}.
}

\subsubsection{Volume law phase and logarithmic correction}

\begin{figure}[t]
\centering
 \includegraphics[width=.5\textwidth]{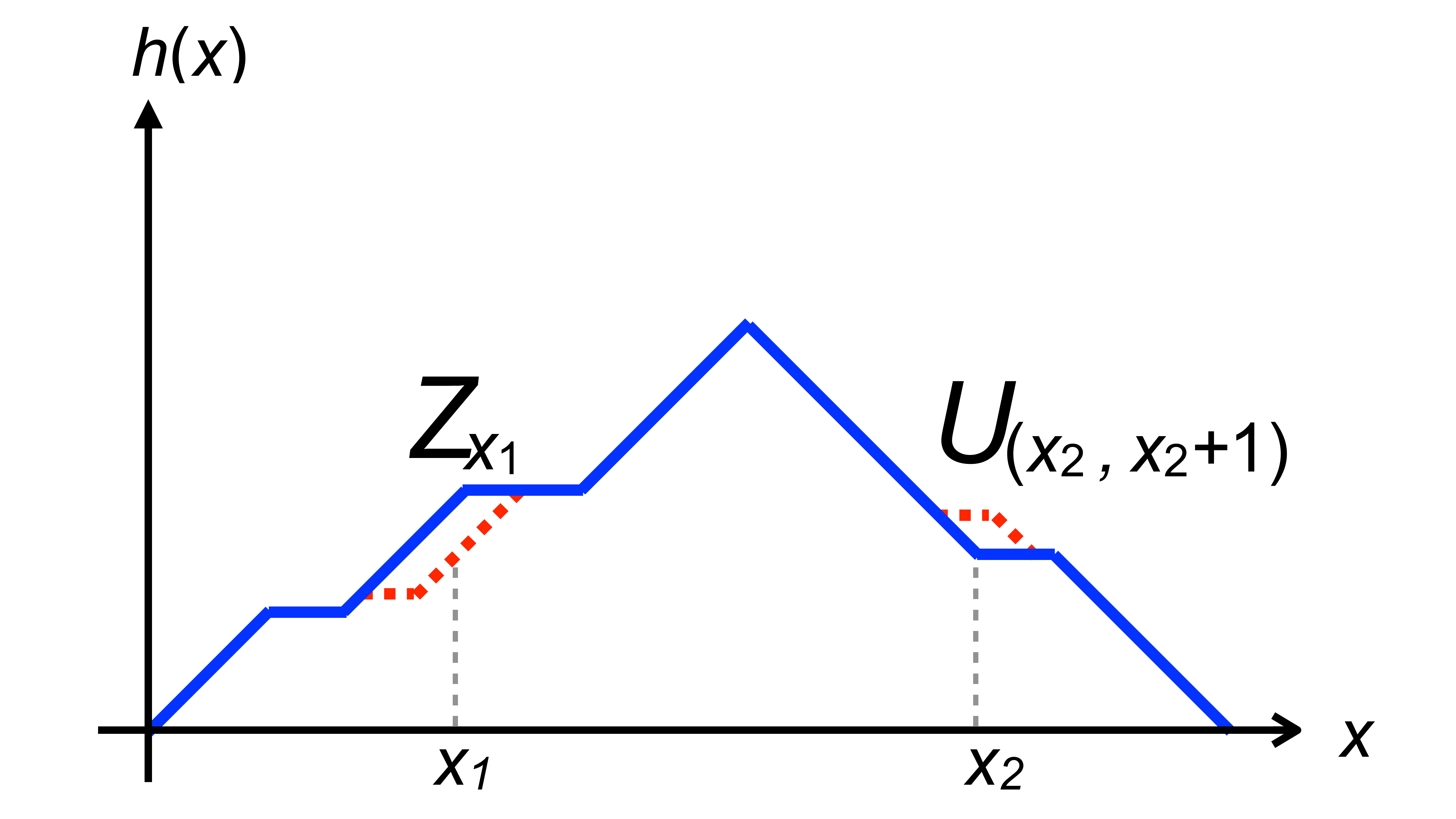}
\caption{The entanglement entropy growth problem can be transformed into a surface growth model.
While the unitary entanglement growth is local, the disentanglement of a local measurement ($Z_{x_1}$) can be non-local.}
\label{fig:schematic_growth}
\end{figure}

We now discuss a general framework incorporating measurements and unitaries that can be used to help better understand and bolster our numerical results.
As above, we emphasize that the steady state entanglement properties of purely unitary circuits are qualitatively different from those circuits with measurements.
In the absence of measurements, the steady state is maximally entangled, i.e. each subset $A$ has an entanglement entropy of $S_A = |A|$.
Measurements on a portion $p$ of all qubits immediately reduces $S_A$ from $|A|$ to $(1-p)|A|$.
This result is a direct consequence of the subadditivity of entanglement.
Thus the maximally entangled state is very susceptible to measurements.
Indeed, if we assume that this $p S_A$ reduction in $S_A$ is true for any volume law entangled state, we would reach the conclusion that no volume law phase should exist~\cite{nandkishore2018hybrid}.

However, this intuition does not carry over to the case for the generic volume law entangled states present with measurements, which, firstly, have a linear slope $s$ smaller than $1-p$, so that the subadditivity bound on entanglement is no longer tight.
{
With less entanglement, 
local measurements would have a weaker effect.
}
Indeed, taking the limit of a trivial product state, a local measurement has only a local effect because of the lack of entanglement.

To illustrate this argument, we consider the following ``surface growth'' picture, as considered in Ref.~\cite{nahum2018hybrid} and shown in Fig.~\ref{fig:schematic_growth}.
Taking open spatial boundary conditions, we define a ``height" function, $h(x)$, to be the entanglement entropy of the subsystem containing the first $x$ qubits,
\envelope{eqnarray}{
    h(x) = S_{A = \{1, 2, \ldots, x\}}.
}
It is convenient to define the average height function,
\envelope{eqnarray}{
    \ovl{h} \coloneqq \frac{1}{L} \sum_x h(x).
}
In the volume law phase, $\ovl{h} \propto L^1$, while in the area law phase $\ovl{h} \propto L^0$, similar to the scaling of the entanglement entropy with subsystem size.
Consider now the effect of the circuit dynamics.  At all times, $\ovl{h}$ grows under unitary time evolution.
After a unitary layer in the circuit, it is expected that,
\envelope{eqnarray}{
    \Delta_U \ovl{h}  \propto L^0 .
}

Recall that each measurement layer has $p L$ measurement gates distributed homogeneously across the $L$ qubits, after which the reduction in $\bar{h}$ is,
\envelope{eqnarray}{
    \label{eq:Delta_M_hbar}
    \Delta_M \ovl{h} &=& \frac{1}{L} \sum_{i=1}^{pL} \sum_{x=1}^L \(h^{(i)}(x) - h^{(i-1)}(x)\) \nn
    &=& \sum_{i=1}^{pL} \delta_M \ovl{h}^{(i)} ,
}
where $h^{(i)}$ is the height function after the first $i$ measurements 
are made, and $\delta_M \ovl{h}^{(i)} \coloneqq \frac{1}{L} \sum_{x=1}^L \(h^{(i)}(x) - h^{(i-1)}(x)\)$ is the reduction of $\ovl{h}$ by the $i$-th measurement.
Each of the $\delta_M \ovl{h}^{(i)}$ has a non-positive expectation value.

At this point, we ignore the correlations and causal relations among measurements within the circuit, and treat $\delta_M \ovl{h}^{(i)}$ for all measurements deep within the circuit as an independent samplings of a single random variable, $\delta_M \ovl{h}$.
This simplification is based on the assumption that in a generic circuit with little structure, the disentanglement of a single measurement should depend only on the entanglement structure of the pre-measurement wavefunction, which fluctuates weakly over time after saturation.

Therefore, Eq.~\eqref{eq:Delta_M_hbar} can be simplified as
\envelope{eqnarray}{
    \Delta_M \ovl{h} = (pL) \avg{\delta_M \ovl{h}},
}
where $\avg{\ldots}$ denotes the expectation value, taken within the ensemble of all measurements after saturation.
Here, $\delta_M \ovl{h}$ quantifies the disentangling ability of a single local measurement.

By definition, within the steady state, the entangling and disentangling effects must balance out, i.e. $\Delta_U \ovl{h} + \Delta_M \ovl{h} = 0$, therefore $\avg{\Delta_M \ovl{h}}  \propto L^0$, or
\envelope{eqnarray}{
    \label{eq:delta_M_hbar_O1overL}
    \avg{\delta_M \ovl{h}} = O\( \frac{1}{L} \).
}
This is a relation that must hold for all $p > 0$, regardless of the steady state entanglement entropy.  In particular, it must hold in any volume law entangled state
in the presence of measurements, despite the fact that
$\avg{\delta_M \ovl{h}} = O(L^0)$ in a maximally entangled state and in a Bell pair state as
discussed in Ref.~\cite{nandkishore2018hybrid}.

Direct numerical evidence for the validity of Eq.~\eqref{eq:delta_M_hbar_O1overL}
for all $p>0$ can be established in our Clifford circuits, as we now discuss.
As detailed in Appendix~\ref{appB}, we compute the normalized distribution function of $\delta_M \ovl{h}$ for the random Clifford circuit.  Specifically, the distribution function of the
``disentanglement length"
$R \equiv - L \times \delta_M \ovl{h}$, which we denote as $\mc{P}(R)$, takes the following schematic form within the volume law and area law phases,
\envelope{eqnarray}{
    \mc{P}(R) \sim 
    \begin{cases}
        R^{-\gamma(p)} , p < p_c, \\
        e^{- R /R_0} R^{-\gamma(p)} , p > p_c, \\
    \end{cases}
}
where $R_0$ is proportional to the correlation length in the area law phase.
Here the power $\gamma(p)$, which varies with $p$ throughout the volume law phase,
grows as we increase $p$, consistent with our intuition that less entanglement implies less disentanglement.
For $p$ very small $\gamma(p)$ appears to approach 2, and is close to 3 when $p=p_c$,
$\gamma(p_c) \approx 3$.   Throughout the volume law phase $\gamma(p)$ is always larger than 2.
Thus, despite the power law distribution of the disentangling scale, $R$, 
in the volume law phase, the {\it average disentangling length}, $\langle R \rangle = \int^{L/2} dR\, R \mc{P}(R)$
is finite for all $p>0$.  
We then conclude that $\avg{\delta_M \ovl{h}} = -  \langle R \rangle/L = O(1/L)$, validating Eq.~\eqref{eq:delta_M_hbar_O1overL}.


When restricted to Clifford circuits, the difference between the maximally entangled state and a general volume law entangled state in the presence of measurements is well illustrated by the stabilizer length distribution.
As we show in Appendix~\ref{appB}, within the clipped gauge, a local measurement (say $Z_x$) replaces one of the $L$ stabilizers with $Z_x$, while rearranging the others in a way that more or less preserve their lengths.
When $p = 0$, the stabilizer distribution function is a delta function at $\ell \approx L/2$.
In other words, there are only long stabilizers but no short ones.
In this case, a local measurement will inevitably replace a long stabilizer with $Z_x$, causing a non-local change in the entanglement structure, as seen from Eq.~\eqref{eq:EE_span}.
On the other hand, when $p > 0$, the power law distribution of ``shorter" stabilizers protects the long stabilizers in the $\ell \approx L/2$ peak from always being replaced by a unit length one ($Z_x$), so that the replacement and rearrangement only happens within the ``shorter" stabilizers, thereby preserving the volume law entropy.
In the (rare) case when a long stabilizer does get replaced by $Z_x$, the power law distribution of ``short" stabilizers can shift to the right under unitary evolution and compensate this reduction, rendering the distribution steady.
In all models that we have studied, the inverse-square power law distribution 
of the ``shorter" stabilizers is present, giving the sub-leading logarithmic correction to the entanglement entropy.
We might thus say that the logarithmic correction is necessary for the stability of the volume law phase.

It seems plausible that the power law distribution in the measurement induced ``disentanglement length", $\mc{P}(R)$, and the power law 
distribution of the ``shorter" stabilizers are related to one another, but the exact relation remains unknown to us.
Although the distribution $\mc{P}(R)$ was computed for the random Clifford circuit it is defined with complete generality, and we believe that both the stability criterion $\gamma > 2$ as well as the logarithmic correction are universal for volume law phases stable against measurements in generic hybrid circuits.

\subsubsection{\label{sec:discuss_nature_of_transition} The nature of the phase transition}

What can we say about the nature of the entanglement phase transition beyond our numerical results?
Ref.~\cite{nahum2018hybrid} showed that in a circuit  
with random Haar unitaries and single qubit projective measurements, the zeroth R{\'e}nyi entropy $S_A^0$ can be mapped to a percolation type problem.
With spatial randomness in the location of the 
measurements, it was thereby concluded that $S_A^0$ exhibits an entanglement transition in the universality class of the first passage percolation (FPP) transition on a square lattice~\cite{Hammersley1965, Chayes1986firstpassage, kesten1986aspects, kesten1987firstpassage}.
In this mapping, $p$ corresponds to the probability for a bond of the lattice to be broken, and the entanglement entropy is mapped to the minimal cut from the temporal boundary at time $T$ in the spacetime manifold of the circuit, known as the ``first passage time''.
Corresponding to the volume law and area law phases, the minimal cut scales with $L$ for small $p$, and is a finite constant for large $p$.
Logarithmic scaling of entropy at the critical point also follows from FPP.

From this perspective, it is perhaps surprising that we have found phase transitions
in models with no randomness in the locations of the measurements, including
the Clifford circuits with spatially (quasi)-periodic measurements, and
random Haar and Floquet circuits with generalized measurements -- the generalized
measurements acting uniformly on each and every qubit.
Indeed, for percolation the randomness in the locations of the measurements is essential in producing fluctuations in the size of ``puddles'' of broken bonds, which then drives the percolation transition.
With spatial periodicity, there should be no transition for $S_A^0$ -- the minimal cut will scale with $L$ for arbitrary $p$.   
Remarkably, this is entirely consistent with our random Haar and Floquet circuit results with generalized measurements for which no transition was found for $S_A^0$,
consistent with volume law entanglement for all $\lambda<1$ (see Sec.~\ref{sec6} and Fig.~\ref{fig:Floq_weak}).

However, the absence of a transition in $S_A^0$ does not preclude transitions in higher R{\'e}nyi entropies; these transitions are seen explicitly for both 
generalized  measurement models in Sec.~\ref{sec6}.  Evidently, $S_A^0$ is very special, and is quite different from higher R{\'e}nyi entropies
with Renyi index $n \ge 1$, which are more physically relevant.
Indeed, in Fig.~\ref{fig:Haar_corr}
the peak in the spin-spin correlation functions are close to those given by higher R{\'e}nyi entropies, but far away from the peak in $S_A^0$.

Even though all R{\'e}nyi entropies in the Clifford circuit are equal, the entanglement entropy has all the virtues of the von Neumann entropy, and is actually close in value (see Fig.~\ref{fig:Floq_projective}).
Therefore, a numerical comparison in terms of critical exponents between the percolation transition and Clifford circuits should also illustrate the difference between the zeroth and higher R{\'e}nyi entropies.
For Clifford circuits with or without spatial and temporal translation symmetries, we consistently find the critical exponents $\nu\approx 1.3$ and $\Delta \approx 2$; the latter even holds beyond Clifford circuits.
These values coincide with those of percolation, as found in Ref.~\cite{nahum2018hybrid}.
However, the coefficient of the critical logarithmic entropy within Clifford circuits, $\alpha(p_c) \approx 1.6$, is much larger than the value predicted by first passage percolation $\alpha(p_c)=\sqrt{3}/\pi \approx 0.55$~\cite{Cardy2000, cardy0103percolation, yao1612firstpassage, nahum2018hybrid} (notice the open boundary condition in Ref.~\cite{nahum2018hybrid}).
Put together, these results indicate that the percolation mapping only works in the limit of Renyi index $n \to 0$, and cannot give a full characterization of the entanglement dynamics.

This point can be analytically understood in the context of effective spin models for R{\'e}nyi entropies with arbitrary $n$~\cite{TianciZhouprivate}.
For $n \ge 1$, in addition to the cost of the minimal path, there is an extra contribution from an ``entropy term'' which counts the number of minimal paths of the same cost.
This term is a relevant perturbation and could drive the critical point away from the percolation transition~\cite{TianciZhouprivate, nahum2018hybrid}.

\subsection{Outlook}

At present, there is little analytical understanding of entanglement dynamics (or of the steady states) in circuits with measurements.
For Clifford circuits, the motion of the end point of the stabilizers under unitary and measurement gates can be approximately modeled  by a simple traffic-flow model (discussed in the appendix), which is related to an asymmetric simple exclusion process (ASEP)~\cite{ASEP}.
Can one find an exactly soluble model that belongs to ASEP and exhibits the same type of entanglement transition as in the full Clifford circuit?
Alternatively, by analogy with the percolation mapping of $S_A^0$ in the random Haar circuit \cite{nahum2018hybrid} with projective measurements, can one find an effective description
for the generic transition in terms of a statistical mechanical model?
 A simplification occurs in the random Haar circuit with large onsite Hilbert space dimension $q \rightarrow \infty$,
which can be mapped to an effective spin model~\cite{zhou1804spin},
again described in terms of bond percolation.
In this limit, $S_A^n$ is independent of $n$. It would be interesting to study the nature of the phase transition when 
 $q$ is finite but large, perhaps as a $1/q$ expansion~\cite{zhou1804spin}.
 Exploring the entanglement transition in higher dimensions, $d > 1$,
 would also be interesting.



In the absence of measurements the dynamics of a random Haar unitary circuit 
exhibits all of the characteristics of quantum chaos~\cite{nahum2017KPZ, nahum2018operator, keyserlingk2018operator}.
With measurements present, it would be interesting to explore signatures of quantum chaos in the entanglement spectrum of the steady state reduced density matrix (e.g. the spectral form factor), especially
in the volume law phase and at the critical point.
Presumably, as one passes through
the entanglement transition into the area law phase
there will be a chaotic to non-chaotic phase transition.


Hints of an underlying CFT at the hybrid circuit entanglement transition,
suggest that there might be a dual holographic description.
If so, the possible role of the universal logarithmic correction to the extensive entropy in the volume law phase might have interesting consequences for the putative ``black hole''.

A more general issue, beyond the hybrid circuit models, concerns the disentangling effects of local measurements 
on various many-body wavefunctions.  For example, the exponent $\gamma$ (when it is defined) appears to be fundamental.
%


Lastly, one might ask whether experimental realizations of the measurement induced entanglement transition are possible.
As we showed numerically, generalized weak measurements in a circuit with spatial and temporal translational symmetry
are sufficient to drive the transition, so one does not require perfect projective measurements or ensemble average.
Eliminating such fine tuning might perhaps lighten the experimental challenges
in accessing the transition.
However, directly measuring the entanglement entropy or the enhanced fluctuation of correlation functions (as discussed in Sec.~\ref{sec4}) requires preparing several copies of the same wavefunction at the end of the circuit evolution.
This is usually exponentially expensive due to the intrinsic randomness in the measurement outcomes, therefore a naive protocol based on postselection is not scalable.
Whether it is possible to access the transition experimentally remains an open question.

    \section*{Acknowledgements}
        We thank Ehud Altman, Hans-Peter B\"{u}chler, Marin Bukov, Xi Dong, Lukasz Fidkowski, Hrant Gharibyan, Daniel Gottesman, Tarun Grover, Chao-Ming Jian, Michael Kolodrubetz, Andreas Ludwig, Adam Nahum, Rahul Nandkishore, Anatoli Polkovnikov, Michael Pretko, Jonathan Ruhman, Steve Shenker, Brian Skinner, Douglas Stanford, Brian Swingle, Graeme Smith, Sagar Vijay, Cenke Xu, Yi-Zhuang You, Tianci Zhou and Peter Zoller for helpful discussions.
        XC is supported by a postdoctoral fellowship from the Gordon and Betty Moore Foundation, under the EPiQS initiative, Grant GBMF4304, at the Kavli Institute for Theoretical Physics.
        YL and MPAF are grateful to the Heising-Simons Foundation for support, to the National Science Foundation for support under Grant No. DMR-1404230, and to the Caltech Institute of Quantum Information and Matter, an NSF Physics Frontiers Center with support of the Gordon and Betty Moore Foundation.
        Use was made of computational facilities purchased with funds from the National Science Foundation (CNS-1725797) and administered by the Center for Scientific Computing (CSC). The CSC is supported by the California NanoSystems Institute and the Materials Research Science and Engineering Center (MRSEC; NSF DMR-1720256) at UC Santa Barbara.

    \appendix
        \section{Brief review of the stabilizer formalism and gauge fixing \label{appA}}

\subsection{Basics}

In this subsection we review the stabilizer formalism and Clifford circuits.
The references for this subsection are Refs.~\cite{gottesman9604hamming, gottesman9807heisenberg, nielsen2010qiqc, aaronson0406chp, nahum2017KPZ}.

\subsubsection{Codewords, stabilizers, and gauge freedom}    

    The defining property of the Clifford circuit is that the pure state wavefunction $\ket{\psi}$ at any time is a \emph{codeword}, the simultaneous +1 eigenstate of $L$ {mutually commuting} and {linear independent (under multiplication)} Pauli string operators
    \envelope{equation}{
        \mc{G} = \{g_1, \ldots, g_L\} \subset \mc{P}_+(L),\ \mc{P}_+(L) = \{g \in \mc{P}(L) : g^2 = 1\},
    }
    among which none of the $g_i$'s is proportional to the identity.
    These Pauli string operators generate the \emph{stabilizer group}~\cite{gottesman9604hamming, calderbank1997quantum} of the codeword, denoted $\mc{S}(\ket{\psi}) = \avg{\mc{G}}$, or simply $\mc{S}$.
    The codeword is uniquely determined given the stabilizer group, and the stabilizer group is uniquely determined given the codeword $\ket{\psi}$,
    \envelope{eqnarray}{
        \mc{S} = \{ g \in \mc{P}_+(L) : g\ket{\psi} = \ket{\psi} \}.
    }
    One can explicitly write down all elements of $\mc{S}$ given $\mc{G}$,
    \envelope{eqnarray}{
        \mc{S} = \left\{ g_1^{p_1} g_2^{p_2} \ldots g_L^{p_L} : (p_1, \ldots, p_L) \in \{0, 1\}^L \right\}.
    }
    In this case, we also write $\mc{G} = \mc{G}(\mc{S})$, which means the same thing as $\mc{S} = \avg{\mc{G}(\mc{S})}$.
    Because of the linear indepence of $\mc{G}$, each element of $\mc{S}$ has a unique representation in this form, hence there is a one-to-one mapping between $\{0, 1\}^L$ and $\mc{S}$.
    It follows that $\mc{S}$ is a finite abelian group of order $|\mc{S}| = 2^L$.

    Being a finite abelian group, and with each element of order 2, $\mc{S}$ can be viewed as an $L$-dimensional vector space on $\mb{Z}_2$, and group multiplication can be viewed as addition in this vector space (ignoring phase factors).
    Thus, an independent generating set $\mc{G}(\mc{S})$ corresponds to a choice of basis for this vector space. 
    Such a choice is not unique, and the freedom in choosing $\mc{G}(\mc{S})$ is referred to as the \emph{gauge freedom} in this paper.

    For the rest of this appendix, we will always take $\mc{G}(\mc{S})$ to be an independent generating basis (thus has $L$ elements), and use the word \emph{stabilizer} for elements of $\mc{G}(\mc{S})$.
    When we talk about a codeword state, we mostly work with its stabilizers, $\mc{G}(\mc{S})$.

\subsubsection{Simulating Clifford circuits \label{sec:A1b}}
We briefly review our simulation of the Clifford circuits with Pauli measurements.
The main result we use is the Gottesman-Knill theorem.

    First consider the action of a unitary operator, $U$.
    For a state $\ket{\psi}$ whose stabilizer group is $\mc{S} = \{ g_1, \ldots, g_{|\mc{S}|} \}$, the state evolves as $\ket{\psi} \mapsto U\ket{\psi}$, while the stabilizer group evolves as
    \envelope{equation}{
        \mc{S} \mapsto \mc{S}^U = \{g_1^U, \ldots, g_{|\mc{S}|}^U\} = \{U g_1 U^\dg, \ldots, U g_{|\mc{S}|} U^\dg \}.
    }
    For the state to remain a codeword under unitary time evolution, the unitaries must be taken from the \emph{Clifford group}, which transforms a Pauli string operator $g$ into $g^U = U g U^\dg$ that is still a Pauli string operator.
    Thus, $\mc{S}^U$ remains a group of Pauli string operators, hence the wavefunction remains a codeword.
    To simulate a circuit under Clifford unitary evolution, one only needs to keep track of $\mc{S}$, or equivalently (and more conveniently) its generating set $\mc{G}(\mc{S})$. Such a simulation only takes polynomial time in $L$.

    It is common knowledge that the Clifford group on two-qubits is generated by $\{\text{CNOT, SWAP, H, P}\}$, where in the standard bases
    \envelope{eqnarray}{
        \text{CNOT} = \envelope{pmatrix}{
            1 & 0 & 0 & 0 \\
            0 & 1 & 0 & 0 \\
            0 & 0 & 0 & 1 \\
            0 & 0 & 1 & 0 \\
        }, && \quad
        \text{SWAP} = \envelope{pmatrix}{
            1 & 0 & 0 & 0 \\
            0 & 0 & 1 & 0 \\
            0 & 1 & 0 & 0 \\
            0 & 0 & 0 & 1 \\
        },\\
        \text{H} = \frac{1}{\sqrt{2}}\envelope{pmatrix}{
            1 & 1 \\
            1 & -1
        },&& \quad
        \text{P} = \envelope{pmatrix}{
            1 & 0 \\
            0 & i
        }.
    }
    The $\text{CNOT}$ gate defined here is also known as $\text{CNOT}_{\text{L}}$, whereas $\text{CNOT}_{\text{R}} = \text{SWAP} \cdot \text{CNOT}_{\text{L}} \cdot \text{SWAP}$.

    Next we consider Pauli measurements, that is, measuring a Pauli string operator $g$.
    Let $\mc{G} = \{g_1, \ldots, g_k, g_{k+1}, \ldots, g_L\}$ be the stabilizers of $\ket{\psi}$ and suppose that $[g_j, g] = 0$ for $j \le k$, and $\{g_j, g\} = 0$ for $j > k$.
    After the measurement, there are two possible outcomes ($1$ or $-1$), hence two possibilities of the measured wavefunction,
    \envelope{eqnarray}{
        \ket{\psi}_{\pm} \propto \frac{1\pm g}{2} \ket{\psi}.
    }
    Their corresponding probabilities can be computed, as detailed in \cite{aaronson0406chp}.
    Remarkbly, the measured state is still a codeword, and its corresponding stabilizer group is generated by the following stabilizers \cite{nielsen2010qiqc}
    \begin{eqnarray}
        \label{eq:stabilizer_after_measurement}
        \mc{G}_{\pm} = \{g_1, \ldots, g_k, g_{k+1} g_{k+2}, \ldots, g_{L-1} g_L, \pm g\}.
    \end{eqnarray}
    Such a simulation can also be performed in polynomial time.

    We use the particular algorithm in \cite{aaronson0406chp} for our simulation of the Clifford circuits, where we take the unitary and measurement gates to be local.

\subsubsection{Generating random Clifford unitaries}
In the random Clifford circuit, the local unitaries are taken from the uniform distribution on the two-qubit Clifford group.
Here we explain the sampling process from the $L$-qubit Clifford group $\mc{C}(L)$~\cite{DanielGottesmanprivate}.
It applies to $L=2$ as a special case.

    First we notice that the Clifford group acts on the Pauli group transitively, and that a Clifford unitary $U$ is determined (up to a sign) by images of the generators of $\mc{P}_+(L)$, conveniently taken to be $\{X_1, Z_1, \ldots, X_L, Z_L\}$.
    Thus, sampling a random Clifford unitary is equivalent to sampling random images of the generators.
    We proceed by induction, and start with assuming that one is able to sample from the uniform distribution on $\mc{C}(k)$.
    Now consider the action of a random Clifford unitary on $\{X_{k+1}, Z_{k+1}\}$.
    Since the random unitary is taken from the uniform distribution, it maps $X_{k+1}$ to all the non-identity elements of $\mc{P}_+(k+1)$ with equal probability. 
    $X_{k+1}^U$ is essentially a random non-trivial Pauli string operator of length $k+1$; there are $2 (4^{k+1} - 1)$ choices, where the factor of $2$ comes from the sign.
    $Z_{k+1}^U$ is also almost random, except that it must also square to 1, and anticommute with $X_{k+1}^U$; there are $2\(2 \times 4^{k}\)$ choices.

    Having randomly chosen $X_{k+1}^U$ and $Z_{k+1}^U$, we can find \emph{one} unitary $U^\p$ (again represented by its action on the generators of $\mc{P}_+(k+1)$) such that $X_{k+1}^{U^\p} = X_{k+1}^{U}$ and $Z_{k+1}^{U^\p} = Z_{k+1}^{U}$ satisfying the following relations,
    \begin{eqnarray}
        (U^\p)^\dg X^{U}_{k+1} U^\p &=& X_{k+1}, \\
        (U^\p)^\dg Z^{U}_{k+1} U^\p &=& Z_{k+1}.
    \end{eqnarray}
    To preserve the commutation relations, we must have for $i \le k$,
    \begin{eqnarray}
        (U^\p)^\dg X^{U}_{i} U^\p = \(\ldots\) \otimes I_{k+1}, \\
        (U^\p)^\dg Z^{U}_{i} U^\p = \(\ldots\) \otimes I_{k+1},
    \end{eqnarray}
    which is equivalent to
    \begin{eqnarray}
        X_i^V = \(\ldots\) \otimes I_{k+1}, \\
        Z_i^V = \(\ldots\) \otimes I_{k+1},
    \end{eqnarray}
    where $V = (U^\p)^\dg U$ is now shown to be in the Clifford group of the first $k$ qubits.
    Thus to sample $U$ from $\mc{C}(k+1)$, we just need to sample $V$ from $\mc{C}(k)$, and multiply it by $U^\p$ (which is determined by $X_{k+1}^U$ and $Z_{k+1}^U$, which are also random), to get a random $U$ from $\mc{C}(k+1)$.
    Since it is easy to generate elements in $\mc{C}(1)$, we know how to generate elements in $\mc{C}(k+1)$, by induction.


    From the above, we get the following recurrence relation
    \envelope{eqnarray}{
        |\mc{C}(L+1)| = 2(4^{L+1}-1) \times (4^{L+1}) \times |\mc{C}(L)|,
    }
    where the first factor corresponds to the number of choices of the image of $X_{L+1}$, and the second factor corresponds to that of $Z_{L+1}$.

\subsubsection{Entanglement entropy from stabilizers}
    Given a pure state wavefunction $\ket{\psi}$, the $n$-th R{\'e}nyi entanglement entropy with respect to a given bipartition $(A, \ovl{A})$ is defined to be (c.f. Eq.~(\ref{eq:Renyi_EE}))
    \envelope{eqnarray}{
        S_A^{n} = \frac{1}{1-n} \log_2 {\rm Tr} \(\rho_A\)^n, \text{ where }\rho_A = {\rm Tr}_{\ovl{A}} \ket{\psi}\bra{\psi}. \nonumber
    }
    When $\ket{\psi}$ is a codeword, the R{\'e}nyi entropies are independent of the R{\'e}nyi index $n$, and is related to its stabilizers through the following relation \cite{hamma2005entanglement1, hamma2005entanglement2,nahum2017KPZ}
    \envelope{eqnarray}{
        \label{eq:EE_to_subgroup_size}
        S_A = |A| - \log_2 |\mc{S}_A|,
    }
    where $\mc{S}_A$ is the subgroup of $\mc{S}$ of all elements that have trivial content ($I$) on $\ovl{A}$.
    Equivalently,
    \envelope{eqnarray}{
        \label{eq:EE_to_subgroup_generator}
        S_A = |A| - |\mc{G}(\mc{S}_A)|,
    }
    where $\mc{G}(\mc{S}_A)$ is an arbitrary generating set of $\mc{S}_A$.

    We recall an alternative formula as derived in Ref.~\cite{nahum2017KPZ}. Define the linear operator $\text{proj}_A$ such that $\text{proj}_A (\mc{S})$ contains all elements from $\mc{S}$ with their contents on $\ovl{A}$ set to identity (``projected out'').
    In this notation we have $|\mc{G}(\mc{S}_A)| = \dim \text{Ker} (\text{proj}_{\ovl{A}})$.
    By a theorem in linear algebra we have $\dim \text{Ker} (\text{proj}_{\ovl{A}}) + \dim \text{Im} (\text{proj}_{\ovl{A}}) = \dim \mc{S} = L$, so that
    \envelope{eqnarray}{
        S_A &=& |A| - \dim \text{Ker} (\text{proj}_{\ovl{A}}) \nn
            &=& |A| - \( L - \dim \text{Im} (\text{proj}_{\ovl{A}}) \) \nn
            &=& \dim \text{Im} (\text{proj}_{\ovl{A}}) - |\ovl{A}|,
    }
    or, interchanging the roles of $A$ and $\ovl{A}$,
    \envelope{eqnarray}{
        \label{eq:EE_to_rank}
        S_A &=& S_{\ovl{A}} = \dim \text{Im} (\text{proj}_A) - |A| \nn
            &=& \text{rank} (\text{proj}_A (\mc{S})) - |A|.
    }

    Given the entanglement entropy, the computation of the bipartite mutual information is immediate.

\subsubsection{Computing Pauli correlation function}

Consider the following $ZZ$ correlator for the state $\psi$,
\envelope{eqnarray}{
    c_{x y} = \bra{\psi} Z_x Z_y \ket{\psi},
}
which can be written as a difference,
\envelope{eqnarray}{
    c_{xy} &=& \bra{\psi} \frac{1 + Z_x Z_y}{2} \ket{\psi} - \bra{\psi} \frac{1 - Z_x Z_y}{2} \ket{\psi}\\
    &=& p_+ - p_-,
}
where the first term is the probability of measuring the Pauli operator $g = Z_x Z_y$ and getting $+$, and the second of getting $-$.
Since the probabilities can be computed~\cite{aaronson0406chp}, the computation of correlation functions of Pauli string operators is straightforward.

\subsection{The clipped gauge}

In this subsection we review the clipped gauge and the clipping algorithm introduced in Ref.~\cite{nahum2017KPZ}, and slightly extend the computation of entanglement entropy within this gauge.

Consider an $L$-qubit codeword $\ket{\psi}$ with stabilizer group $\mc{S}$, where $\mc{S} = \avg{\mc{G}(\mc{S})}$.
{
For a stabilizer $g \in \mc{G}(\mc{S})$, we define $\mt{l}(g)$ to be the position of the left endpoint, and $\mt{r}(g)$ to be the position of the right endpoint, as in Eqs.~(\ref{eq:Pauli_lend}, \ref{eq:Pauli_rend}),
\envelope{eqnarray}{
    \mt{l}(g) &=& \min\{x : \text{$g$ acts non-trivially on site $x$} \}, \\
    \mt{r}(g) &=& \max\{x : \text{$g$ acts non-trivially on site $x$} \},
}
where $x$ is the coordinate of the site, which takes values in $\{1, 2, \ldots, L\}$.
For systems with open spatial boundary conditions, there is a natural coordinate system: we label the sites sequencially, from the left boundary to the right one.
For systems with periodic spatial boundary conditions, there is an arbitrariness in choosing the origin of the coordinate system, and there is no absolute distinction between left and right.
To resolve this arbitrariness we will assume that the origin is chosen and fixed (by hand), so that the functions $\mt{l}(g)$ and $\mt{r}(g)$ are well-defined.
}

We further define $\rho_\mt{l}$ and $\rho_\mt{r}$, the \emph{densities of left and right endpoints}, to be
\begin{eqnarray}
    \rho_\mt{l} (x) &=& \sum_{x=1}^L \delta_{\mt{l}(g_i), x}, \\
    \rho_\mt{r} (x) &=& \sum_{x=1}^L \delta_{\mt{r}(g_i), x}.
\end{eqnarray}
The total number of left and right endpoints are conserved, and $\sum_x \rho_\mt{l} (x) = \sum_x \rho_\mt{r} (x) = L$.
It was shown in Ref.~\cite{nahum2017KPZ} that it is always possible to ``gauge fix'' a stabilizer basis $\mc{G}$ in an arbitrary gauge into the \emph{clipped gauge}, where
\envelope{itemize}{
\item $\rho_\mt{l}(x) + \rho_\mt{r}(x) = 2$, for all sites $x$.
\item For each site with $\rho_\mt{l}(x) = 2$ or $\rho_\mt{r}(x) = 2$, the two stabilizers that end at $x$ must have different content on $x$.
}

\subsubsection{Clipping algorithm}

    We here give an explicit algorithm for gauge fixing an arbitrary stabilizer basis $\mc{G}$ into the clipped gauge $\mc{G}^c$, such that $\avg{\mc{G}} = \avg{\mc{G}^c}$.
    We use the word ``clipping'' for this process.
    Such a process was given in Ref.~\cite{nahum2017KPZ}.

    {\bf Clipping algorithm part 1}.
    Given a stabilizer group $\mc{S}$, there exists an generating set $\mc{G}$ of $\mc{S}$ such that 
    \begin{itemize}
        \item $\forall x, \rho_\mt{l}(x) \le 2$;
        \item If $\rho_\mt{l}(x) = 2$, the two Pauli operators at the left endpoints must be different.
    \end{itemize}
    We call this the \emph{pre-gauge} condition.
    It is different from the gauge condition in that it does not refer to the right endpoints of the stabilizers.

    [\emph{Sketch}:
    Recall that elements of $\mc{G}$ can be viewed as basis vectors of the $L$-dimensional vector space, $\mc{S}$.
    For concreteness, we construct an $L \times 2L$ matrix $M$ on $\mb{Z}_2$, for which the $i$-th row corresponds to $g_i$, where each Pauli matrix is represented by two bits,
    \envelope{eqnarray}{
        I \mapsto 00,\ X \mapsto 10,\ Y \mapsto 11,\ Z \mapsto 01.
    }
    Then we perform Gaussian elimination (row reduction) on $M$ to reduce it into the row echelon form~\cite{GaussianEliminationWikipedia}.
    The resultant matrix, with each row viewed as a stabilizer, satisfy the pre-gauge condition.]

    {\bf Clipping algorithm part 2}. A generating set $\mc{G}$ that satisfies the pre-gauge constraint in part 1 can be transformed into the clipped gauge while preserving $\rho_\mt{l}$.

    [\emph{Sketch}: This is achieved by performing another Gaussian elimination based on the resulting matrix of the previous algorithm, focusing the right endpoints, from the right to the left.
    In doing so, one has to always eliminate the longer stabilizer by the shorter one.
    One can check that $\rho_\mt{l}$ is not changed under this process.
    That the stabilizers commute with each other guarantees that after the algorithm terminates, each site has no more than 2 endpoints, and both left and right endpoints satisfy the pre-gauge constraint in part 1.
    It follows that the resultant $\mc{G}$ is in the clipped gauge.]

\subsubsection{From clipped gauge to $\mc{B}(\mc{G})$}

        Consider the following quantity (which we call {\bf bigrams}) defined for the generating set $\mc{G}$ in the clipped gauge,
        \begin{eqnarray}
            \mc{B}(\mc{G}) \equiv \{ \(\mt{l}(g_1), \mt{r}(g_1)\), \ldots, \(\mt{l}(g_L), \mt{r}(g_L)\) \}.
        \end{eqnarray}
        $\mc{B}(\mc{G})$ is a set of $L$ ordered pairs.

        {\bf Proposition 1.} If $\langle \mc{G} \rangle = \langle \mc{G}^\p \rangle$, where $\mc{G}$ and $\mc{G}^\p$ are both independent and in the clipped gauge, then $\mc{B}(\mc{G}) = \mc{B}(\mc{G}^\p)$.

        But before we prove Proposition 1, it is helpful to state the following

        {\bf Lemma.} Let $\mc{G}$ be in the clipped gauge. For an arbitrary product of the stabilizers,
        \begin{eqnarray}
            g = g_{i_1} \ldots g_{i_k},
        \end{eqnarray}
        where $g_{i_j} \in \mc{G}$, and $\{i_1, \ldots, i_k\}$ are mutually distinct, we have
        \begin{eqnarray}
            \mt{l}(g) &=& \min\, \{\mt{l}(g_{i_1}), \ldots, \mt{l}(g_{i_k}) \}, \\
            \mt{r}(g) &=& \max   \{\mt{r}(g_{i_1}), \ldots, \mt{r}(g_{i_k}) \}.
        \end{eqnarray}
        Intuitively, this is saying that the ``span'' of the product would be the outer envelope of its factors.

        \emph{Proof of the Lemma}:
        \begin{figure}[t]
            \centering
            \includegraphics[width=.49\textwidth]{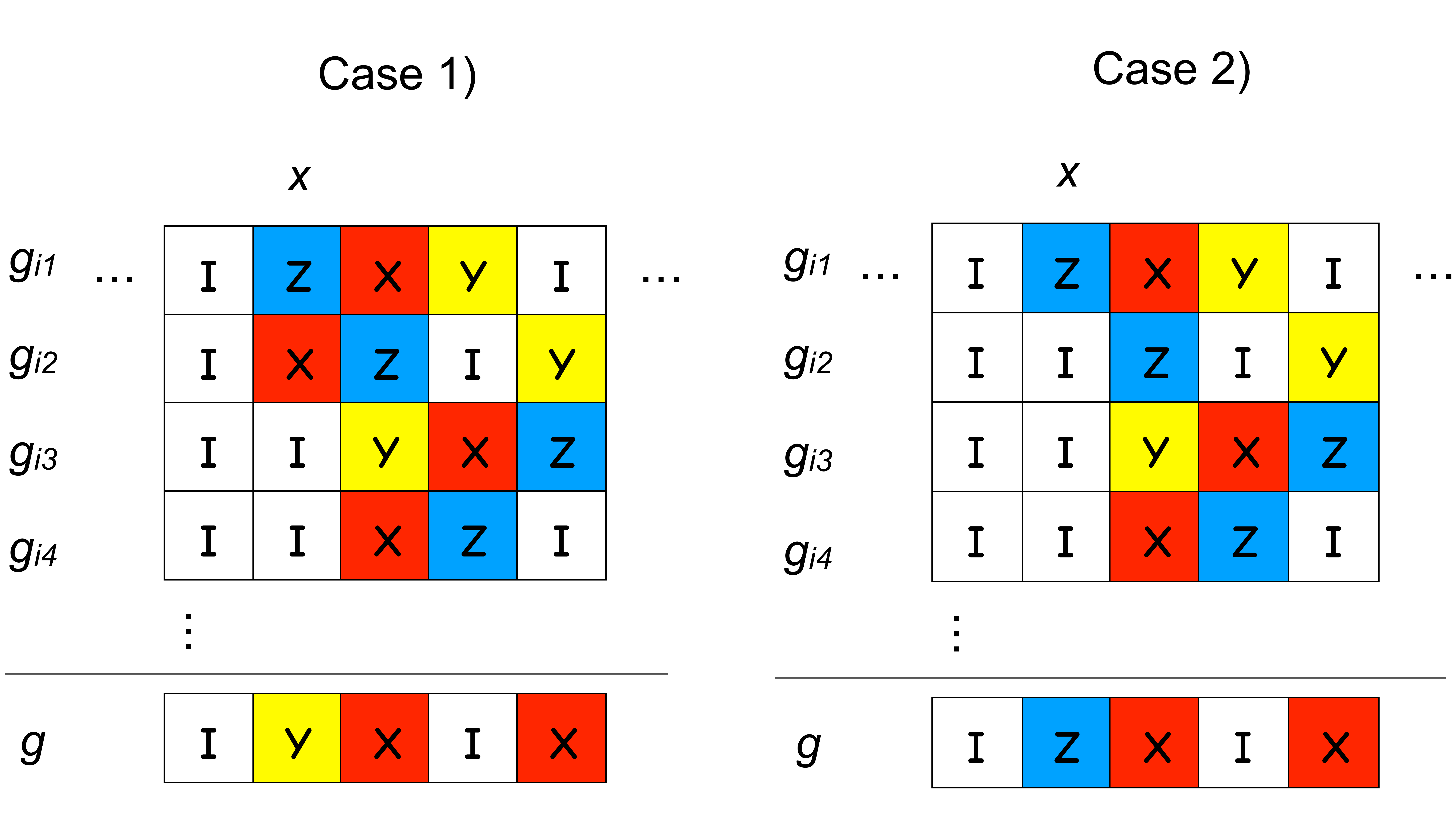}
            \caption{Illustration of the two cases in the proof of the Lemma.}
            \label{fig:lemma}
        \end{figure}
        Without loss of generality, let $\mt{l}(g_{i_1}) \le \mt{l}(g_{i_2}) \le \ldots \le \mt{l}(g_{i_k})$. According to the clipped gauge condition we have two possibilities (see Fig.~\ref{fig:lemma}),
        \begin{enumerate}
            \item $x = \mt{l}(g_{i_1}) = \mt{l}(g_{i_2}) < \mt{l}(g_{i_3}) \le \mt{l}(g_{i_4}) \le \ldots \le \mt{l}(g_{i_k})$. In this case, the clipped gauge condition guarantees that the $g_{i_1}$ and $g_{i_2}$ have different but nontrivial ($X$, $Y$, or $Z$) contents on $x$, and $g_{i_j}$ has trivial content (I) on site $x$, for $j \ge 3$. The product $g$ would then have nontrivial content on $x$, but trivial content for $y < x$.
            \item $x = \mt{l}(g_{i_1}) < \mt{l}(g_{i_2}) \le \mt{l}(g_{i_3}) \le \mt{l}(g_{i_4}) \le \ldots \le \mt{l}(g_{i_k})$. In this case, only $g_{i_1}$ has nontrivial ($X$, $Y$, or $Z$) content on $x$, and $g_{i_j}$ has trivial content (I) on site $x$, for $j \ge 2$. The product $g$ would then have nontrivial content on $x$, but trivial content for $y < x$.
        \end{enumerate}
        Thus $\mt{l}(g) = \mt{l}(g_{i_1})$ as claimed.
        A similar reasoning gives $\mt{r}(g)$. \hfill $\Box$


        \emph{Proof of Proposition 1}: First recall that $\rho_{\mt{l}/\mt{r}}$ in the clipped gauge are completely fixed by the entanglement entropy (which is a gauge invariant quantity) through the following relation \cite{nahum2017KPZ},
        \begin{eqnarray}
            \label{eq:EE_to_rho_L}
            S_A(x) = \sum_{y \le x} \(\rho_\mt{l}(y) - 1\) = \sum_{y > x} \( \rho_\mt{r}(y) - 1\).
        \end{eqnarray}

        To reach our conclusion, we are going to show that, for two arbitrary generating sets $\mc{G}$ and $\mc{G}^\p$ (both in the clipped gauge), the lengths of the $\rho_\mt{l}(x)$ stabilizers that start at site $x$ are the same for $\mc{G}$ and $\mc{G}^\p$, for all sites $x$.

        \begin{enumerate}
        \item
        First, the case $\rho_\mt{l}(x) = 0$ is trivial.
        \item
        Second, consider the case where $\rho_\mt{l}(x) = 1$. Let $g_i \in \mc{G}$ and $g_i^\p \in \mc{G}^\p$, where $\mt{l}(g_i) = \mt{l}(g_i^\p) = x$. Since both $\mc{G}$ and $\mc{G}^\p$ are independent generating sets, $g_i$ has a unique representation as products of elements from $\mc{G}^\p$, and conversely, $g_i^\p$ has a unique representation as products of elements from $\mc{G}$. That is,
        \begin{eqnarray}
            g_i = \prod_{j=1}^L \(g_j^\p\)^{p_j^\p},\quad
            g_i^\p = \prod_{j=1}^L \(g_j\)^{p_j},
        \end{eqnarray}
        where $p_j, p_j^\p$ take values in $\{0, 1\}$. Since $\mt{l}(g_i) = \mt{l}(g_i^\p)$, we know $p_i = p_i^\p = 1$ from the Lemma. Then, again from the Lemma,
        \begin{eqnarray}
            \mt{r}(g_i) \ge \mt{r}(g_i^\p), \\
            \mt{r}(g_i^\p) \ge \mt{r}(g_i).
        \end{eqnarray}
        Hence $\mt{r}(g_i) = \mt{r}(g_i^\p)$, and $g_i$ and $g_i^\p$ have the same lengths.

        \item
        Finally, consider the case where $\rho_\mt{l}(x) = 2$, and let $g_i, g_j \in \mc{G}$, $g_i^\p, g_j^\p \in \mc{G}^\p$, where $\mt{l}(g_i) = \mt{l}(g_j) = \mt{l}(g_i^\p) = \mt{l}(g_j^\p) = x$.
        We again have
        \begin{eqnarray}
            g_i = \prod_{k=1}^L \(g_k^\p\)^{p_k^\p}, \quad
            g_j = \prod_{k=1}^L \(g_k^\p\)^{q_k^\p}, \\
            g_i^\p = \prod_{k=1}^L \(g_k\)^{p_k},\quad
            g_j^\p = \prod_{k=1}^L \(g_k\)^{q_k}.
        \end{eqnarray}
        Without loss of generality, assume $\mt{r}(g_i) \le \mt{r}(g_j)$ and $\mt{r}(g_i^\p) \le \mt{r}(g_j^\p)$. From the Lemma, we know that
        \begin{eqnarray}
            p_i^\p + p_j^\p \ge 1,\quad
            p_i + p_j \ge 1, \\
            q_i^\p + q_j^\p \ge 1,\quad
            q_i + q_j \ge 1.
        \end{eqnarray}
        That is, $g_i$ must has a least one factor of either $g_i^\p$ or $g_j^\p$, to have its left endpoint at $x$. So from the Lemma we have
        \begin{eqnarray}
            \mt{r}(g_i) \ge \min \{ \mt{r}(g_i^\p), \mt{r}(g_j^\p) \} = \mt{r}(g_i^\p).
        \end{eqnarray}
        Similarly,
        \begin{eqnarray}
            \mt{r}(g_i^\p) \ge \min \{ \mt{r}(g_i), \mt{r}(g_j) \} = \mt{r}(g_i).
        \end{eqnarray}
        Hence $\mt{r}(g_i) = \mt{r}(g_i^\p)$.

        Again, without loss of generality, assume $\mt{r}(g_j) \le \mt{r}(g_j^\p)$, thus $\mt{r}(g_i) \le \mt{r}(g_j) \le \mt{r}(g_j^\p)$.

        We observe that $p_j^\p + q_j^\p \ge 1$; otherwise $p_j^\p = q_j^\p = 0$, and we must have $p_i^\p = q_i^\p = 1$, which implies that $g_i$ and $g_j$ have the same content on $x$, in contradiction with the clipping condition. Thus, from the Lemma, we must have at least one of the following,
        \begin{enumerate}
            \item $\mt{r}(g_i) \ge \mt{r}(g_j^\p)$, in which case
            \begin{eqnarray}
                \mt{r}(g_i) = \mt{r}(g_i^\p) = \mt{r}(g_j) = \mt{r}(g_j^\p).
            \end{eqnarray}
            \item $\mt{r}(g_j) \ge \mt{r}(g_j^\p)$, in which case
            \begin{eqnarray}
                \mt{r}(g_i) = \mt{r}(g_i^\p), \quad \mt{r}(g_j) = \mt{r}(g_j^\p).
            \end{eqnarray}
        \end{enumerate}
        Therefore, the stabilizers starting at $x$ have the same length in $\mc{G}$ and in $\mc{G}^\p$.
        \end{enumerate}

        The above arguments work for every site $x$. We have thus proven the Proposition. \hfill $\Box$


        We immediately have the

        {\bf Corollary.} Let $\mt{len}(g) \equiv \mt{r}(g) - \mt{l}(g)$, and
        \begin{eqnarray}
        \mathfrak{D}_{\mc{G}}(\ell) = \frac{1}{L} \sum_{i=1}^L \delta_{\mt{len}(g_i), \ell},
        \end{eqnarray}
        where $\mc{G} = \{g_1, \ldots, g_L\}$. For $\mc{G}$ and $\mc{G}^\p$ satisfying the conditions in the Proposition, we have
        \begin{eqnarray}
            \mathfrak{D}_{\mc{G}} = \mathfrak{D}_{\mc{G}^\p}.
        \end{eqnarray}
        Thus, the length distribution of stabilizers in the clipping gauge is well defined.

\subsubsection{From $\mc{B}(\mc{G})$ to entanglement entropy}


            Define the following subset of $\mc{G}$:
            \begin{eqnarray}
                \mc{G}_A = \{g \in \mc{G} : g \text{ is supported only on } A \}.
            \end{eqnarray}

            {\bf Proposition 2.} Let $\mc{G}$ be a generating set of $\mc{S}$ in the clipped gauge, and $A$ be a \emph{contiguous} subregion of the system. Then $\mc{S}_A$, defined in Eq.~\eqref{eq:EE_to_subgroup_size} as the subgroup of $\mc{S}$ of all the stabilizers that are only supported on $A$, is generated by $\mc{G}_A$.

            \emph{Proof}: Let $g_A$ be an arbitrary element of $\mc{S}_A$. It has the following represention,
            \begin{eqnarray}
                g_A = \prod_{i=1}^L \(g_i\)^{p_i},
            \end{eqnarray}
            where we recall that $\mc{G} = \{g_1, \ldots, g_L\}$, and $p_i = 0, 1$. Suppose $g_i$ is supported on both $A$ and $\ovl{A}$. Either $\mt{l}(g_i) \in \ovl{A}$ or $\mt{r}(g_i) \in \ovl{A}$. From the Lemma, we see that $p_i = 0$, otherwise $g_A$ will have support on $\ovl{A}$, in contradiction with the assumption that $g_A \in \mc{S}_A$.
            Thus, $p_i = 1$ implies that $g_i$ is supported only on $A$.

            We have shown that $\mc{S}_A = \langle \mc{G}_A \rangle$. \hfill $\Box$

            Noticing that $\mc{G}_A$ is also independent, from Eq.~\eqref{eq:EE_to_subgroup_generator} we have the following

            {\bf Corollary.} The entanglement of a contiguous subregion $A$ is given by $S_A = |A| - |\mc{G}_A|$.

            \begin{figure}[t]
                \centering
                \includegraphics[width=.49\textwidth]{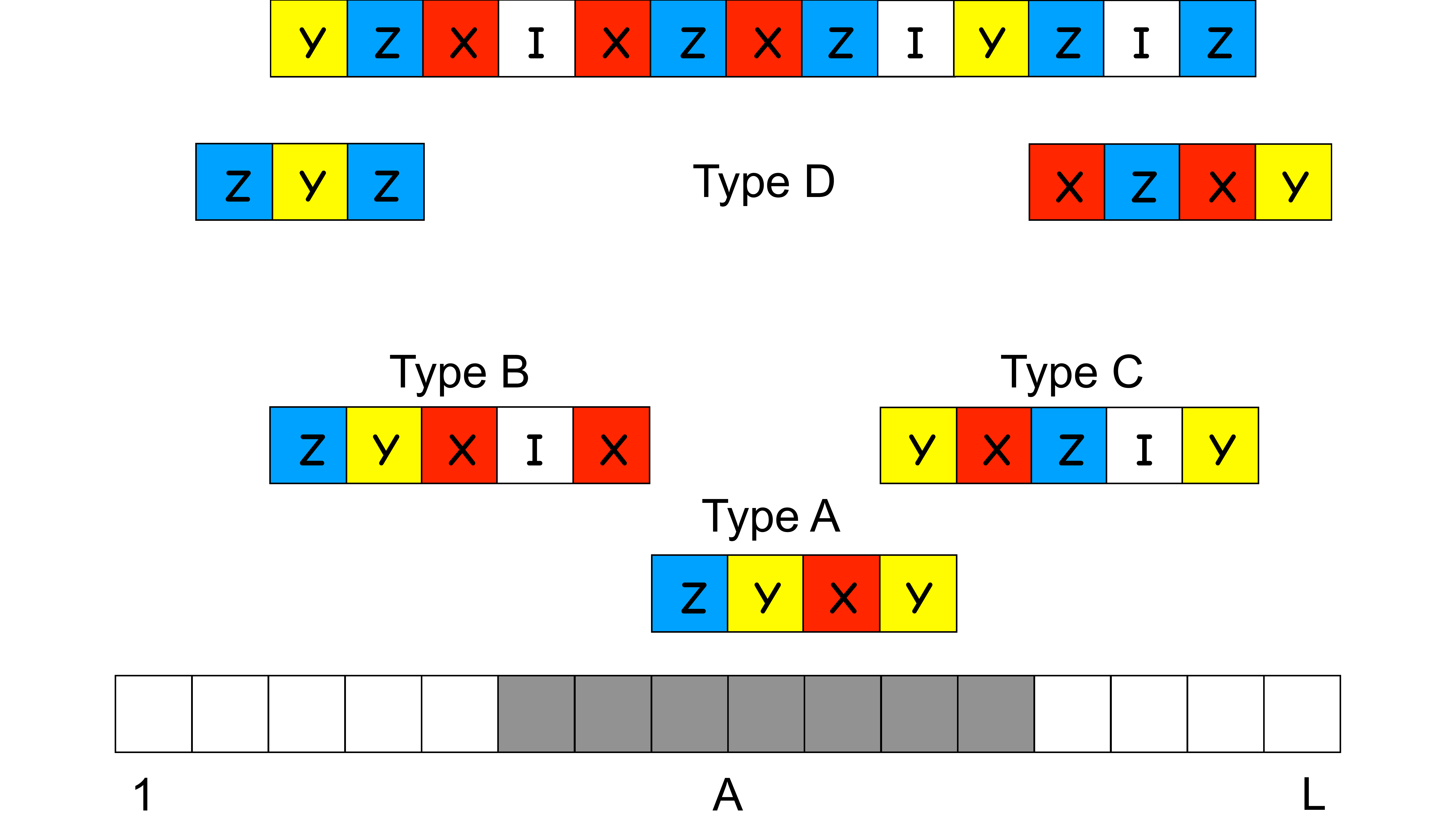}
                \caption{The $4$ types of stabilizers.}
                \label{fig:subregion}
            \end{figure}

            From now on, we will assume that $A$ is contiguous, unless otherwise specified
            \footnote{
            In stating these results, the requirement that $A$ is contiguous is important. Consider the following example of $L=3$,
            \begin{eqnarray}
                \mc{G} = \{ XXI, IXZ, YZY\}.
            \end{eqnarray}
            This set is in the clipped gauge.
            Let $A = \{1, 3\}$. $S_A$ can be shown to be $1$, while $|\mc{G}_A| = 0 \neq |A| - S_A = 1$. Thus, this simple formula \emph{cannot} be readily used for computation of the mutual information,
            \begin{eqnarray}
                I_{A, B} = S_A + S_B - S_{A \cup B},
            \end{eqnarray}
            where $A$ and $B$ are qubits that could be far away.
            }.

            All the stabilizers in $\mc{G}$ can be divided into $4$ types (see Fig.~\ref{fig:subregion}),
            \begin{enumerate}[(A)]
                \item Those that are contained in $A$. These constitute $\mc{G}_A$. Let there be $a = |\mc{G}_A|$ of them.
                \item Those that have their right endpoint in $A$, but left endpoint outside $A$. Let there be $b$ of them.
                \item Those that have their left endpoint in $A$, but right endpoint outside $A$. Let there be $c$ of them.
                \item Those that have their left and endpoints outside $A$. Let there be $d$ of them.
            \end{enumerate}
            Counting the number of endpoints in subregion $A$, we have
            \begin{eqnarray}
                2|A| = 2a + b + c.
            \end{eqnarray}
            Thus
            \begin{eqnarray}
                S_A = |A| - |\mc{G}_A| = |A| - a = \frac{1}{2} (b+c).
            \end{eqnarray}
            When $A$ contains the first site, $b = 0$, it reduces to the familiar formula Eq.~\eqref{eq:EE_to_rho_L}.
            Surprisingly, the entanglement entropy of $A$ depends only on the endpoints of the stabilizers, but not the contents of the stabilizers, as in the more general formulae Eqs.~(\ref{eq:EE_to_subgroup_size}, \ref{eq:EE_to_subgroup_generator}, \ref{eq:EE_to_rank}).
            This simplicity is only present in the clipped gauge.

            Several comments are in order.
            \begin{enumerate}
            \item
            This formula works for any $\mc{G}$ that is in the clipped gauge. It provides another proof that $\mc{B}(\mc{G})$, hence $\mf{D}_\mc{G}$, are well-defined in the clipped gauge.

            Here is an algorithm for getting $\mc{B}(\mc{G})$ from $S_A$ for all contiguous subregions (segments) $A$. At the beginning of the algorithm, we define the variables $a_{[l, r]} = |\mc{G}_{[l, r]}|$ for all segments $[l, r]$, and let $\mc{B} = \{\}$.
            In the $w$-th stage of the algorithm, we look at \emph{all} segments $[x,y]$ of length $w$ $(w = y-x+1)$. $a_{[x,y]} > 0$ means that there are $a_{[x, y]}$ stabilizers that start at $x$ and end at $y$, and we add $a_{[x, y]}$ copies of $(x, y)$ to $\mc{B}$. Then we subtract $a_{[x^\p, y^\p]}$ by the amount of $a_{[x, y]}$, for all $[x^\p, y^\p] \supset [x, y]$.
            This marks the end of the $w$-th stage.

            The algorithm terminates after $L$ stages. The resultant $\mc{B}$ gives the correct $\mc{B}(\mc{G})$. Hence, it is a quantity that is uniquely determined by entanglement entropy (assuming clipped gauge).




            \item
            It has the intuitive interpretation that the entanglement is half the number of stabilizers that span the boundaries of the subregion. In certain limits the formula reduces to simply counting the number of entangled Bell pairs across the boundary, which is an example we know and like. However, the Bell pair picture fails to characterize multipartite entanglement because of the trivial internal structure of the stabilizers.
            \end{enumerate}

        \section{Entanglement dynamics under Clifford unitary-projective evolution \label{appB}}

In this section, we try to give a simple picture for the entanglement entropy for contiguous subregions starting from the 1st site, which we define as the height function,
\envelope{eqnarray}{
    h(x) \coloneqq S_{A = \{1,\ldots,x\}}. 
}
This is the same function considered in Ref.~\cite{nahum2017KPZ} and shown in Fig.~\ref{fig:schematic_growth}. 

\begin{figure}[t]
    \includegraphics[width=.49\textwidth]{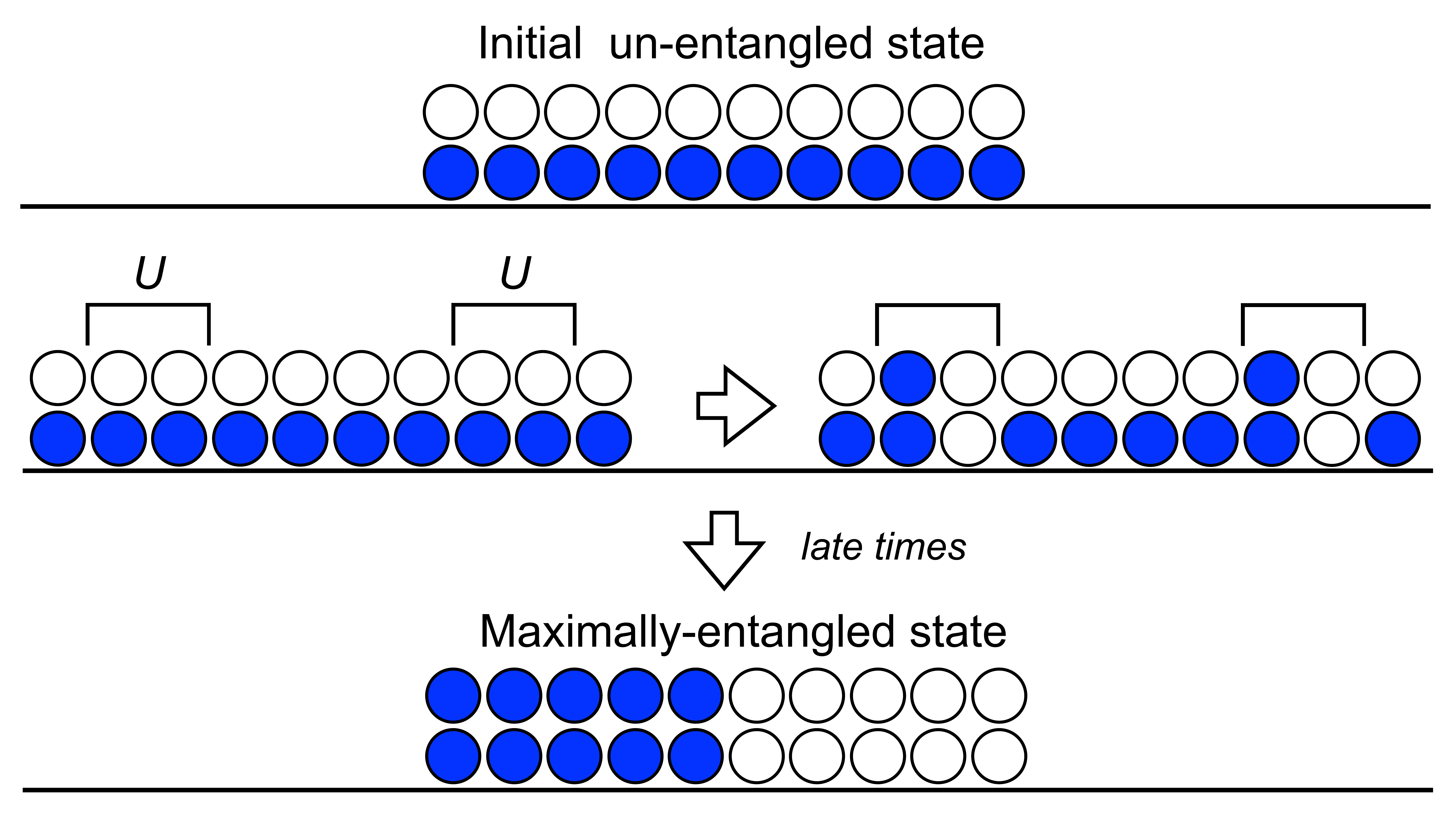}
    \caption{Schematic illustration of the particle movement under purely unitary evolution, from a trivial product state to a maximally entangled state.}
    \label{fig:particle_under_unitary}
\end{figure}

Alternatively, based on Eq.~\eqref{eq:EE_to_rho_L}, we can also consider dynamics of $\rho_\mt{l}$ within the clipped gauge, which encodes the same information as the height function.
We will use the pictorial representation in Fig.~\ref{fig:particle_under_unitary}, where each blue dot represents a left endpoint, and each white dot represents a right endpoint.
We will view the left endpoints as ``particles'' and the right ones as ``holes''.
Recall that the clipped gauge requires that the total number of dots on each site is $2$.

For the convenience of discussion, we consider systems with open boundary condition in this appendix.

\subsection{Unitary dynamics}

Consider a local unitary on qubits $x$ and $x+1$, as in Fig.~\ref{fig:particle_under_unitary}.
According to Eq.~\eqref{eq:EE_to_rho_L},
\begin{equation}
    \rho_\mt{l}(x) + \rho_\mt{l}(x+1) - 2 = h(x+1) - h(x-1).
\end{equation}
The local unitary on the bond $(x, x+1)$ does not change $h(x+1)$ or $h(x-1)$, thus,
restricting to the clipped gauge before and after the gate, the quantity $\rho_\mt{l}(x) + \rho_\mt{l}(x+1)$ remains the same as before the unitary gate. Moreover, $\rho_\mt{l}(y)$ is left invariant by $U_{x,x+1}$ for $y \neq x, x+1$ for a similar reason. Hence the following

{\bf Observation:} a local unitary gate on qubits $(x,x+1)$ can only redistribute particles on sites $x$ and $x+1$, while leaving particles on other sites untouched, as illustrated in Fig.~\ref{fig:particle_under_unitary}.

If the unitary is taken from the Haar measure, and we take the local Hilbert space dimension $q$ to infinity, the entanglement growth is governed by the following equation \cite{nahum2017KPZ}
\envelope{eqnarray}{
    h(x, t+1) = \min \{h(x-1, t), h(x+1, t)\} + 1.
}
This is the 
crystal growth model. 
Since $\rho_\mt{l}$ is the derivative of $h(x)$, under the action of a random Haar unitary, the particles within the range of action will drift to the left as much as they can with the filling constraint $\rho_\mt{l}(x) \le 2$, while particles outside the range of action stay where they are. 

The difference between Clifford unitaries and random Haar unitaries is that instead of ballistic movement, the particles experience the biased diffusion with filling constraint.
This is captured by the KPZ equation derived in Ref.~\cite{nahum2017KPZ}.
Without further justification, we assume that this is the correct picture for entanglement growth under Clifford dynamics.

At long times $t \to \infty$, all the particles will clump to the left half of the system, corresponding to a maximally entangled state (see Fig.~\ref{fig:particle_under_unitary}). The fluctuation of $h(x)$ around the maximal value is expected to be small~\cite{nahum2017KPZ}.

\subsection{Measurement dynamics}

Here we consider one-qubit Pauli-$Z$ measurements and their effects on $\rho_\mt{l}$.

First recall the transformation of $\mc{G}$ under the effect of a measurement of $Z_x$ in Eq.~\eqref{eq:stabilizer_after_measurement}.
Let $\mc{G} = \{g_1, \ldots, g_k, g_{k+1}, \ldots, g_L\}$ be in the clipped gauge and suppose that $[g_j, Z_x] = 0$ for $j \le k$, and $\{g_j, Z_x\} = 0$ for $j > k$.
The stabilizer group of the measured wavefunction is generated by
\begin{eqnarray}
    \label{eq:stabilizer_after_measurement_prime}
    \mc{G}^\p = \{g_1, \ldots, g_k, g_{k+1} g_{k+2}, \ldots, g_{L-1} g_L, Z_x\}.
\end{eqnarray}
This set does not necessarily respects the clipped gauge; some clipping is necessary.
In Appendix~\ref{appA} we see that $\rho_\mt{l}$ is determined by just the pre-gauge condition, and is left invariant by the second Gaussian elimination.
Since we are focusing on the $\rho_\mt{l}$ dynamics, it suffices to check only the pre-gauge condition.

Observe that since $x$ is disentangled from the rest of the system after the measurement, $Z_x$ will remain in $\mc{G}^\p$ after clipping.

In Eq.~\eqref{eq:stabilizer_after_measurement_prime}, the ordering of the stabilizers is not essential; different orderings correspond to the same wavefunction.
For convenience, we assume that $g_{k+1}, g_{k+2}, \ldots, g_L$ are ordered in such a way that their left endpoints are non-decreasing,
\begin{eqnarray}
    \mt{l}(g_{k+1}) \le \mt{l}(g_{k+2}) \le \ldots \le \mt{l}(g_{L}).
\end{eqnarray}
The clipped gauge guarantees that $g_{j} g_{j+1}$ has the same left endpoint as $g_{j}$, for $j > k$.
Thus, comparing $\rho_\mt{l}$ for $\mc{G}$ and $\mc{G}^\p$, the net effect of a measurement $Z_x$ is the following,
\begin{eqnarray}
    \label{eq:hop_before_clip}
    \rho_\mt{l}( \mt{l}(g_L) ) \to \rho_\mt{l}( \mt{l}(g_L) ) - 1, \ 
    \rho_\mt{l}( x ) \to \rho_\mt{l}( x ) + 1.
\end{eqnarray}
If we now run the clipping algorithm and check for the pre-gauge (i.e. the first Gaussian elimination), it will find that the pre-gauge constraints are satisfied for all $y < x$.
The first site that might violate this constraint is $x$.
The clipping algorithm would then check the constraint and move the left-endpoints to the right of $x$ (the row elimination process), if necessary.

\begin{figure}[t]
    \includegraphics[width=.49\textwidth]{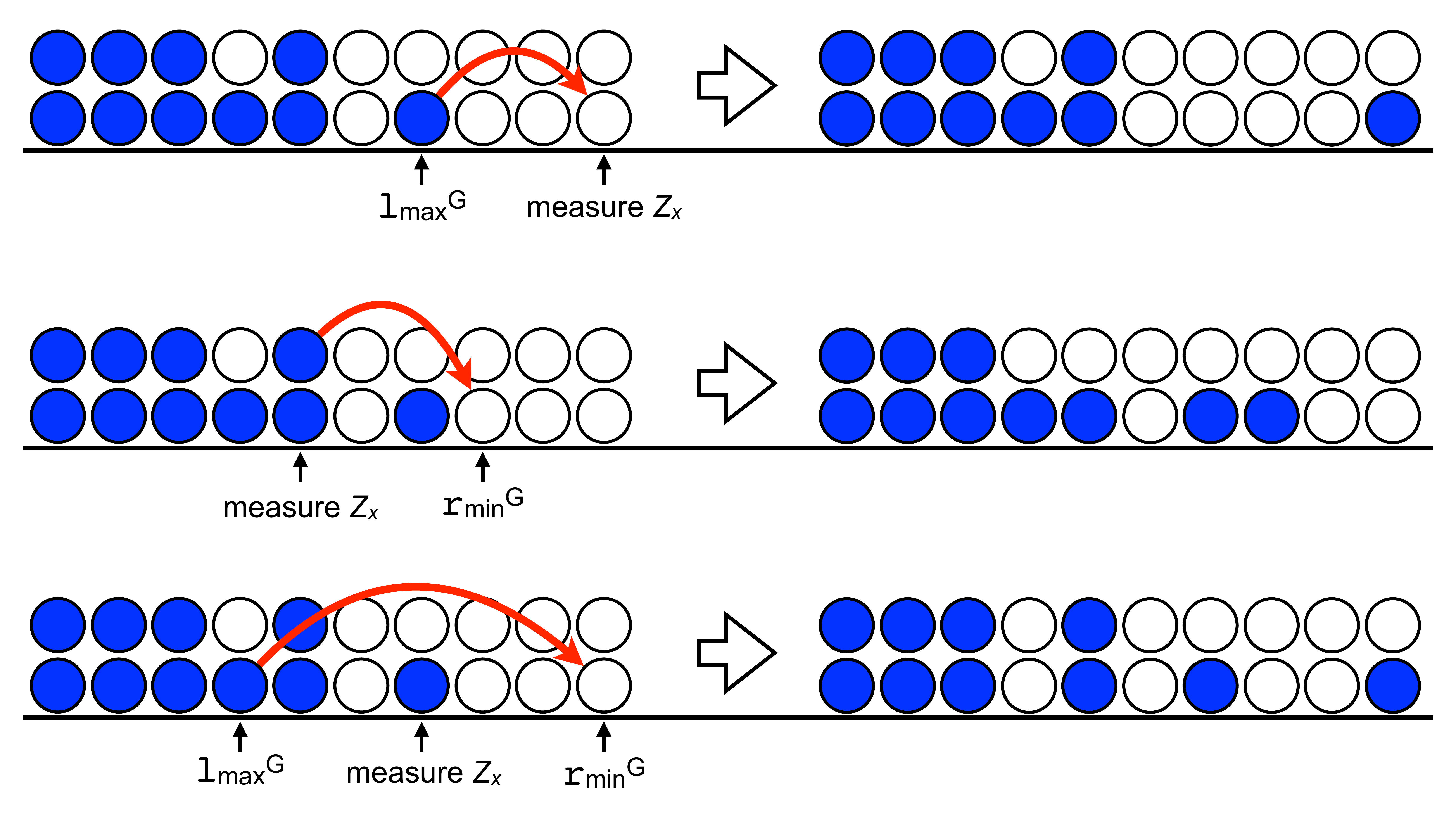}
    \caption{Illustration of the hopping processes of the particles under a local measurement at site $x$.}
    \label{fig:hop_when_clipped}
\end{figure}

The coordinate $\mt{l}(g_L)$ where a particle gets removed is the right-most left-endpoint among stabilizers that anticommute with $Z_x$, which we define to be
\envelope{eqnarray}{
    \mt{l}_{\text{max}}^\mc{G} \coloneqq \max \{ \mt{l}(g) : g \in \mc{G}, \mc{G} \text{ clipped, and } \{g, Z_x\} = 0. \}\nn
}
We define a similar quantity which will prove to be useful,
\envelope{eqnarray}{
    \mt{r}_{\text{min}}^\mc{G} \coloneqq \min \{ \mt{r}(g) : g \in \mc{G}, \mc{G} \text{ clipped, and } \{g, Z_x\} = 0. \}\nn
}
Using this notation, we can further deduce the change of $\rho_\mt{l}$ under a local measurement. There are three cases (see Fig.~\ref{fig:hop_when_clipped}),
\begin{enumerate}
    \item $\rho_\mt{l}(x) = 0$ before measurement. It follows that $\mt{l}_{\text{max}}^\mc{G} < x$, and $\rho_\mt{l}(\mt{l}_{\text{max}}^\mc{G}) \ge 1$. After the operation in Eq.~\eqref{eq:hop_before_clip}, the pre-gauge constraint is satisfied everywhere, and the algorithm terminates. 
    The height $h(w)$ is reduced by 1 for $w \in [\mt{l}_{\text{max}}^\mc{G}, x)$.

    \item $\rho_\mt{l}(x) = 2$ before measurement. It follows that $\mt{l}_{\text{max}}^\mc{G} = x$.
    After clipping, $\rho_\mt{l}(x)$ is reduced by $1$, and that reduction is compensated by the increase of $\rho_\mt{l}(y)$ for some $y>x$, for which $\rho_\mt{l}(y) \le 1$ before the measurement.

    If we view this processes from the perspective of $\rho_\mt{r}$, it would have the particle-hole symmetric dynamics, where the symmetry operation is
    \begin{eqnarray}
        \label{eq:ph_symm}
        x \to L-x, \quad \rho \to 2-\rho.
    \end{eqnarray}
    Consequently, the position $y$ is equal to $\mt{r}_{\text{min}}^\mc{G}$, and the height $h(w)$ is reduced by 1 for $w \in [x, \mt{r}_{\text{min}}^\mc{G})$.

    \item $\rho_\mt{l}(x) = 1$ before measurement.
    If this stabilizer has $X$ or $Y$ on site $x$,  the measurement has no effect on $\rho_\mt{l}(x)$.
    If this stabilizer has $Z$ on site $x$, the measurement will first hop a particle from site $\mt{l}_{\text{max}}^\mc{G} < x$ to $x$, then hop a particle from $x$ to $\mt{r}_{\text{min}}^\mc{G} > x$, as described in the previous two cases.
    The height $h(w)$ is reduced by 1 for $w \in [\mt{l}_{\text{max}}^\mc{G}, \mt{r}_{\text{min}}^\mc{G})$.
\end{enumerate}

Given these observations, we see that the effect of a local measurement at $x$, in the particle picture, is to hop exactly one particle across $x$ via clipping.
Thus we have an apparently simple picture for the entanglement dynamics in the unitary-measurement Clifford circuit in terms of the particles, which are drifted to the left in a local fashion under unitary gates, and ``hopped'' to the right under measurements in a non-local fashion.

What remains unspecified is the hopping distance, $R$, that is, the distance between the initial and final positions of the moving particle.
This quantity takes the values $x - \mt{l}_{\text{max}}^\mc{G}$, $\mt{r}_{\text{min}}^\mc{G} - x$, and $\mt{r}_{\text{min}}^\mc{G} - \mt{l}_{\text{max}}^\mc{G}$ in the three cases above, respectively.
For concreteness, consider the following function,
\envelope{eqnarray}{
    H \coloneqq \sum_w h(w).
}
From the discussion above, it is easy to see that the change in $H$ after a time cycle is
\envelope{eqnarray}{
    \Delta H = O(L) + \sum_{k=1}^{p L} (-R_k),
}
where $R_k$ is the distance of the hopping in the $k$-th measurement, and the $O(L)$ terms comes from the unitary gates.
We replace the second term by its mean value,
\envelope{eqnarray}{
    \Delta H = O(L) - p L \avg{R}.
}
Within the steady state, the two terms must cancel out, so that $\avg{R} = O(1)$.

\begin{figure}[t]
    \centering
    \includegraphics[width=.5\textwidth]{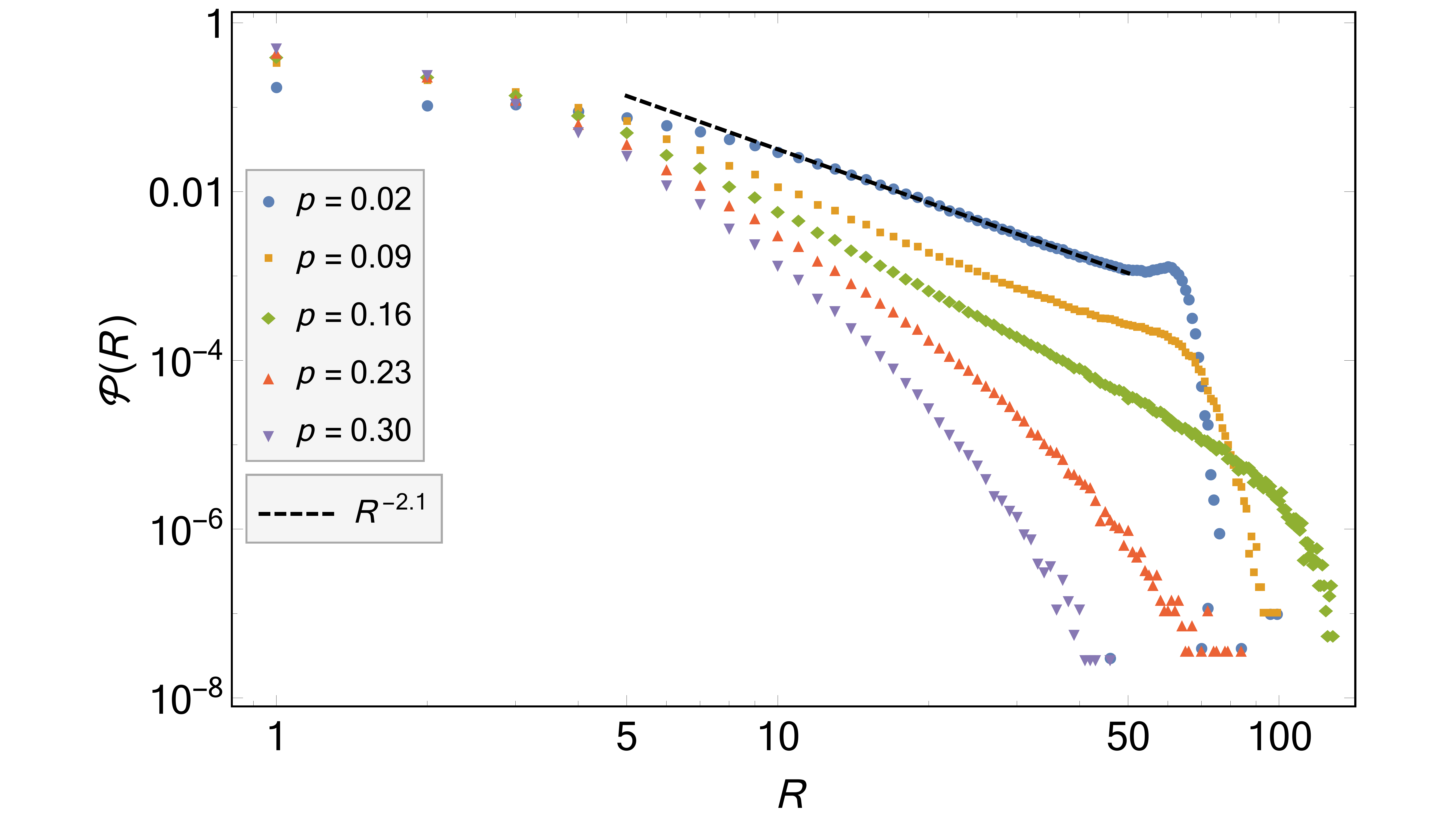}
    \caption{The normalized distribution funciton of $R$, on a log-log scale.}
    \label{fig:P(R)}
\end{figure}

In Fig.~\ref{fig:P(R)}, we plot the normalized distribution function of $R$, denoted $\mc{P}(R)$, for several different values of $p$ within a system of size $L = 128$, within the random Clifford circuit.
Within the volume law phase $p < p_c$, the distribution function takes the form of a power law decaying function whose magnitude does not depend on the system size (as we verify but not shown), $\mc{P}(R) \sim R^{-\gamma}$ up to $R \sim L/2$.
Within the area law phase the distribution is short ranged.
Schematically,
\envelope{eqnarray}{
    \mc{P}\(R\) \sim 
    \begin{cases}
        \frac{1}{R^\gamma}, p < p_c, \\
        \frac{e^{-R/R_0}}{R^\gamma}, p > p_c, \\
    \end{cases}
}
{where $\gamma$, which varies throughout the volume phase, always satisfies $\gamma > 2$, and} $R_0$ is a finite length scale.
{As of now, we have not understood this power law distribution, and leave it for future work.
Nevertheless, the expectation values of the hopping distance can be readily computed,
$\langle R \rangle = \int^{L/2} dR\, R \mc{P}(R)$.}
In the volume law phase, the mean value of $R$ is finite (as $L \rightarrow \infty$) since $\gamma > 2$, while in the area law phase this value is finite regardless.

Notice that the quantity $\delta_M \ovl{h}$, defined in Sec.~\ref{sec7}, is proportional to the hopping distance $R$ within {the Clifford context, $ \delta_M \ovl{h} = - R/L$,
so that $\langle \delta_M \bar{h} \rangle = O(1/L)$.}

\subsection{Toy particle traffic-flow model}

\begin{figure}[t]
    \centering
    \includegraphics[width=.49\textwidth]{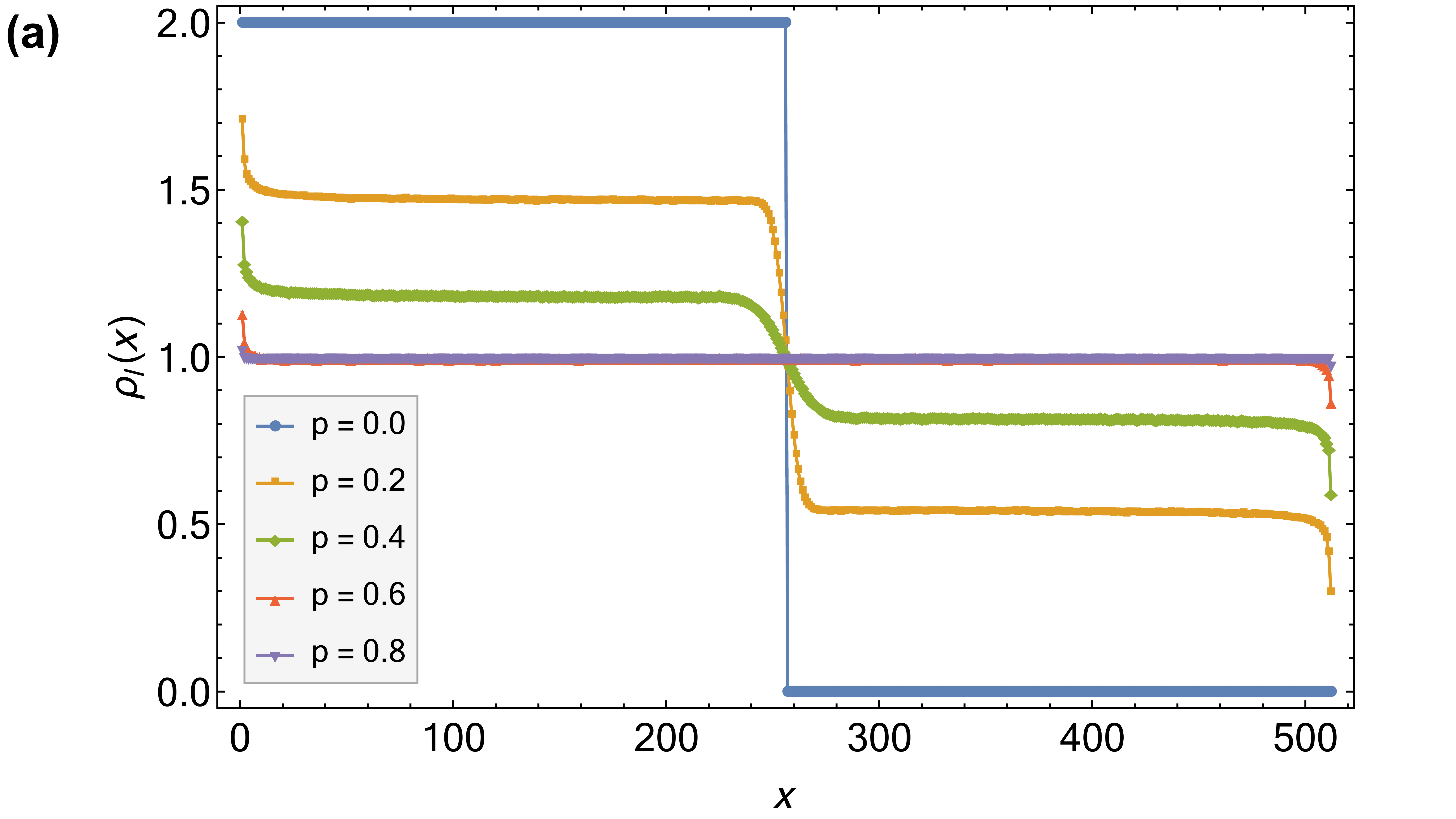}
    \includegraphics[width=.49\textwidth]{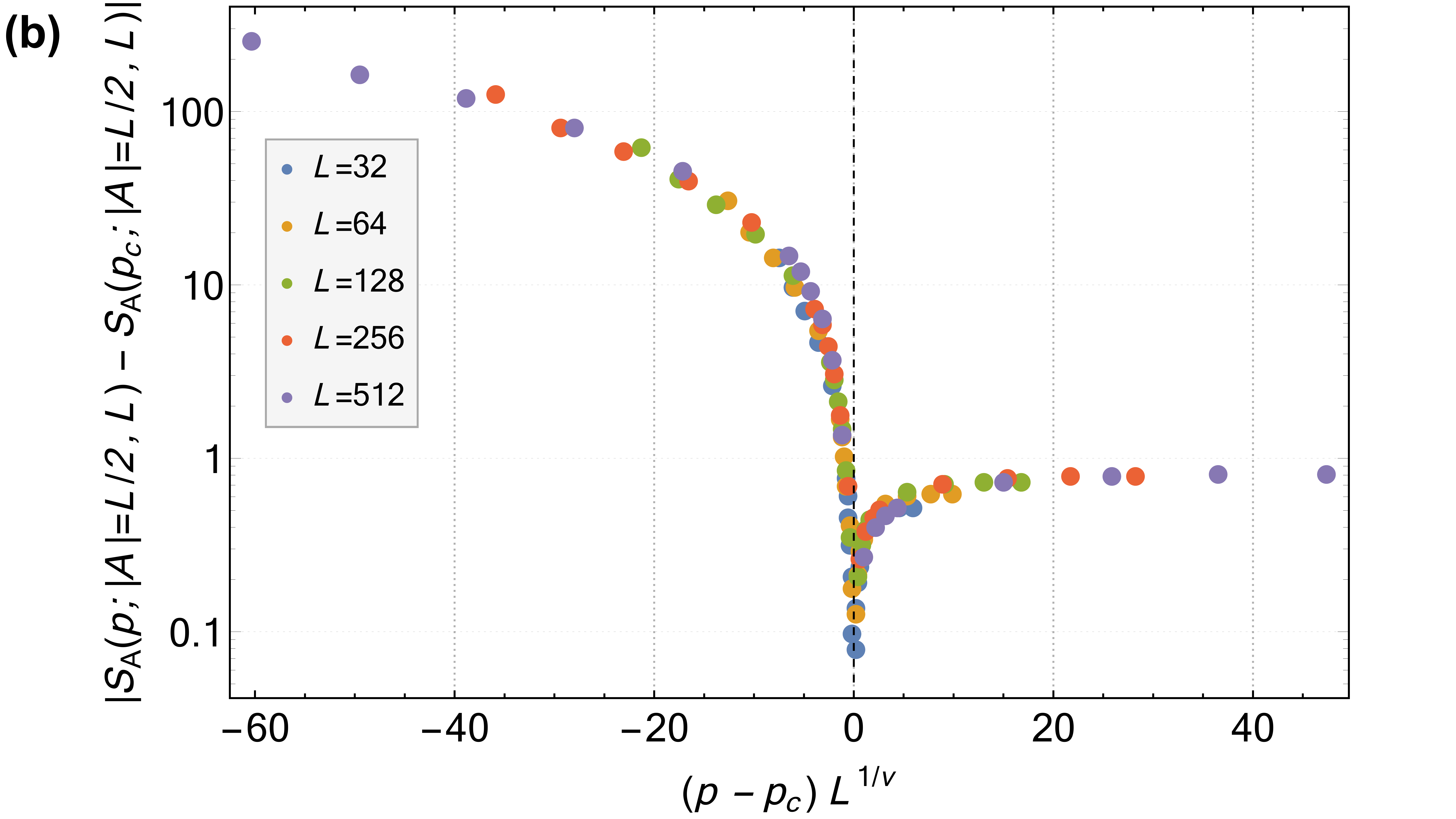}
    \includegraphics[width=.49\textwidth]{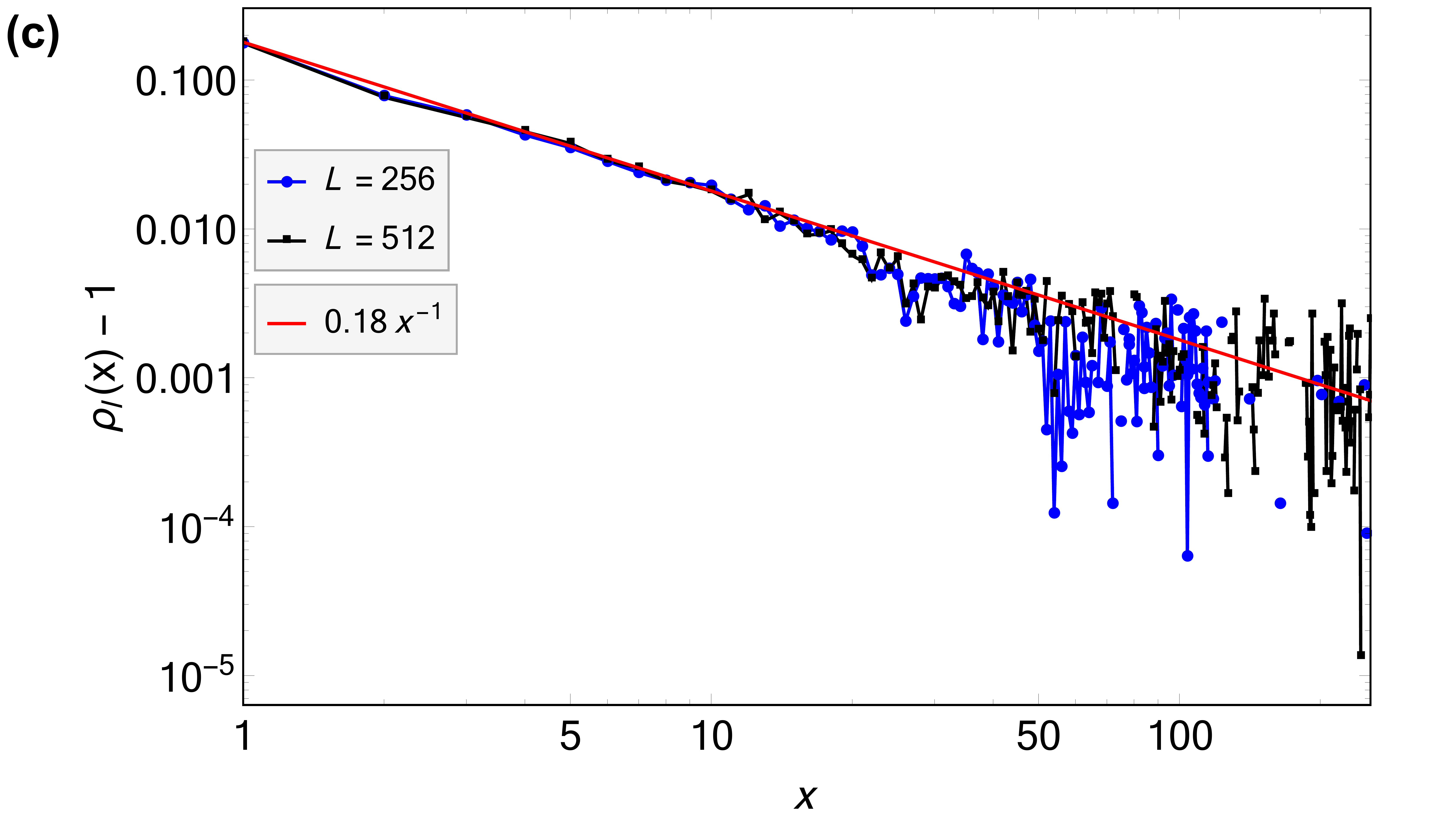}
    \caption{(a) The average steady state particle density for different values of $p$ with fixed $L = 512$.
    As seen in Eq.~\eqref{eq:EE_to_rho_L}, the volume law phase corresponds to a plateau in $\rho_\mt{l}(x)$ with height greater than $1$, while the transition is signified by a continuous decrease of this height to $1$.
    There is clearly a particle hole symmetry in $\rho_\mt{l}(x)$ (see Eq.~\eqref{eq:ph_symm}) at all values of $p$.
    (b) Collapse of the entanglement entropy using the scaling form in Eq.~\eqref{eq:EE_scaling_form}, where we choose $\nu = 1.33$ and $p_c = 0.56$.
    (c) The particle density at the critical point. The data can be fit to a slope $-1$ on a log-log scale, suggesting logarithmic scaling of entanglement entropy (see Eq.~\eqref{eq:EE_to_rho_L}), reproducing the result of the full Clifford dynamics.
    }
    \label{fig:traffic_result}
\end{figure}

The apparent simplicity of the dynamical rules governing the particles motion
in the Clifford circuit studied in the previous subsection is somewhat misleading; to faithfully simulate the particle dynamics, the knowledge of particle densities are not enough, and one has to specify the internal contents of the stabilizers (so as to obtain $\mt{l}_{\text{max}}^\mc{G}$ and $\mt{r}_{\text{min}}^\mc{G}$).
In this subsection, we design an effective toy model which we term the ``traffic-flow model'', that aims to capture the essence of the Clifford particle dynamics without resorting to 
a full stabilizer simulation.  
As we shall see, the particles motion is designed to mimic the motion of the stabilizers left-endpoints under both unitaries and measurements, as described in detail in the previous subsection.

Specifically, we start with a one dimensional system of $L$ sites with open boundary condition, and initially put in $L$ particles (mimicking the left endpoints of the stabilizers), one on each site, as in a product state; the total number of particles is conserved.
At all times, we impose the constraint that on any site there are at most two particles, equivalent to the clipping condition.

To imitate the random Clifford circuit,
we choose the particle motion under unitary gates to be ballistic and uni-directional (to the left), instead of diffusive.
The particle motion under ``measurements'' is chosen to satisfy the following simple rules:
\envelope{enumerate}{
    \item
    When $\rho_\mt{l}(x) = 0$, choose the closest particle to the left of $x$ at $y < x$, and hop it from $y$ to $x$.
    \item
    When $\rho_\mt{l}(x) = 2$, choose the closest hole to the right of $x$ at $z > x$, and hop one particle from $x$ to $z$.
    \item
    When $\rho_\mt{l}(x) = 1$, leave the particle density untouched.
    \item
    After each measurement, the measured qubit is taken out of the system, until the layer (with $pL$ measurements) terminates.
    This is because the measurements within the same layer commute with each other, so that a site that is already measured cannot serve as $\mt{l}^{\mc{G}}_{\text{max}}$ or $\mt{r}^{\mc{G}}_{\text{min}}$ for subsequent measurements.
    Moreover, the temporal ordering of the measurements is inessential given this rule, as expected.
}

In effect, we are replacing $\mt{l}^\mc{G}_\text{max}$ and $\mt{r}^\mc{G}_\text{min}$ above with possibilities that are closest to $x$.
This choice is of course an over-simplification, and is not faithful to real Clifford dynamics. 
In particular, the hopping distance distribution is strictly short-ranged (data not shown), and does not have the power law form.
However, as we will see below, this toy model captures some universal features of the random Clifford circuit.

We numerically simulate this classical model and present the results in Fig.~\ref{fig:traffic_result}.
The function $\rho_\mt{l}$ shows a volume law to area law transition, with similar critical exponents and logarithmic scaling of entanglement at the critical point (although the coefficient of the logarithmic function is significantly smaller than $\alpha(p_c)$ we found in earlier sections).
Thus the rules of our toy model are partially justified.

The traffic-flow model provides a different perspective for studying entanglement dynamics.
While our  ``traffic rules'' are over-simplified, one might still hope to design a set of rules that faithfully represents the particle dynamics under the full Clifford evolution.
In fact, this framework could be more versatile than what is already envisioned, and tweaking with the rules might result in  a whole class of different entanglement dynamics, not necessarily within the same universality class as the Clifford ones.
We leave these studies to future works.

    \bibliography{refs}

\begin{thebibliography}{66}%
\makeatletter
\providecommand \@ifxundefined [1]{%
 \@ifx{#1\undefined}
}%
\providecommand \@ifnum [1]{%
 \ifnum #1\expandafter \@firstoftwo
 \else \expandafter \@secondoftwo
 \fi
}%
\providecommand \@ifx [1]{%
 \ifx #1\expandafter \@firstoftwo
 \else \expandafter \@secondoftwo
 \fi
}%
\providecommand \natexlab [1]{#1}%
\providecommand \enquote  [1]{``#1''}%
\providecommand \bibnamefont  [1]{#1}%
\providecommand \bibfnamefont [1]{#1}%
\providecommand \citenamefont [1]{#1}%
\providecommand \href@noop [0]{\@secondoftwo}%
\providecommand \href [0]{\begingroup \@sanitize@url \@href}%
\providecommand \@href[1]{\@@startlink{#1}\@@href}%
\providecommand \@@href[1]{\endgroup#1\@@endlink}%
\providecommand \@sanitize@url [0]{\catcode `\\12\catcode `\$12\catcode
  `\&12\catcode `\#12\catcode `\^12\catcode `\_12\catcode `\%12\relax}%
\providecommand \@@startlink[1]{}%
\providecommand \@@endlink[0]{}%
\providecommand \url  [0]{\begingroup\@sanitize@url \@url }%
\providecommand \@url [1]{\endgroup\@href {#1}{\urlprefix }}%
\providecommand \urlprefix  [0]{URL }%
\providecommand \Eprint [0]{\href }%
\providecommand \doibase [0]{http://dx.doi.org/}%
\providecommand \selectlanguage [0]{\@gobble}%
\providecommand \bibinfo  [0]{\@secondoftwo}%
\providecommand \bibfield  [0]{\@secondoftwo}%
\providecommand \translation [1]{[#1]}%
\providecommand \BibitemOpen [0]{}%
\providecommand \bibitemStop [0]{}%
\providecommand \bibitemNoStop [0]{.\EOS\space}%
\providecommand \EOS [0]{\spacefactor3000\relax}%
\providecommand \BibitemShut  [1]{\csname bibitem#1\endcsname}%
\let\auto@bib@innerbib\@empty
\bibitem [{\citenamefont {{Deutsch}}(1991)}]{Deutsch1991}%
  \BibitemOpen
  \bibfield  {author} {\bibinfo {author} {\bibfnamefont {J.~M.}\ \bibnamefont
  {{Deutsch}}},\ }\bibfield  {title} {\enquote {\bibinfo {title} {{Quantum
  statistical mechanics in a closed system}},}\ }\href {\doibase
  10.1103/PhysRevA.43.2046} {\bibfield  {journal} {\bibinfo  {journal} {\pra}\
  }\textbf {\bibinfo {volume} {43}},\ \bibinfo {pages} {2046--2049} (\bibinfo
  {year} {1991})}\BibitemShut {NoStop}%
\bibitem [{\citenamefont {{Srednicki}}(1994)}]{Srednicki1994}%
  \BibitemOpen
  \bibfield  {author} {\bibinfo {author} {\bibfnamefont {Mark}\ \bibnamefont
  {{Srednicki}}},\ }\bibfield  {title} {\enquote {\bibinfo {title} {{Chaos and
  quantum thermalization}},}\ }\href {\doibase 10.1103/PhysRevE.50.888}
  {\bibfield  {journal} {\bibinfo  {journal} {\pre}\ }\textbf {\bibinfo
  {volume} {50}},\ \bibinfo {pages} {888--901} (\bibinfo {year} {1994})},\
  \Eprint {http://arxiv.org/abs/cond-mat/9403051} {arXiv:cond-mat/9403051
  [cond-mat]} \BibitemShut {NoStop}%
\bibitem [{\citenamefont {{Calabrese}}\ and\ \citenamefont
  {{Cardy}}(2005)}]{Calabrese_2005}%
  \BibitemOpen
  \bibfield  {author} {\bibinfo {author} {\bibfnamefont {Pasquale}\
  \bibnamefont {{Calabrese}}}\ and\ \bibinfo {author} {\bibfnamefont {John}\
  \bibnamefont {{Cardy}}},\ }\bibfield  {title} {\enquote {\bibinfo {title}
  {{Evolution of entanglement entropy in one-dimensional systems}},}\ }\href
  {\doibase 10.1088/1742-5468/2005/04/P04010} {\bibfield  {journal} {\bibinfo
  {journal} {Journal of Statistical Mechanics: Theory and Experiment}\ }\textbf
  {\bibinfo {volume} {2005}},\ \bibinfo {pages} {04010} (\bibinfo {year}
  {2005})},\ \Eprint {http://arxiv.org/abs/cond-mat/0503393}
  {arXiv:cond-mat/0503393 [cond-mat.stat-mech]} \BibitemShut {NoStop}%
\bibitem [{\citenamefont {{Calabrese}}\ and\ \citenamefont
  {{Cardy}}(2007)}]{Calabrese_2007}%
  \BibitemOpen
  \bibfield  {author} {\bibinfo {author} {\bibfnamefont {Pasquale}\
  \bibnamefont {{Calabrese}}}\ and\ \bibinfo {author} {\bibfnamefont {John}\
  \bibnamefont {{Cardy}}},\ }\bibfield  {title} {\enquote {\bibinfo {title}
  {{Quantum quenches in extended systems}},}\ }\href {\doibase
  10.1088/1742-5468/2007/06/P06008} {\bibfield  {journal} {\bibinfo  {journal}
  {Journal of Statistical Mechanics: Theory and Experiment}\ }\textbf {\bibinfo
  {volume} {2007}},\ \bibinfo {pages} {06008} (\bibinfo {year} {2007})},\
  \Eprint {http://arxiv.org/abs/0704.1880} {arXiv:0704.1880
  [cond-mat.stat-mech]} \BibitemShut {NoStop}%
\bibitem [{\citenamefont {{Rigol}}\ \emph {et~al.}(2008)\citenamefont
  {{Rigol}}, \citenamefont {{Dunjko}},\ and\ \citenamefont
  {{Olshanii}}}]{rigol2008}%
  \BibitemOpen
  \bibfield  {author} {\bibinfo {author} {\bibfnamefont {Marcos}\ \bibnamefont
  {{Rigol}}}, \bibinfo {author} {\bibfnamefont {Vanja}\ \bibnamefont
  {{Dunjko}}}, \ and\ \bibinfo {author} {\bibfnamefont {Maxim}\ \bibnamefont
  {{Olshanii}}},\ }\bibfield  {title} {\enquote {\bibinfo {title}
  {{Thermalization and its mechanism for generic isolated quantum systems}},}\
  }\href {\doibase 10.1038/nature06838} {\bibfield  {journal} {\bibinfo
  {journal} {\nat}\ }\textbf {\bibinfo {volume} {452}},\ \bibinfo {pages}
  {854--858} (\bibinfo {year} {2008})},\ \Eprint
  {http://arxiv.org/abs/0708.1324} {arXiv:0708.1324 [cond-mat.stat-mech]}
  \BibitemShut {NoStop}%
\bibitem [{\citenamefont {{Kim}}\ and\ \citenamefont {{Huse}}(2013)}]{Kim2013}%
  \BibitemOpen
  \bibfield  {author} {\bibinfo {author} {\bibfnamefont {Hyungwon}\
  \bibnamefont {{Kim}}}\ and\ \bibinfo {author} {\bibfnamefont {David~A.}\
  \bibnamefont {{Huse}}},\ }\bibfield  {title} {\enquote {\bibinfo {title}
  {{Ballistic Spreading of Entanglement in a Diffusive Nonintegrable
  System}},}\ }\href {\doibase 10.1103/PhysRevLett.111.127205} {\bibfield
  {journal} {\bibinfo  {journal} {\prl}\ }\textbf {\bibinfo {volume} {111}},\
  \bibinfo {eid} {127205} (\bibinfo {year} {2013})},\ \Eprint
  {http://arxiv.org/abs/1306.4306} {arXiv:1306.4306 [quant-ph]} \BibitemShut
  {NoStop}%
\bibitem [{\citenamefont {{Mezei}}\ and\ \citenamefont
  {{Stanford}}(2017)}]{Mezei2017}%
  \BibitemOpen
  \bibfield  {author} {\bibinfo {author} {\bibfnamefont {M{\'a}rk}\
  \bibnamefont {{Mezei}}}\ and\ \bibinfo {author} {\bibfnamefont {Douglas}\
  \bibnamefont {{Stanford}}},\ }\bibfield  {title} {\enquote {\bibinfo {title}
  {{On entanglement spreading in chaotic systems}},}\ }\href {\doibase
  10.1007/JHEP05(2017)065} {\bibfield  {journal} {\bibinfo  {journal} {Journal
  of High Energy Physics}\ }\textbf {\bibinfo {volume} {2017}},\ \bibinfo {eid}
  {65} (\bibinfo {year} {2017})},\ \Eprint {http://arxiv.org/abs/1608.05101}
  {arXiv:1608.05101 [hep-th]} \BibitemShut {NoStop}%
\bibitem [{\citenamefont {{Nandkishore}}\ and\ \citenamefont
  {{Huse}}(2015)}]{Nandkishore2015}%
  \BibitemOpen
  \bibfield  {author} {\bibinfo {author} {\bibfnamefont {Rahul}\ \bibnamefont
  {{Nandkishore}}}\ and\ \bibinfo {author} {\bibfnamefont {David~A.}\
  \bibnamefont {{Huse}}},\ }\bibfield  {title} {\enquote {\bibinfo {title}
  {{Many-Body Localization and Thermalization in Quantum Statistical
  Mechanics}},}\ }\href {\doibase 10.1146/annurev-conmatphys-031214-014726}
  {\bibfield  {journal} {\bibinfo  {journal} {Annual Review of Condensed Matter
  Physics}\ }\textbf {\bibinfo {volume} {6}},\ \bibinfo {pages} {15--38}
  (\bibinfo {year} {2015})},\ \Eprint {http://arxiv.org/abs/1404.0686}
  {arXiv:1404.0686 [cond-mat.stat-mech]} \BibitemShut {NoStop}%
\bibitem [{\citenamefont {{Abanin}}\ \emph {et~al.}(2018)\citenamefont
  {{Abanin}}, \citenamefont {{Altman}}, \citenamefont {{Bloch}},\ and\
  \citenamefont {{Serbyn}}}]{2018arXiv180411065A}%
  \BibitemOpen
  \bibfield  {author} {\bibinfo {author} {\bibfnamefont {Dmitry~A.}\
  \bibnamefont {{Abanin}}}, \bibinfo {author} {\bibfnamefont {Ehud}\
  \bibnamefont {{Altman}}}, \bibinfo {author} {\bibfnamefont {Immanuel}\
  \bibnamefont {{Bloch}}}, \ and\ \bibinfo {author} {\bibfnamefont {Maksym}\
  \bibnamefont {{Serbyn}}},\ }\bibfield  {title} {\enquote {\bibinfo {title}
  {{Ergodicity, Entanglement and Many-Body Localization}},}\ }\href@noop {}
  {\bibfield  {journal} {\bibinfo  {journal} {arXiv e-prints}\ ,\ \bibinfo
  {eid} {arXiv:1804.11065}} (\bibinfo {year} {2018})},\ \Eprint
  {http://arxiv.org/abs/1804.11065} {arXiv:1804.11065 [cond-mat.dis-nn]}
  \BibitemShut {NoStop}%
\bibitem [{\citenamefont {{Cao}}\ \emph {et~al.}(2018)\citenamefont {{Cao}},
  \citenamefont {{Tilloy}},\ and\ \citenamefont {{De
  Luca}}}]{cao2018monitoring}%
  \BibitemOpen
  \bibfield  {author} {\bibinfo {author} {\bibfnamefont {Xiangyu}\ \bibnamefont
  {{Cao}}}, \bibinfo {author} {\bibfnamefont {Antoine}\ \bibnamefont
  {{Tilloy}}}, \ and\ \bibinfo {author} {\bibfnamefont {Andrea}\ \bibnamefont
  {{De Luca}}},\ }\bibfield  {title} {\enquote {\bibinfo {title} {{Entanglement
  and transport of a fermion chain under continuous monitoring}},}\ }\href@noop
  {} {\bibfield  {journal} {\bibinfo  {journal} {arXiv e-prints}\ ,\ \bibinfo
  {eid} {arXiv:1804.04638}} (\bibinfo {year} {2018})},\ \Eprint
  {http://arxiv.org/abs/1804.04638} {arXiv:1804.04638 [cond-mat.stat-mech]}
  \BibitemShut {NoStop}%
\bibitem [{\citenamefont {{Misra}}\ and\ \citenamefont
  {{Sudarshan}}(1977)}]{Misra1977zeno}%
  \BibitemOpen
  \bibfield  {author} {\bibinfo {author} {\bibfnamefont {B.}~\bibnamefont
  {{Misra}}}\ and\ \bibinfo {author} {\bibfnamefont {E.~C.~G.}\ \bibnamefont
  {{Sudarshan}}},\ }\bibfield  {title} {\enquote {\bibinfo {title} {{The Zeno's
  paradox in quantum theory}},}\ }\href {\doibase 10.1063/1.523304} {\bibfield
  {journal} {\bibinfo  {journal} {Journal of Mathematical Physics}\ }\textbf
  {\bibinfo {volume} {18}},\ \bibinfo {pages} {756--763} (\bibinfo {year}
  {1977})}\BibitemShut {NoStop}%
\bibitem [{\citenamefont {{Chan}}\ \emph
  {et~al.}(2018{\natexlab{a}})\citenamefont {{Chan}}, \citenamefont
  {{Nandkishore}}, \citenamefont {{Pretko}},\ and\ \citenamefont
  {{Smith}}}]{nandkishore2018hybrid}%
  \BibitemOpen
  \bibfield  {author} {\bibinfo {author} {\bibfnamefont {Amos}\ \bibnamefont
  {{Chan}}}, \bibinfo {author} {\bibfnamefont {Rahul~M.}\ \bibnamefont
  {{Nandkishore}}}, \bibinfo {author} {\bibfnamefont {Michael}\ \bibnamefont
  {{Pretko}}}, \ and\ \bibinfo {author} {\bibfnamefont {Graeme}\ \bibnamefont
  {{Smith}}},\ }\bibfield  {title} {\enquote {\bibinfo {title} {{Weak
  measurements limit entanglement to area law (with possible log
  corrections)}},}\ }\href@noop {} {\bibfield  {journal} {\bibinfo  {journal}
  {arXiv e-prints}\ ,\ \bibinfo {eid} {arXiv:1808.05949}} (\bibinfo {year}
  {2018}{\natexlab{a}})},\ \Eprint {http://arxiv.org/abs/1808.05949}
  {arXiv:1808.05949 [cond-mat.stat-mech]} \BibitemShut {NoStop}%
\bibitem [{\citenamefont {{Skinner}}\ \emph {et~al.}(2018)\citenamefont
  {{Skinner}}, \citenamefont {{Ruhman}},\ and\ \citenamefont
  {{Nahum}}}]{nahum2018hybrid}%
  \BibitemOpen
  \bibfield  {author} {\bibinfo {author} {\bibfnamefont {Brian}\ \bibnamefont
  {{Skinner}}}, \bibinfo {author} {\bibfnamefont {Jonathan}\ \bibnamefont
  {{Ruhman}}}, \ and\ \bibinfo {author} {\bibfnamefont {Adam}\ \bibnamefont
  {{Nahum}}},\ }\bibfield  {title} {\enquote {\bibinfo {title}
  {{Measurement-Induced Phase Transitions in the Dynamics of Entanglement}},}\
  }\href@noop {} {\bibfield  {journal} {\bibinfo  {journal} {arXiv e-prints}\
  ,\ \bibinfo {eid} {arXiv:1808.05953}} (\bibinfo {year} {2018})},\ \Eprint
  {http://arxiv.org/abs/1808.05953} {arXiv:1808.05953 [cond-mat.stat-mech]}
  \BibitemShut {NoStop}%
\bibitem [{\citenamefont {{Li}}\ \emph {et~al.}(2018)\citenamefont {{Li}},
  \citenamefont {{Chen}},\ and\ \citenamefont {{Fisher}}}]{li2018hybrid}%
  \BibitemOpen
  \bibfield  {author} {\bibinfo {author} {\bibfnamefont {Yaodong}\ \bibnamefont
  {{Li}}}, \bibinfo {author} {\bibfnamefont {Xiao}\ \bibnamefont {{Chen}}}, \
  and\ \bibinfo {author} {\bibfnamefont {Matthew P.~A.}\ \bibnamefont
  {{Fisher}}},\ }\bibfield  {title} {\enquote {\bibinfo {title} {{Quantum Zeno
  effect and the many-body entanglement transition}},}\ }\href {\doibase
  10.1103/PhysRevB.98.205136} {\bibfield  {journal} {\bibinfo  {journal}
  {Physical Review B}\ }\textbf {\bibinfo {volume} {98}},\ \bibinfo {eid}
  {205136} (\bibinfo {year} {2018})},\ \Eprint
  {http://arxiv.org/abs/1808.06134} {arXiv:1808.06134 [quant-ph]} \BibitemShut
  {NoStop}%
\bibitem [{\citenamefont {Mehta}(2004)}]{mehta2004matrices}%
  \BibitemOpen
  \bibfield  {author} {\bibinfo {author} {\bibfnamefont {Madan~Lal}\
  \bibnamefont {Mehta}},\ }\href@noop {} {\emph {\bibinfo {title} {Random
  Matrices}}},\ \bibinfo {edition} {3rd}\ ed.\ (\bibinfo {year}
  {2004})\BibitemShut {NoStop}%
\bibitem [{\citenamefont {Forrester}(2010)}]{loggasrandommatrices}%
  \BibitemOpen
  \bibfield  {author} {\bibinfo {author} {\bibfnamefont {P.~J.}\ \bibnamefont
  {Forrester}},\ }\href {http://www.jstor.org/stable/j.ctt7t5vq} {\emph
  {\bibinfo {title} {Log-Gases and Random Matrices (LMS-34)}}}\ (\bibinfo
  {publisher} {Princeton University Press},\ \bibinfo {year}
  {2010})\BibitemShut {NoStop}%
\bibitem [{\citenamefont {{Gottesman}}(1996)}]{gottesman9604hamming}%
  \BibitemOpen
  \bibfield  {author} {\bibinfo {author} {\bibfnamefont {Daniel}\ \bibnamefont
  {{Gottesman}}},\ }\bibfield  {title} {\enquote {\bibinfo {title} {{Class of
  quantum error-correcting codes saturating the quantum Hamming bound}},}\
  }\href {\doibase 10.1103/PhysRevA.54.1862} {\bibfield  {journal} {\bibinfo
  {journal} {Physical Review A}\ }\textbf {\bibinfo {volume} {54}},\ \bibinfo
  {pages} {1862--1868} (\bibinfo {year} {1996})},\ \Eprint
  {http://arxiv.org/abs/quant-ph/9604038} {arXiv:quant-ph/9604038 [quant-ph]}
  \BibitemShut {NoStop}%
\bibitem [{\citenamefont {{Gottesman}}(1998)}]{gottesman9807heisenberg}%
  \BibitemOpen
  \bibfield  {author} {\bibinfo {author} {\bibfnamefont {Daniel}\ \bibnamefont
  {{Gottesman}}},\ }\bibfield  {title} {\enquote {\bibinfo {title} {{The
  Heisenberg Representation of Quantum Computers}},}\ }\href@noop {} {\bibfield
   {journal} {\bibinfo  {journal} {arXiv e-prints}\ ,\ \bibinfo {eid}
  {quant-ph/9807006}} (\bibinfo {year} {1998})},\ \Eprint
  {http://arxiv.org/abs/quant-ph/9807006} {arXiv:quant-ph/9807006 [quant-ph]}
  \BibitemShut {NoStop}%
\bibitem [{\citenamefont {{Aaronson}}\ and\ \citenamefont
  {{Gottesman}}(2004)}]{aaronson0406chp}%
  \BibitemOpen
  \bibfield  {author} {\bibinfo {author} {\bibfnamefont {Scott}\ \bibnamefont
  {{Aaronson}}}\ and\ \bibinfo {author} {\bibfnamefont {Daniel}\ \bibnamefont
  {{Gottesman}}},\ }\bibfield  {title} {\enquote {\bibinfo {title} {{Improved
  simulation of stabilizer circuits}},}\ }\href {\doibase
  10.1103/PhysRevA.70.052328} {\bibfield  {journal} {\bibinfo  {journal}
  {Physical Review A}\ }\textbf {\bibinfo {volume} {70}},\ \bibinfo {eid}
  {052328} (\bibinfo {year} {2004})},\ \Eprint
  {http://arxiv.org/abs/quant-ph/0406196} {arXiv:quant-ph/0406196 [quant-ph]}
  \BibitemShut {NoStop}%
\bibitem [{\citenamefont {{Nielsen}}\ and\ \citenamefont
  {{Chuang}}(2010)}]{nielsen2010qiqc}%
  \BibitemOpen
  \bibfield  {author} {\bibinfo {author} {\bibfnamefont {M.~A.}\ \bibnamefont
  {{Nielsen}}}\ and\ \bibinfo {author} {\bibfnamefont {I.~L.}\ \bibnamefont
  {{Chuang}}},\ }\href@noop {} {\emph {\bibinfo {title} {{Quantum Computation
  and Quantum Information}}}}\ (\bibinfo {year} {2010})\BibitemShut {NoStop}%
\bibitem [{\citenamefont {{Wiseman}}(1996)}]{Wiseman1996}%
  \BibitemOpen
  \bibfield  {author} {\bibinfo {author} {\bibfnamefont {H.~M.}\ \bibnamefont
  {{Wiseman}}},\ }\bibfield  {title} {\enquote {\bibinfo {title} {{Quantum
  trajectories and quantum measurement theory}},}\ }\href {\doibase
  10.1088/1355-5111/8/1/015} {\bibfield  {journal} {\bibinfo  {journal}
  {Quantum and Semiclassical Optics}\ }\textbf {\bibinfo {volume} {8}},\
  \bibinfo {pages} {205--222} (\bibinfo {year} {1996})},\ \Eprint
  {http://arxiv.org/abs/quant-ph/0302080} {arXiv:quant-ph/0302080 [quant-ph]}
  \BibitemShut {NoStop}%
\bibitem [{\citenamefont {Breuer}\ and\ \citenamefont
  {Petruccione}(2002)}]{Breuer2002theory}%
  \BibitemOpen
  \bibfield  {author} {\bibinfo {author} {\bibfnamefont {Heinz-Peter}\
  \bibnamefont {Breuer}}\ and\ \bibinfo {author} {\bibfnamefont {Francesco}\
  \bibnamefont {Petruccione}},\ }\href@noop {} {\emph {\bibinfo {title} {The
  theory of open quantum systems}}}\ (\bibinfo  {publisher} {Oxford University
  Press},\ \bibinfo {year} {2002})\BibitemShut {NoStop}%
\bibitem [{\citenamefont {{Nahum}}\ \emph {et~al.}(2017)\citenamefont
  {{Nahum}}, \citenamefont {{Ruhman}}, \citenamefont {{Vijay}},\ and\
  \citenamefont {{Haah}}}]{nahum2017KPZ}%
  \BibitemOpen
  \bibfield  {author} {\bibinfo {author} {\bibfnamefont {Adam}\ \bibnamefont
  {{Nahum}}}, \bibinfo {author} {\bibfnamefont {Jonathan}\ \bibnamefont
  {{Ruhman}}}, \bibinfo {author} {\bibfnamefont {Sagar}\ \bibnamefont
  {{Vijay}}}, \ and\ \bibinfo {author} {\bibfnamefont {Jeongwan}\ \bibnamefont
  {{Haah}}},\ }\bibfield  {title} {\enquote {\bibinfo {title} {{Quantum
  Entanglement Growth under Random Unitary Dynamics}},}\ }\href {\doibase
  10.1103/PhysRevX.7.031016} {\bibfield  {journal} {\bibinfo  {journal}
  {Physical Review X}\ }\textbf {\bibinfo {volume} {7}},\ \bibinfo {eid}
  {031016} (\bibinfo {year} {2017})},\ \Eprint
  {http://arxiv.org/abs/1608.06950} {arXiv:1608.06950 [cond-mat.stat-mech]}
  \BibitemShut {NoStop}%
\bibitem [{\citenamefont {Hammersley}\ and\ \citenamefont
  {Welsh}(1965)}]{Hammersley1965}%
  \BibitemOpen
  \bibfield  {author} {\bibinfo {author} {\bibfnamefont {J.~M.}\ \bibnamefont
  {Hammersley}}\ and\ \bibinfo {author} {\bibfnamefont {D.~J.~A.}\ \bibnamefont
  {Welsh}},\ }\enquote {\bibinfo {title} {First-passage percolation,
  subadditive processes, stochastic networks, and generalized renewal
  theory},}\ in\ \href {\doibase 10.1007/978-3-642-99884-3_7} {\emph {\bibinfo
  {booktitle} {Bernoulli 1713 Bayes 1763 Laplace 1813: Anniversary Volume}}},\
  \bibinfo {editor} {edited by\ \bibinfo {editor} {\bibfnamefont {Jerzy}\
  \bibnamefont {Neyman}}\ and\ \bibinfo {editor} {\bibfnamefont {Lucien~M.}\
  \bibnamefont {Le~Cam}}}\ (\bibinfo  {publisher} {Springer Berlin
  Heidelberg},\ \bibinfo {address} {Berlin, Heidelberg},\ \bibinfo {year}
  {1965})\ pp.\ \bibinfo {pages} {61--110}\BibitemShut {NoStop}%
\bibitem [{\citenamefont {Chayes}\ \emph {et~al.}(1986)\citenamefont {Chayes},
  \citenamefont {Chayes},\ and\ \citenamefont
  {Durrett}}]{Chayes1986firstpassage}%
  \BibitemOpen
  \bibfield  {author} {\bibinfo {author} {\bibfnamefont {J.~T.}\ \bibnamefont
  {Chayes}}, \bibinfo {author} {\bibfnamefont {L.}~\bibnamefont {Chayes}}, \
  and\ \bibinfo {author} {\bibfnamefont {R.}~\bibnamefont {Durrett}},\
  }\bibfield  {title} {\enquote {\bibinfo {title} {Critical behavior of the
  two-dimensional first passage time},}\ }\href {\doibase 10.1007/BF01020583}
  {\bibfield  {journal} {\bibinfo  {journal} {Journal of Statistical Physics}\
  }\textbf {\bibinfo {volume} {45}},\ \bibinfo {pages} {933--951} (\bibinfo
  {year} {1986})}\BibitemShut {NoStop}%
\bibitem [{\citenamefont {Kesten}(1986)}]{kesten1986aspects}%
  \BibitemOpen
  \bibfield  {author} {\bibinfo {author} {\bibfnamefont {Harry}\ \bibnamefont
  {Kesten}},\ }\bibfield  {title} {\enquote {\bibinfo {title} {Aspects of first
  passage percolation},}\ }in\ \href@noop {} {\emph {\bibinfo {booktitle}
  {{\'E}cole d'{\'e}t{\'e} de probabilit{\'e}s de Saint Flour XIV-1984}}}\
  (\bibinfo  {publisher} {Springer},\ \bibinfo {year} {1986})\ pp.\ \bibinfo
  {pages} {125--264}\BibitemShut {NoStop}%
\bibitem [{\citenamefont {Kesten}(1987)}]{kesten1987firstpassage}%
  \BibitemOpen
  \bibfield  {author} {\bibinfo {author} {\bibfnamefont {Harry}\ \bibnamefont
  {Kesten}},\ }\bibfield  {title} {\enquote {\bibinfo {title} {Percolation
  theory and first-passage percolation},}\ }\href {\doibase
  10.1214/aop/1176991975} {\bibfield  {journal} {\bibinfo  {journal} {Ann.
  Probab.}\ }\textbf {\bibinfo {volume} {15}},\ \bibinfo {pages} {1231--1271}
  (\bibinfo {year} {1987})}\BibitemShut {NoStop}%
\bibitem [{\citenamefont {Li}\ and\ \citenamefont {Fisher}()}]{LiMPAF2018Cat}%
  \BibitemOpen
  \bibfield  {author} {\bibinfo {author} {\bibfnamefont {Yaodong}\ \bibnamefont
  {Li}}\ and\ \bibinfo {author} {\bibfnamefont {Matthew P.~A.}\ \bibnamefont
  {Fisher}},\ }\href@noop {} {\ }\Eprint {http://arxiv.org/abs/unpublished}
  {unpublished} \BibitemShut {NoStop}%
\bibitem [{\citenamefont {{Klappenecker}}\ and\ \citenamefont
  {{Roetteler}}(2000)}]{Klappenecker2002stabilizer}%
  \BibitemOpen
  \bibfield  {author} {\bibinfo {author} {\bibfnamefont {Andreas}\ \bibnamefont
  {{Klappenecker}}}\ and\ \bibinfo {author} {\bibfnamefont {Martin}\
  \bibnamefont {{Roetteler}}},\ }\bibfield  {title} {\enquote {\bibinfo {title}
  {{Beyond Stabilizer Codes II: Clifford Codes}},}\ }\href@noop {} {\bibfield
  {journal} {\bibinfo  {journal} {arXiv e-prints}\ ,\ \bibinfo {eid}
  {quant-ph/0010076}} (\bibinfo {year} {2000})},\ \Eprint
  {http://arxiv.org/abs/quant-ph/0010076} {arXiv:quant-ph/0010076 [quant-ph]}
  \BibitemShut {NoStop}%
\bibitem [{\citenamefont {{Linden}}\ \emph {et~al.}(2013)\citenamefont
  {{Linden}}, \citenamefont {{Mat{\'u}{\v{s}}}}, \citenamefont {{Ruskai}},\
  and\ \citenamefont {{Winter}}}]{Linden2013stabilizer}%
  \BibitemOpen
  \bibfield  {author} {\bibinfo {author} {\bibfnamefont {Noah}\ \bibnamefont
  {{Linden}}}, \bibinfo {author} {\bibfnamefont {Franti{\v{s}}ek}\ \bibnamefont
  {{Mat{\'u}{\v{s}}}}}, \bibinfo {author} {\bibfnamefont {Mary~Beth}\
  \bibnamefont {{Ruskai}}}, \ and\ \bibinfo {author} {\bibfnamefont {Andreas}\
  \bibnamefont {{Winter}}},\ }\bibfield  {title} {\enquote {\bibinfo {title}
  {{The Quantum Entropy Cone of Stabiliser States}},}\ }\href@noop {}
  {\bibfield  {journal} {\bibinfo  {journal} {arXiv e-prints}\ ,\ \bibinfo
  {eid} {arXiv:1302.5453}} (\bibinfo {year} {2013})},\ \Eprint
  {http://arxiv.org/abs/1302.5453} {arXiv:1302.5453 [quant-ph]} \BibitemShut
  {NoStop}%
\bibitem [{\citenamefont {{DiVincenzo}}\ \emph {et~al.}(2001)\citenamefont
  {{DiVincenzo}}, \citenamefont {{Leung}},\ and\ \citenamefont
  {{Terhal}}}]{DiVincenzo2002hiding}%
  \BibitemOpen
  \bibfield  {author} {\bibinfo {author} {\bibfnamefont {David~P.}\
  \bibnamefont {{DiVincenzo}}}, \bibinfo {author} {\bibfnamefont {Debbie~W.}\
  \bibnamefont {{Leung}}}, \ and\ \bibinfo {author} {\bibfnamefont
  {Barbara~M.}\ \bibnamefont {{Terhal}}},\ }\bibfield  {title} {\enquote
  {\bibinfo {title} {{Quantum Data Hiding}},}\ }\href@noop {} {\bibfield
  {journal} {\bibinfo  {journal} {arXiv e-prints}\ ,\ \bibinfo {eid}
  {quant-ph/0103098}} (\bibinfo {year} {2001})},\ \Eprint
  {http://arxiv.org/abs/quant-ph/0103098} {arXiv:quant-ph/0103098 [quant-ph]}
  \BibitemShut {NoStop}%
\bibitem [{\citenamefont {{Fattal}}\ \emph {et~al.}(2004)\citenamefont
  {{Fattal}}, \citenamefont {{Cubitt}}, \citenamefont {{Yamamoto}},
  \citenamefont {{Bravyi}},\ and\ \citenamefont
  {{Chuang}}}]{Fattal2004stabilizer}%
  \BibitemOpen
  \bibfield  {author} {\bibinfo {author} {\bibfnamefont {David}\ \bibnamefont
  {{Fattal}}}, \bibinfo {author} {\bibfnamefont {Toby~S.}\ \bibnamefont
  {{Cubitt}}}, \bibinfo {author} {\bibfnamefont {Yoshihisa}\ \bibnamefont
  {{Yamamoto}}}, \bibinfo {author} {\bibfnamefont {Sergey}\ \bibnamefont
  {{Bravyi}}}, \ and\ \bibinfo {author} {\bibfnamefont {Isaac~L.}\ \bibnamefont
  {{Chuang}}},\ }\bibfield  {title} {\enquote {\bibinfo {title} {{Entanglement
  in the stabilizer formalism}},}\ }\href@noop {} {\bibfield  {journal}
  {\bibinfo  {journal} {arXiv e-prints}\ ,\ \bibinfo {eid} {quant-ph/0406168}}
  (\bibinfo {year} {2004})},\ \Eprint {http://arxiv.org/abs/quant-ph/0406168}
  {arXiv:quant-ph/0406168 [quant-ph]} \BibitemShut {NoStop}%
\bibitem [{\citenamefont {{Hamma}}\ \emph
  {et~al.}(2005{\natexlab{a}})\citenamefont {{Hamma}}, \citenamefont
  {{Ionicioiu}},\ and\ \citenamefont {{Zanardi}}}]{hamma2005entanglement1}%
  \BibitemOpen
  \bibfield  {author} {\bibinfo {author} {\bibfnamefont {Alioscia}\
  \bibnamefont {{Hamma}}}, \bibinfo {author} {\bibfnamefont {Radu}\
  \bibnamefont {{Ionicioiu}}}, \ and\ \bibinfo {author} {\bibfnamefont {Paolo}\
  \bibnamefont {{Zanardi}}},\ }\bibfield  {title} {\enquote {\bibinfo {title}
  {{Bipartite entanglement and entropic boundary law in lattice spin
  systems}},}\ }\href {\doibase 10.1103/PhysRevA.71.022315} {\bibfield
  {journal} {\bibinfo  {journal} {Physical Review A}\ }\textbf {\bibinfo
  {volume} {71}},\ \bibinfo {eid} {022315} (\bibinfo {year}
  {2005}{\natexlab{a}})},\ \Eprint {http://arxiv.org/abs/quant-ph/0409073}
  {arXiv:quant-ph/0409073 [quant-ph]} \BibitemShut {NoStop}%
\bibitem [{\citenamefont {{Hamma}}\ \emph
  {et~al.}(2005{\natexlab{b}})\citenamefont {{Hamma}}, \citenamefont
  {{Ionicioiu}},\ and\ \citenamefont {{Zanardi}}}]{hamma2005entanglement2}%
  \BibitemOpen
  \bibfield  {author} {\bibinfo {author} {\bibfnamefont {Alioscia}\
  \bibnamefont {{Hamma}}}, \bibinfo {author} {\bibfnamefont {Radu}\
  \bibnamefont {{Ionicioiu}}}, \ and\ \bibinfo {author} {\bibfnamefont {Paolo}\
  \bibnamefont {{Zanardi}}},\ }\bibfield  {title} {\enquote {\bibinfo {title}
  {{Ground state entanglement and geometric entropy in the Kitaev model [rapid
  communication]}},}\ }\href {\doibase 10.1016/j.physleta.2005.01.060}
  {\bibfield  {journal} {\bibinfo  {journal} {Physics Letters A}\ }\textbf
  {\bibinfo {volume} {337}},\ \bibinfo {pages} {22--28} (\bibinfo {year}
  {2005}{\natexlab{b}})},\ \Eprint {http://arxiv.org/abs/quant-ph/0406202}
  {arXiv:quant-ph/0406202 [quant-ph]} \BibitemShut {NoStop}%
\bibitem [{\citenamefont {{Nahum}}\ \emph {et~al.}(2018)\citenamefont
  {{Nahum}}, \citenamefont {{Vijay}},\ and\ \citenamefont
  {{Haah}}}]{nahum2018operator}%
  \BibitemOpen
  \bibfield  {author} {\bibinfo {author} {\bibfnamefont {Adam}\ \bibnamefont
  {{Nahum}}}, \bibinfo {author} {\bibfnamefont {Sagar}\ \bibnamefont
  {{Vijay}}}, \ and\ \bibinfo {author} {\bibfnamefont {Jeongwan}\ \bibnamefont
  {{Haah}}},\ }\bibfield  {title} {\enquote {\bibinfo {title} {{Operator
  Spreading in Random Unitary Circuits}},}\ }\href {\doibase
  10.1103/PhysRevX.8.021014} {\bibfield  {journal} {\bibinfo  {journal}
  {Physical Review X}\ }\textbf {\bibinfo {volume} {8}},\ \bibinfo {eid}
  {021014} (\bibinfo {year} {2018})},\ \Eprint
  {http://arxiv.org/abs/1705.08975} {arXiv:1705.08975 [cond-mat.str-el]}
  \BibitemShut {NoStop}%
\bibitem [{\citenamefont {{Chandran}}\ and\ \citenamefont
  {{Laumann}}(2015)}]{chandran1501semiclassical}%
  \BibitemOpen
  \bibfield  {author} {\bibinfo {author} {\bibfnamefont {Anushya}\ \bibnamefont
  {{Chandran}}}\ and\ \bibinfo {author} {\bibfnamefont {C.~R.}\ \bibnamefont
  {{Laumann}}},\ }\bibfield  {title} {\enquote {\bibinfo {title}
  {{Semiclassical limit for the many-body localization transition}},}\ }\href
  {\doibase 10.1103/PhysRevB.92.024301} {\bibfield  {journal} {\bibinfo
  {journal} {Physical Review B}\ }\textbf {\bibinfo {volume} {92}},\ \bibinfo
  {eid} {024301} (\bibinfo {year} {2015})},\ \Eprint
  {http://arxiv.org/abs/1501.01971} {arXiv:1501.01971 [cond-mat.dis-nn]}
  \BibitemShut {NoStop}%
\bibitem [{Note1()}]{Note1}%
  \BibitemOpen
  \bibinfo {note} {We also notice a small hump at $\ell \approx L$. This part
  of the distribution is a boundary effect due to the periodic boundary
  condition, and the height of the hump decays as $1/L$ as we go to the
  thermodynamic limit. Moreover, from Eq.~\protect \textup {\hbox
  {\mathsurround \z@ \protect \normalfont (\ignorespaces \ref
  {eq:EE_span}\unskip \@@italiccorr )}}, these long stabilizers of length $\sim
  L$ barely contribute to the entanglement entropy. Thus we ignore this
  unimportant hump.}\BibitemShut {Stop}%
\bibitem [{\citenamefont {{Vasseur}}\ \emph {et~al.}(2018)\citenamefont
  {{Vasseur}}, \citenamefont {{Potter}}, \citenamefont {{You}},\ and\
  \citenamefont {{Ludwig}}}]{vasseur2018rtn}%
  \BibitemOpen
  \bibfield  {author} {\bibinfo {author} {\bibfnamefont {Romain}\ \bibnamefont
  {{Vasseur}}}, \bibinfo {author} {\bibfnamefont {Andrew~C.}\ \bibnamefont
  {{Potter}}}, \bibinfo {author} {\bibfnamefont {Yi-Zhuang}\ \bibnamefont
  {{You}}}, \ and\ \bibinfo {author} {\bibfnamefont {Andreas W.~W.}\
  \bibnamefont {{Ludwig}}},\ }\bibfield  {title} {\enquote {\bibinfo {title}
  {{Entanglement Transitions from Holographic Random Tensor Networks}},}\
  }\href@noop {} {\bibfield  {journal} {\bibinfo  {journal} {arXiv e-prints}\
  ,\ \bibinfo {eid} {arXiv:1807.07082}} (\bibinfo {year} {2018})},\ \Eprint
  {http://arxiv.org/abs/1807.07082} {arXiv:1807.07082 [cond-mat.stat-mech]}
  \BibitemShut {NoStop}%
\bibitem [{\citenamefont {Buff}\ \emph {et~al.}(1965)\citenamefont {Buff},
  \citenamefont {Lovett},\ and\ \citenamefont
  {Stillinger}}]{buff1965capillary}%
  \BibitemOpen
  \bibfield  {author} {\bibinfo {author} {\bibfnamefont {F.~P.}\ \bibnamefont
  {Buff}}, \bibinfo {author} {\bibfnamefont {R.~A.}\ \bibnamefont {Lovett}}, \
  and\ \bibinfo {author} {\bibfnamefont {F.~H.}\ \bibnamefont {Stillinger}},\
  }\bibfield  {title} {\enquote {\bibinfo {title} {Interfacial density profile
  for fluids in the critical region},}\ }\href {\doibase
  10.1103/PhysRevLett.15.621} {\bibfield  {journal} {\bibinfo  {journal} {Phys.
  Rev. Lett.}\ }\textbf {\bibinfo {volume} {15}},\ \bibinfo {pages} {621--623}
  (\bibinfo {year} {1965})}\BibitemShut {NoStop}%
\bibitem [{\citenamefont {Weeks}(1977)}]{weeks1977capillary}%
  \BibitemOpen
  \bibfield  {author} {\bibinfo {author} {\bibfnamefont {John~D.}\ \bibnamefont
  {Weeks}},\ }\bibfield  {title} {\enquote {\bibinfo {title} {Structure and
  thermodynamics of the liquid–vapor interface},}\ }\href {\doibase
  10.1063/1.435276} {\bibfield  {journal} {\bibinfo  {journal} {The Journal of
  Chemical Physics}\ }\textbf {\bibinfo {volume} {67}},\ \bibinfo {pages}
  {3106--3121} (\bibinfo {year} {1977})},\ \Eprint
  {http://arxiv.org/abs/https://doi.org/10.1063/1.435276}
  {https://doi.org/10.1063/1.435276} \BibitemShut {NoStop}%
\bibitem [{\citenamefont {{Wolf}}\ \emph {et~al.}(2008)\citenamefont {{Wolf}},
  \citenamefont {{Verstraete}}, \citenamefont {{Hastings}},\ and\ \citenamefont
  {{Cirac}}}]{WVHC}%
  \BibitemOpen
  \bibfield  {author} {\bibinfo {author} {\bibfnamefont {Michael~M.}\
  \bibnamefont {{Wolf}}}, \bibinfo {author} {\bibfnamefont {Frank}\
  \bibnamefont {{Verstraete}}}, \bibinfo {author} {\bibfnamefont {Matthew~B.}\
  \bibnamefont {{Hastings}}}, \ and\ \bibinfo {author} {\bibfnamefont
  {J.~Ignacio}\ \bibnamefont {{Cirac}}},\ }\bibfield  {title} {\enquote
  {\bibinfo {title} {{Area Laws in Quantum Systems: Mutual Information and
  Correlations}},}\ }\href {\doibase 10.1103/PhysRevLett.100.070502} {\bibfield
   {journal} {\bibinfo  {journal} {\prl}\ }\textbf {\bibinfo {volume} {100}},\
  \bibinfo {eid} {070502} (\bibinfo {year} {2008})},\ \Eprint
  {http://arxiv.org/abs/0704.3906} {arXiv:0704.3906 [quant-ph]} \BibitemShut
  {NoStop}%
\bibitem [{\citenamefont {{Calabrese}}\ and\ \citenamefont
  {{Cardy}}(2009)}]{Calabrese:2009qy}%
  \BibitemOpen
  \bibfield  {author} {\bibinfo {author} {\bibfnamefont {Pasquale}\
  \bibnamefont {{Calabrese}}}\ and\ \bibinfo {author} {\bibfnamefont {John}\
  \bibnamefont {{Cardy}}},\ }\bibfield  {title} {\enquote {\bibinfo {title}
  {{Entanglement entropy and conformal field theory}},}\ }\href {\doibase
  10.1088/1751-8113/42/50/504005} {\bibfield  {journal} {\bibinfo  {journal}
  {Journal of Physics A Mathematical General}\ }\textbf {\bibinfo {volume}
  {42}},\ \bibinfo {eid} {504005} (\bibinfo {year} {2009})},\ \Eprint
  {http://arxiv.org/abs/0905.4013} {arXiv:0905.4013 [cond-mat.stat-mech]}
  \BibitemShut {NoStop}%
\bibitem [{\citenamefont {Di~Francesco}\ \emph {et~al.}(1997)\citenamefont
  {Di~Francesco}, \citenamefont {Mathieu},\ and\ \citenamefont
  {Senechal}}]{DiFrancesco:1997nk}%
  \BibitemOpen
  \bibfield  {author} {\bibinfo {author} {\bibfnamefont {P.}~\bibnamefont
  {Di~Francesco}}, \bibinfo {author} {\bibfnamefont {P.}~\bibnamefont
  {Mathieu}}, \ and\ \bibinfo {author} {\bibfnamefont {D.}~\bibnamefont
  {Senechal}},\ }\href {\doibase 10.1007/978-1-4612-2256-9} {\emph {\bibinfo
  {title} {{Conformal Field Theory}}}},\ Graduate Texts in Contemporary
  Physics\ (\bibinfo  {publisher} {Springer-Verlag},\ \bibinfo {address} {New
  York},\ \bibinfo {year} {1997})\BibitemShut {NoStop}%
\bibitem [{\citenamefont {{Kim}}\ \emph {et~al.}(2014)\citenamefont {{Kim}},
  \citenamefont {{Ikeda}},\ and\ \citenamefont {{Huse}}}]{Kim2014}%
  \BibitemOpen
  \bibfield  {author} {\bibinfo {author} {\bibfnamefont {Hyungwon}\
  \bibnamefont {{Kim}}}, \bibinfo {author} {\bibfnamefont {Tatsuhiko~N.}\
  \bibnamefont {{Ikeda}}}, \ and\ \bibinfo {author} {\bibfnamefont {David~A.}\
  \bibnamefont {{Huse}}},\ }\bibfield  {title} {\enquote {\bibinfo {title}
  {{Testing whether all eigenstates obey the eigenstate thermalization
  hypothesis}},}\ }\href {\doibase 10.1103/PhysRevE.90.052105} {\bibfield
  {journal} {\bibinfo  {journal} {\pre}\ }\textbf {\bibinfo {volume} {90}},\
  \bibinfo {eid} {052105} (\bibinfo {year} {2014})},\ \Eprint
  {http://arxiv.org/abs/1408.0535} {arXiv:1408.0535 [cond-mat.stat-mech]}
  \BibitemShut {NoStop}%
\bibitem [{\citenamefont {{Chan}}\ \emph
  {et~al.}(2018{\natexlab{b}})\citenamefont {{Chan}}, \citenamefont {{De
  Luca}},\ and\ \citenamefont {{Chalker}}}]{chalker2017analytic}%
  \BibitemOpen
  \bibfield  {author} {\bibinfo {author} {\bibfnamefont {Amos}\ \bibnamefont
  {{Chan}}}, \bibinfo {author} {\bibfnamefont {Andrea}\ \bibnamefont {{De
  Luca}}}, \ and\ \bibinfo {author} {\bibfnamefont {J.~T.}\ \bibnamefont
  {{Chalker}}},\ }\bibfield  {title} {\enquote {\bibinfo {title} {{Solution of
  a Minimal Model for Many-Body Quantum Chaos}},}\ }\href {\doibase
  10.1103/PhysRevX.8.041019} {\bibfield  {journal} {\bibinfo  {journal}
  {Physical Review X}\ }\textbf {\bibinfo {volume} {8}},\ \bibinfo {eid}
  {041019} (\bibinfo {year} {2018}{\natexlab{b}})},\ \Eprint
  {http://arxiv.org/abs/1712.06836} {arXiv:1712.06836 [cond-mat.stat-mech]}
  \BibitemShut {NoStop}%
\bibitem [{\citenamefont {{Chan}}\ \emph
  {et~al.}(2018{\natexlab{c}})\citenamefont {{Chan}}, \citenamefont {{De
  Luca}},\ and\ \citenamefont {{Chalker}}}]{chalker2018analytic}%
  \BibitemOpen
  \bibfield  {author} {\bibinfo {author} {\bibfnamefont {Amos}\ \bibnamefont
  {{Chan}}}, \bibinfo {author} {\bibfnamefont {Andrea}\ \bibnamefont {{De
  Luca}}}, \ and\ \bibinfo {author} {\bibfnamefont {J.~T.}\ \bibnamefont
  {{Chalker}}},\ }\bibfield  {title} {\enquote {\bibinfo {title} {{Spectral
  Statistics in Spatially Extended Chaotic Quantum Many-Body Systems}},}\
  }\href {\doibase 10.1103/PhysRevLett.121.060601} {\bibfield  {journal}
  {\bibinfo  {journal} {\prl}\ }\textbf {\bibinfo {volume} {121}},\ \bibinfo
  {eid} {060601} (\bibinfo {year} {2018}{\natexlab{c}})},\ \Eprint
  {http://arxiv.org/abs/1803.03841} {arXiv:1803.03841 [cond-mat.stat-mech]}
  \BibitemShut {NoStop}%
\bibitem [{\citenamefont {{Kos}}\ \emph {et~al.}(2018)\citenamefont {{Kos}},
  \citenamefont {{Ljubotina}},\ and\ \citenamefont
  {{Prosen}}}]{prosen2017analytic}%
  \BibitemOpen
  \bibfield  {author} {\bibinfo {author} {\bibfnamefont {Pavel}\ \bibnamefont
  {{Kos}}}, \bibinfo {author} {\bibfnamefont {Marko}\ \bibnamefont
  {{Ljubotina}}}, \ and\ \bibinfo {author} {\bibfnamefont {Toma{\v{z}}}\
  \bibnamefont {{Prosen}}},\ }\bibfield  {title} {\enquote {\bibinfo {title}
  {{Many-Body Quantum Chaos: Analytic Connection to Random Matrix Theory}},}\
  }\href {\doibase 10.1103/PhysRevX.8.021062} {\bibfield  {journal} {\bibinfo
  {journal} {Physical Review X}\ }\textbf {\bibinfo {volume} {8}},\ \bibinfo
  {eid} {021062} (\bibinfo {year} {2018})},\ \Eprint
  {http://arxiv.org/abs/1712.02665} {arXiv:1712.02665 [nlin.CD]} \BibitemShut
  {NoStop}%
\bibitem [{\citenamefont {{Bertini}}\ \emph
  {et~al.}(2018{\natexlab{a}})\citenamefont {{Bertini}}, \citenamefont
  {{Kos}},\ and\ \citenamefont {{Prosen}}}]{prosen2018analytic}%
  \BibitemOpen
  \bibfield  {author} {\bibinfo {author} {\bibfnamefont {Bruno}\ \bibnamefont
  {{Bertini}}}, \bibinfo {author} {\bibfnamefont {Pavel}\ \bibnamefont
  {{Kos}}}, \ and\ \bibinfo {author} {\bibfnamefont {Tomaz}\ \bibnamefont
  {{Prosen}}},\ }\bibfield  {title} {\enquote {\bibinfo {title} {{Exact
  Spectral Form Factor in a Minimal Model of Many-Body Quantum Chaos}},}\
  }\href@noop {} {\bibfield  {journal} {\bibinfo  {journal} {ArXiv e-prints}\
  ,\ \bibinfo {eid} {arXiv:1805.00931}} (\bibinfo {year}
  {2018}{\natexlab{a}})},\ \Eprint {http://arxiv.org/abs/1805.00931}
  {arXiv:1805.00931 [nlin.CD]} \BibitemShut {NoStop}%
\bibitem [{\citenamefont {{Zhang}}\ \emph {et~al.}(2015)\citenamefont
  {{Zhang}}, \citenamefont {{Kim}},\ and\ \citenamefont
  {{Huse}}}]{ZhangKimHuse_2015}%
  \BibitemOpen
  \bibfield  {author} {\bibinfo {author} {\bibfnamefont {Liangsheng}\
  \bibnamefont {{Zhang}}}, \bibinfo {author} {\bibfnamefont {Hyungwon}\
  \bibnamefont {{Kim}}}, \ and\ \bibinfo {author} {\bibfnamefont {David~A.}\
  \bibnamefont {{Huse}}},\ }\bibfield  {title} {\enquote {\bibinfo {title}
  {{Thermalization of entanglement}},}\ }\href {\doibase
  10.1103/PhysRevE.91.062128} {\bibfield  {journal} {\bibinfo  {journal}
  {\pre}\ }\textbf {\bibinfo {volume} {91}},\ \bibinfo {eid} {062128} (\bibinfo
  {year} {2015})},\ \Eprint {http://arxiv.org/abs/1501.01315} {arXiv:1501.01315
  [cond-mat.stat-mech]} \BibitemShut {NoStop}%
\bibitem [{\citenamefont {{Bertini}}\ \emph
  {et~al.}(2018{\natexlab{b}})\citenamefont {{Bertini}}, \citenamefont
  {{Kos}},\ and\ \citenamefont {{Prosen}}}]{Bertini2018}%
  \BibitemOpen
  \bibfield  {author} {\bibinfo {author} {\bibfnamefont {Bruno}\ \bibnamefont
  {{Bertini}}}, \bibinfo {author} {\bibfnamefont {Pavel}\ \bibnamefont
  {{Kos}}}, \ and\ \bibinfo {author} {\bibfnamefont {Tomaz}\ \bibnamefont
  {{Prosen}}},\ }\bibfield  {title} {\enquote {\bibinfo {title} {{Entanglement
  spreading in a minimal model of maximal many-body quantum chaos}},}\
  }\href@noop {} {\bibfield  {journal} {\bibinfo  {journal} {arXiv e-prints}\
  ,\ \bibinfo {eid} {arXiv:1812.05090}} (\bibinfo {year}
  {2018}{\natexlab{b}})},\ \Eprint {http://arxiv.org/abs/1812.05090}
  {arXiv:1812.05090 [cond-mat.stat-mech]} \BibitemShut {NoStop}%
\bibitem [{\citenamefont {{Chen}}\ \emph {et~al.}(2017)\citenamefont {{Chen}},
  \citenamefont {{Zhou}}, \citenamefont {{Huse}},\ and\ \citenamefont
  {{Fradkin}}}]{Chen2016}%
  \BibitemOpen
  \bibfield  {author} {\bibinfo {author} {\bibfnamefont {Xiao}\ \bibnamefont
  {{Chen}}}, \bibinfo {author} {\bibfnamefont {Tianci}\ \bibnamefont {{Zhou}}},
  \bibinfo {author} {\bibfnamefont {David~A.}\ \bibnamefont {{Huse}}}, \ and\
  \bibinfo {author} {\bibfnamefont {Eduardo}\ \bibnamefont {{Fradkin}}},\
  }\bibfield  {title} {\enquote {\bibinfo {title} {{Out-of-time-order
  correlations in many-body localized and thermal phases}},}\ }\href {\doibase
  10.1002/andp.201600332} {\bibfield  {journal} {\bibinfo  {journal} {Annalen
  der Physik}\ }\textbf {\bibinfo {volume} {529}},\ \bibinfo {pages} {1600332}
  (\bibinfo {year} {2017})},\ \Eprint {http://arxiv.org/abs/1610.00220}
  {arXiv:1610.00220 [cond-mat.str-el]} \BibitemShut {NoStop}%
\bibitem [{\citenamefont {{Hosur}}\ \emph {et~al.}(2016)\citenamefont
  {{Hosur}}, \citenamefont {{Qi}}, \citenamefont {{Roberts}},\ and\
  \citenamefont {{Yoshida}}}]{hosur:2015ylk}%
  \BibitemOpen
  \bibfield  {author} {\bibinfo {author} {\bibfnamefont {Pavan}\ \bibnamefont
  {{Hosur}}}, \bibinfo {author} {\bibfnamefont {Xiao-Liang}\ \bibnamefont
  {{Qi}}}, \bibinfo {author} {\bibfnamefont {Daniel~A.}\ \bibnamefont
  {{Roberts}}}, \ and\ \bibinfo {author} {\bibfnamefont {Beni}\ \bibnamefont
  {{Yoshida}}},\ }\bibfield  {title} {\enquote {\bibinfo {title} {{Chaos in
  quantum channels}},}\ }\href {\doibase 10.1007/JHEP02(2016)004} {\bibfield
  {journal} {\bibinfo  {journal} {Journal of High Energy Physics}\ }\textbf
  {\bibinfo {volume} {2016}},\ \bibinfo {eid} {4} (\bibinfo {year} {2016})},\
  \Eprint {http://arxiv.org/abs/1511.04021} {arXiv:1511.04021 [hep-th]}
  \BibitemShut {NoStop}%
\bibitem [{\citenamefont {{Mezzadri}}(2006)}]{mezzadri2006haar}%
  \BibitemOpen
  \bibfield  {author} {\bibinfo {author} {\bibfnamefont {F.}~\bibnamefont
  {{Mezzadri}}},\ }\bibfield  {title} {\enquote {\bibinfo {title} {{How to
  generate random matrices from the classical compact groups}},}\ }\href@noop
  {} {\bibfield  {journal} {\bibinfo  {journal} {ArXiv Mathematical Physics
  e-prints}\ } (\bibinfo {year} {2006})},\ \Eprint
  {http://arxiv.org/abs/math-ph/0609050} {math-ph/0609050} \BibitemShut
  {NoStop}%
\bibitem [{\citenamefont {{von Keyserlingk}}\ \emph {et~al.}(2018)\citenamefont
  {{von Keyserlingk}}, \citenamefont {{Rakovszky}}, \citenamefont
  {{Pollmann}},\ and\ \citenamefont {{Sondhi}}}]{keyserlingk2018operator}%
  \BibitemOpen
  \bibfield  {author} {\bibinfo {author} {\bibfnamefont {C.~W.}\ \bibnamefont
  {{von Keyserlingk}}}, \bibinfo {author} {\bibfnamefont {Tibor}\ \bibnamefont
  {{Rakovszky}}}, \bibinfo {author} {\bibfnamefont {Frank}\ \bibnamefont
  {{Pollmann}}}, \ and\ \bibinfo {author} {\bibfnamefont {S.~L.}\ \bibnamefont
  {{Sondhi}}},\ }\bibfield  {title} {\enquote {\bibinfo {title} {{Operator
  Hydrodynamics, OTOCs, and Entanglement Growth in Systems without Conservation
  Laws}},}\ }\href {\doibase 10.1103/PhysRevX.8.021013} {\bibfield  {journal}
  {\bibinfo  {journal} {Physical Review X}\ }\textbf {\bibinfo {volume} {8}},\
  \bibinfo {eid} {021013} (\bibinfo {year} {2018})},\ \Eprint
  {http://arxiv.org/abs/1705.08910} {arXiv:1705.08910 [cond-mat.str-el]}
  \BibitemShut {NoStop}%
\bibitem [{Note2()}]{Note2}%
  \BibitemOpen
  \bibinfo {note} {Notice that for R{\'e}nyi indices greater than $1$, there is
  no subadditivity of entanglement, and the mutual information is not
  necessarily non-negative, although in our data the mean values are never
  negative.}\BibitemShut {Stop}%
\bibitem [{Note3()}]{Note3}%
  \BibitemOpen
  \bibinfo {note} {{We have explicitly excluded the zeroth R\'{e}nyi entropy in
  our conjecture, which is quite singular and appear to be in a different
  universality class; see Sec.~\ref {sec:discuss_nature_of_transition} for more
  discussions.}}\BibitemShut {Stop}%
\bibitem [{\citenamefont {{Cardy}}(2000)}]{Cardy2000}%
  \BibitemOpen
  \bibfield  {author} {\bibinfo {author} {\bibfnamefont {John}\ \bibnamefont
  {{Cardy}}},\ }\bibfield  {title} {\enquote {\bibinfo {title} {{Linking
  Numbers for Self-Avoiding Loops and Percolation: Application to the Spin
  Quantum Hall Transition}},}\ }\href {\doibase 10.1103/PhysRevLett.84.3507}
  {\bibfield  {journal} {\bibinfo  {journal} {\prl}\ }\textbf {\bibinfo
  {volume} {84}},\ \bibinfo {pages} {3507--3510} (\bibinfo {year} {2000})},\
  \Eprint {http://arxiv.org/abs/cond-mat/9911457} {arXiv:cond-mat/9911457
  [cond-mat.stat-mech]} \BibitemShut {NoStop}%
\bibitem [{\citenamefont {{Cardy}}(2001)}]{cardy0103percolation}%
  \BibitemOpen
  \bibfield  {author} {\bibinfo {author} {\bibfnamefont {John}\ \bibnamefont
  {{Cardy}}},\ }\bibfield  {title} {\enquote {\bibinfo {title} {{Conformal
  Invariance and Percolation}},}\ }\href@noop {} {\bibfield  {journal}
  {\bibinfo  {journal} {arXiv e-prints}\ ,\ \bibinfo {eid} {math-ph/0103018}}
  (\bibinfo {year} {2001})},\ \Eprint {http://arxiv.org/abs/math-ph/0103018}
  {arXiv:math-ph/0103018 [math-ph]} \BibitemShut {NoStop}%
\bibitem [{\citenamefont {{Jiang}}\ and\ \citenamefont
  {{Yao}}(2016)}]{yao1612firstpassage}%
  \BibitemOpen
  \bibfield  {author} {\bibinfo {author} {\bibfnamefont {Jianping}\
  \bibnamefont {{Jiang}}}\ and\ \bibinfo {author} {\bibfnamefont {Chang-Long}\
  \bibnamefont {{Yao}}},\ }\bibfield  {title} {\enquote {\bibinfo {title}
  {{Critical first-passage percolation starting on the boundary}},}\
  }\href@noop {} {\bibfield  {journal} {\bibinfo  {journal} {arXiv e-prints}\
  ,\ \bibinfo {eid} {arXiv:1612.01803}} (\bibinfo {year} {2016})},\ \Eprint
  {http://arxiv.org/abs/1612.01803} {arXiv:1612.01803 [math.PR]} \BibitemShut
  {NoStop}%
\bibitem [{\citenamefont {Zhou}()}]{TianciZhouprivate}%
  \BibitemOpen
  \bibfield  {author} {\bibinfo {author} {\bibfnamefont {Tianci}\ \bibnamefont
  {Zhou}},\ }\href@noop {} {\ }\Eprint {http://arxiv.org/abs/private
  communications} {private communications} \BibitemShut {NoStop}%
\bibitem [{\citenamefont {Halpin-Healy}\ and\ \citenamefont
  {Zhang}(1995)}]{ASEP}%
  \BibitemOpen
  \bibfield  {author} {\bibinfo {author} {\bibfnamefont {Timothy}\ \bibnamefont
  {Halpin-Healy}}\ and\ \bibinfo {author} {\bibfnamefont {Yi-Cheng}\
  \bibnamefont {Zhang}},\ }\bibfield  {title} {\enquote {\bibinfo {title}
  {{Kinetic roughening phenomena, stochastic growth, directed polymers and all
  that. Aspects of multidisciplinary statistical mechanics}},}\ }\href
  {\doibase https://doi.org/10.1016/0370-1573(94)00087-J} {\bibfield  {journal}
  {\bibinfo  {journal} {Physics Reports}\ }\textbf {\bibinfo {volume} {254}},\
  \bibinfo {pages} {215 -- 414} (\bibinfo {year} {1995})}\BibitemShut {NoStop}%
\bibitem [{\citenamefont {{Zhou}}\ and\ \citenamefont
  {{Nahum}}(2018)}]{zhou1804spin}%
  \BibitemOpen
  \bibfield  {author} {\bibinfo {author} {\bibfnamefont {Tianci}\ \bibnamefont
  {{Zhou}}}\ and\ \bibinfo {author} {\bibfnamefont {Adam}\ \bibnamefont
  {{Nahum}}},\ }\bibfield  {title} {\enquote {\bibinfo {title} {{Emergent
  statistical mechanics of entanglement in random unitary circuits}},}\
  }\href@noop {} {\bibfield  {journal} {\bibinfo  {journal} {arXiv e-prints}\
  ,\ \bibinfo {eid} {arXiv:1804.09737}} (\bibinfo {year} {2018})},\ \Eprint
  {http://arxiv.org/abs/1804.09737} {arXiv:1804.09737 [cond-mat.stat-mech]}
  \BibitemShut {NoStop}%
\bibitem [{\citenamefont {Calderbank}\ \emph {et~al.}(1997)\citenamefont
  {Calderbank}, \citenamefont {Rains}, \citenamefont {Shor},\ and\
  \citenamefont {Sloane}}]{calderbank1997quantum}%
  \BibitemOpen
  \bibfield  {author} {\bibinfo {author} {\bibfnamefont {A.~Robert}\
  \bibnamefont {Calderbank}}, \bibinfo {author} {\bibfnamefont {Eric~M.}\
  \bibnamefont {Rains}}, \bibinfo {author} {\bibfnamefont {Peter~W.}\
  \bibnamefont {Shor}}, \ and\ \bibinfo {author} {\bibfnamefont {Neil J.~A.}\
  \bibnamefont {Sloane}},\ }\bibfield  {title} {\enquote {\bibinfo {title}
  {{Quantum Error Correction and Orthogonal Geometry}},}\ }\href {\doibase
  10.1103/PhysRevLett.78.405} {\bibfield  {journal} {\bibinfo  {journal}
  {\prl}\ }\textbf {\bibinfo {volume} {78}},\ \bibinfo {pages} {405--408}
  (\bibinfo {year} {1997})},\ \Eprint {http://arxiv.org/abs/quant-ph/9605005}
  {arXiv:quant-ph/9605005 [quant-ph]} \BibitemShut {NoStop}%
\bibitem [{\citenamefont {Gottesman}()}]{DanielGottesmanprivate}%
  \BibitemOpen
  \bibfield  {author} {\bibinfo {author} {\bibfnamefont {Daniel}\ \bibnamefont
  {Gottesman}},\ }\href@noop {} {\ }\Eprint {http://arxiv.org/abs/private
  communications} {private communications} \BibitemShut {NoStop}%
\bibitem [{\citenamefont {{Wikipedia
  contributors}}(2018)}]{GaussianEliminationWikipedia}%
  \BibitemOpen
  \bibfield  {author} {\bibinfo {author} {\bibnamefont {{Wikipedia
  contributors}}},\ }\href
  {https://en.wikipedia.org/w/index.php?title=Gaussian_elimination&oldid=871552857}
  {\enquote {\bibinfo {title} {{Gaussian elimination --- {Wikipedia}{,} The
  Free Encyclopedia}},}\ } (\bibinfo {year} {2018}),\ \bibinfo {note} {[Online;
  accessed 15-January-2019]}\BibitemShut {NoStop}%
\bibitem [{Note4()}]{Note4}%
  \BibitemOpen
  \bibinfo {note} {In stating these results, the requirement that $A$ is
  contiguous is important. Consider the following example of $L=3$, \begin
  {eqnarray} \protect \mathcal {G} = \protect \{ XXI, IXZ, YZY\protect \}. \end
  {eqnarray} This set is in the clipped gauge. Let $A = \protect \{1, 3\protect
  \}$. $S_A$ can be shown to be $1$, while $|\protect \mathcal {G}_A| = 0 \not
  =|A| - S_A = 1$. Thus, this simple formula \protect \emph {cannot} be readily
  used for computation of the mutual information, \begin {eqnarray} I_{A, B} =
  S_A + S_B - S_{A \cup B}, \end {eqnarray} where $A$ and $B$ are qubits that
  could be far away.}\BibitemShut {Stop}%
\end{thebibliography}%


\end{document}